# An Unsupervised Learning Approach For A Reliable Profiling Of Cyber Threat Actors Reported Globally Based On Complete Contextual Information Of Cyber Attacks

**Sawera Shahid[1], Umara Noor[2], Zahid Rashid[3]**

[1]Department of Computer Science, Faculty of Computing and Information Technology, International Islamic University, Islamabad, Pakistan

[2]Department of Software Engineering, Faculty of Computing and Information Technology, International Islamic University, Islamabad, Pakistan

[3]Technology Management Economics and Policy Program, College of Engineering, Seoul National University, 1 Gwanak-Ro, Gwanak-Gu, 08826 Seoul, South Korea

sawera.mscs1109@iiu.edu.pk, umara.zahid@iiu.edu.pk, rashidzahid@snu.ac.kr[a]

***ABSTRACT***

*Cyber-attacks are rapidly increasing with the advancement of technology and there is no protection for our information. To prevent future cyberattacks it is critical to promptly recognize cyberattacks and establish strong defence mechanisms against them. To respond to cybersecurity threats immediately, it is essential to examine the attacker's skills, knowledge, and behaviors with the goal of evaluating their impact on the system and comprehending the traits associated with these attacks. Creating a profile of cyber threat actors based on their traits or patterns of behavior can help to create effective defenses against cyberattacks in advance. In the current literature, multiple supervised machine learning-based approaches considered a smaller number of features for attacker profiling that are reported in textual cyber threat incident documents although these profiles have been developed based on the security expert's own perception, we cannot rely on them. Supervised machine learning approaches strictly depend upon the structure data set. This usually leads to a two-step process where we first have to establish a structured data set before we can analyze it and then employ it to construct defense mechanisms, which takes time. In this paper, an unsupervised efficient agglomerative hierarchal clustering technique is proposed for profiling cybercriminal groups based on their comprehensive contextual threat information in order to address the aforementioned issues. The main objective of this report is to identify the relationship between cyber threat actors based on their common features, aggregate them, and also profile cyber-criminal groups. The performance of this approach is evaluated using multiple parameters such as the silhouette coefficient and Davies Bouldin index. A proposed framework achieves a 0.095 silhouette score and a 2.212 Davies Bouldin index. The outcomes of this study; the profile of cybercriminal groups will benefit security analysts and security firms that offer security services in creating effective defence strategies to proactively counter these cyberattacks.*

## 1. Introduction

Today, the internet has become unavoidable in our daily lives. "The internet is becoming the town square for the global village of tomorrow". It is not just a need or luxury; it has become a household necessity. With the rapid advancement of technology, cybersecurity has become one of the major concerns for all organizations, governments, and other stakeholders. Cyber incidents can have serious consequences. The theft of private, financial, or other sensitive data and cyber-attacks that damage computer systems are capable of causing lasting harm to anyone engaged in personal or commercial online transactions. Such risks are increasingly faced by businesses, consumers, and all other users of the internet if we look up the statistics of that present moment after every 39 seconds, there is a new attack launched somewhere on the web[1]. In 2022 432 million cyber-attacks were reported due to these cyber-attacks people are facing $1,90000 financial loss after every second [2]. The Twitter hack incident is one of the most well-known attacks these days, with nearly 125 Twitter accounts being hacked[3]. Also, in June 2022 a former Amazon employee, Paige Thompson, working for Amazon web services(AWS) exploited her knowledge of cloud server vulnerabilities and stole the personal information of over 100 million people[4]. The consequences of these attacks affect not only the organizations but also the entire world. These types of attacks are occasionally carried out by advanced persistent threat actors typically nation-state or state-sponsored groups, which gain unauthorized access to a computer network and remain undetected for an extended period. The possession of specific characteristics by APT groups makes their detection very difficult. Previous attempts and studies suggest and use the concept of cyber threat attribution[5-14] in which they associate targeted cyber-attacks against a cyber threat actor based on some characteristics[15]. Cyber security experts and security repositories have identified several cyber threat groups. They also classify them on the basis of attack patterns however, this is based on their security experts' own perceptions. We don't have this reliable information. We cannot rely on it. In a cyber threat environment, different sorts of cyber threat actors exist. There is a chance that the security community has identified two cyber criminals individually, but they could be the same cyber threat actor. They have some sort of connection or interaction. We want to break the loop or put an end to this grouping concept. Unfortunately, in the cyber world, we cannot physically reach the cybercriminal. The identity of the cybercriminal is not very crucial. The reason is once an incident happens after the identity of the cybercriminal is revealed but we are unable to restore the harm inflicted by that attack. So, from a defence perspective, the concept of attributing the cyber threat based on some characteristics is not really that important. To prevent future cyber-attacks in the cyber security world there is a need to timely respond to cybersecurity attacks and employ defence or mitigation strategies against them. In order to react promptly to cybersecurity assaults requires observing the attacker's knowledge, skills, and behaviors to examine their influence over the system and understand the characteristics associated with these attacks. Profiling as stated by[16] involves the analysis of personal characteristics or behavioral patterns which allows an investigator to generalize about a person or a crime scene. Creating a profile of cyber threat actors based on their complete context of attack can assist in developing effective security mechanisms against cyberattacks in advance. An interesting association between these cyber threat actors can be easily identified by grouping them together based on their common features by using the clustering technique. We have developed profiles of cyber-criminal groups that reflect a particular pattern of incidents on the basis of complete contextual information. These profiles will be used by the security community in order to develop effective defence mechanisms against future attacks. Further studying the literature, we have observed that multiple supervised machine learning-based approaches considered a small number of features(behavioral patterns, incidents contextual information ) for attacker profiling that are reported in textual cyber threat intelligence documents. Cyber threat intelligence (CTI) is an area of cybersecurity that focuses on the collection and analysis of information about current and potential attacks that threaten the safety of an organization[17]. This is an informal way of sharing a piece of information about cyber-attacks. Cyber threat intelligence is usually conveyed by structured data like structured threat information expression(STIX)[18]; It is a threat representation language that is designed to share a complete threat incident context

using eight core elements; indicator, observable, incident, TTPs, exploit target, campaign, threat actor and course of action[19] and unstructured data like security blogs analyzed by cyber security analyst and reports that contain complete contextual threat information. Supervised machine learning-based approaches are costly in terms of structuring as well as labeling datasets. These approaches strictly depend upon the structured data set usually it becomes a two-step process, first of all, we have to create a structured data set and then we have to analyze it in order to develop effective mitigation strategies which takes time, but we are readily available with unstructured textual cyber threat incidents report so for this purpose there is need to use unsupervised learning method. The main objectives of this research study are to identify the relationship between cyber threat actors based on their common features and aggregate them using a clustering-based approach named as efficient agglomerative hierarchal clustering[20].We also Profile cyber-criminal groups based on their contextual threat information. One of the major challenges that our world is facing nowadays is the occurrence of cyber-attacks in a massive amount. But unfortunately, existing techniques and defence mechanisms have failed to overcome this challenge. This study will help to prevent future cyber-attacks in the cyber security environment. The findings of this study; profiles of cyber criminals will help the security analyst as well as security organizations that provide security services to develop effective defence strategies to proactively deter these cyber-attacks. To do this, we intend to address the following research questions (RQs):
RQ1: Which sources we will consider for collecting our data?

RQ2: Which technique can be used to profile cyber-criminal groups using textual cyber threat incident documents?
RQ3: Which other characteristics of cybercriminals can be considered for profiling?

Our research has the following contributions:

1) To find interesting associations between cyber threat actors, we collected unstructured cyber threat incident documents of 129 cyber threat actors that globally will cover the events from May 2012 to June 2023.

2) This work is the first of its kind that presents a circumstance for emphasizing the profiling of cybercriminal groups. Previous work has mainly focused on profiling cyber threat criminals and associating a targeted cyber-attack against a threat actor based on limited features.

3) This research work will help the security analyst or security organization to develop effective defence mechanisms or mitigation strategies against the profiles of cybercriminals groups.

The structure of the paper is arranged as follows: The research study's background is covered in the second section, and certain research scholars' related work is the subject of the third section. Our research methodology is contained in the fourth component. A synopsis of the results is given in the fifth section, cybercriminal profiles are covered in the sixth, and research findings are covered in the seventh. The conclusion and future work are presented in the final section.

## 2. Background

In the current literature, multiple techniques are proposed for cyber-criminal profiling. They considered limited incident contextual information. Their threat information is confined to a specific domain(malware, vulnerability) while phases of carrying out a complete attack are missing. They exclusively concentrate on profiling individual cyber threat actors; they are not profiling cybercriminal groups. In this research study our primary goal is to profile cyber-criminal groups based on complete contextual threat incident information. This information contains eight core elements; indicator, observable, incident, TTPs, exploit target, campaign, threat actor, and course of action[19]. The CTI documents present in the data corpus will be of different kinds. The kinds of CTI are classified under four categories i.e., strategic, operational, tactical, and technical threat intelligence. The strategic CTI is about the identification and impact analysis of risks in the form of high-level information consumed by decision-makers. The operational CTI is about the latest vulnerabilities and zero-day attacks collected from closed hacker forums or the dark web. The tactical CTI is about TTPs of Cyber threat actors. They are the modus operandi of cyber threat actors and an integral part of their training. The TTPs represent the behaviour of CTA when interacting with the victim's resources such as operating system and network while the technical CTI is about the indicators employed at technical resources such as firewalls and intrusion detection systems.

## 3. Related Work

In current literature, multiple supervised machine learning-based approaches considered a smaller number of features; behavioral patterns, and incident contextual information for attacker profiling that are reported in unstructured textual cyber threat incident documents. However, supervised-based approaches are costly in terms of structuring as well as labeling data. The Cybersecurity Information Sharing Act (CISA), approved by the US Congress in 2015[21], mandates that businesses (including financial institutions) involved in cybersecurity events share cyber threat intelligence (CTI) with other important parties, particularly their own clients[22].If an association can impart its own knowledge of what it experienced about a threatening event in the context of a cyber threat indicator, another association may be able to employ this information to develop appropriate defence mechanisms against both present and future threats. The dependence on deterministic and heuristic-based petitions, which are incapable of defending against dynamic and intricate threats, is a significant constraint of prevailing CTI sharing information approaches[23] because existing cyber threat intelligence consists primarily of malignant IPs, URLs, and term hashes; however, this represents only a tiny fraction of the threat intelligence while on the other hand, these indicators are also counterproductive for recognizing cyber threat actor(CTA) as they are reconfigured each time by the CTA to circumvent firewall and intrusion detection system. Besides that, the above category of threat intelligence offers no background info on the intrusion. As a result, dependence on this set of information to investigate cyber-attack exercises and categorize occurrences becomes incredibly difficult. According to Liao et al., threat intelligence is scattered across numerous locations, time frames, and intelligence artifacts[24]. Additionally, ENISA revealed that unified threat intelligence (TI) was typically too complex or vast to be put to use [25]. The work of security analysts, who must select usable intelligence, is hampered by this information overload, among other detrimental effects [26]. Both of these problems make the analyst's job more difficult and raise the possibility that crucial information may be missed.

Azevedo et al. suggest expanding the resources from which they gather intelligence (enhancing coverage) and analyzing this information (filtering and aggregation) in accordance with predetermined criteria in order to overcome this conundrum. Intelligence will be the outcome of their interpretation. They also serve as a way for improving open-source threat intelligence(OSINT) processing to yield sufficient threat information in the form of enriched indicators of compromise (IoCs )by representing each threat in a single enriched IoC, which reduces the amount of data that security analysts must evaluate. This improved intelligence is obtained by correlating and merging IoCs from various OSINT feeds that contain data on the same danger, grouping those data into clusters, and then portraying the threat information contained in those clusters in a single enriched IoC. However, they do not specifically state what kind of enriched IOCs they are extracting [27] and information is manually gathered. As the amount of CTI data is continually increasing and only one publicly accessible threat source has roughly 1 billion threat indicators, so manual extraction is hence not viable [28] since "Time is money."Several supervised learning-based methods have been put out in recent years to efficiently retrieve threat intelligence. This research was presented by Kadoguchi et al.[7] uses machine learning models to many dark web forum posts in order to locate forum posts containing threat information they mostly focus on locating forum posts about virus offers. They successfully classified it with 97% accuracy, but they only take malware threats into account the phases of carrying out a full attack are missing, and the ideal parameters are manually adjusted through trial and error. Therefore, a technique for automatically locating the ideal parameters based on the data to be categorized is necessary. A system based on ARM was proposed by Noor et al. [23] for profiling regularities in cyber threat actor (CTA) TTPs. It is made up of the feature extractor, feature selector, and ARM miner modules. However, the proposed methodology required manual feature extraction and data set transformation into unstructured form, requiring human intervention. Conversely, the association will only be chosen based on TTPs. In [29], they presented the domain-oriented topic discovery based on the features extraction and topic clustering approach (DTD-FETC), which mainly consists of modules for data preparation, feature extraction, and topic clustering. They look at the open-source web for a certain domain and categorize new topics as they emerge. Additionally, they gather and organize web content for specific security threat themes like malware and vulnerabilities. The system then uses named entity identification, topic word feature extraction, and keyword feature extraction to extract different kinds of features. The feature vector for the article is created by fusing the three different sorts of features. Finally, the system applies an updated hierarchical clustering technique to cluster the feature vectors of articles in each period, enabling real-time topic identification of historical or emergent topics. According to the experimental findings, DTD-F1, FETC's precision, and recall are, respectively, 0.996, 0.998, and d 0.992. However, their methodology requires more computing time than the single feature extraction method, and the data set is limited to a certain domain (Malware, vulnerability), these entities are unable to fully characterize a security event. They also lack the phases of carrying out a full assault. Using process mining techniques, Rodrguez et al.'s[30] framework for profiling automated attackers (malware) and building models from their behaviour was put out. They investigate the relationship between malware, and an attacker's data, and the methods for identifying and simulating its activity. In essence, they concentrate on malware profiling rather than creating profiles for cyber threat attackers. In addition, Rodrguez et al. do not offer any empirical dataor real-world testing for their proposed paradigm. An approach is put out in[31]that makes use of multilevel machine learning models to identify the expression of two traits- risk aversion and expertise level—based on the adversary's tool usage and tool selection.

In order to gather the needed empirical evidence for the development of such attacker personas, Brynielsson et al. [32] proposed a framework in which they used attacker personas as a technique for attacker profiling and the design of specifically tailored cyber defence exercises. They also offer an analysis of how attacker personas can be used to improve situational awareness within the cyber domain. The main goal of the research that went intothe development of these profiles is to increase threat awareness through a better understanding of one's own strengths and weaknesses. However, the validity of this information is still in doubt because it's possible for respondents to either under or overestimate their own knowledge and skills, which would result in an inaccurate picture. An ontological method is suggested in [33]for automatically identifying threat actor types based on their personas, understanding their nature, and recording polymorphism and changes in their behavior and attributes across time. They specifically showed how a collection of characterization traits can enhance knowledge of threat actors and, when combined, can enumerate different types of threat actors. They showed how, by incorporating this knowledge into an ontology, it was possible to automatically use deductive reasoning to infer the character of a perpetrator while hiding the relations and semantics that support the inference. The work's main objective is to categorize the various cyber threat actors. Based on the limitations of the existing literature, we draw the conclusion that classification-basedapproaches are unreliable for attacker profiling because the topic, as well as the contextual information of threat intelligence, changes every day and there aren't enough publicly available labeled datasets. As a result, we suggested an unsupervised clustering-based method that does not require any form of labeling data set for figuring out the relationships between cyber threat actors. The legitimacy and credibility of a data source, however, are crucial factors in the analysis process, making it an important indicator. Most research relieson Internet discussion boards and open-source websites [23] without considering the reliability of the data. We used cyber threat incident records, which have data that has been examined by cyber security professionals, to get around this restriction.

## 4. Research Methodology

To overcome the above problems, we will propose a new technique for profiling cyber-criminal groups based on their complete contextual threat information. To aggregate cyber threat incidents /actors based on their common characteristics present in them, we proposed a clustering-based approach for determining the relationship between cyber threat actors that does not require any kind of labeling data set. In this part, the research methodology is covered. Three separate phases exist. 1) Data gathering 2) Data preparation 3) The suggested framework. The three subphases of data preparation are data pre-processing (tokenization, stemming, and Stopwords removal), calculating Tf-Idf vectorizer, and similarity measures. The details of data collection are covered in the first section. This section goes over the specifics of data collection and the sources from which data is gathered. Implementation of pre-processing functions, including tokenization, stemming, and stop word removal, a measure of similarity among each document and the other document in the corpus can be generated by measuring Cosine and Jaccard similarity measures against the TF-IDF matrix, which is covered in the second part. In addition, the corpus is converted into vector space using the TF-IDF vectorizer, while the third section illustrates all of the necessary details related to the proposed framework. The whole research methodology is implemented using the Python programming language. The overview of the proposed technique is shown in Figure 1 and the detailed methodology is explained in the following sections.

### 4.1 Data Acquisition

To find interesting associations or links between cyber threat actors we collected unstructured cyber threat incident documents of 129 cyber threat actors that globally will cover the events from May 2012 to June 2023. Using a customized search engine, reports on cyber threat intelligence are gathered [34]. This platform was created specifically to identifycyber threat actors, highly persistent attacks, and other malware in December 2015. Cyber threat intelligence materials are also gathered and stored in a public repository [35]. Using search engines and several repositories, 1200 reports related to cyber threat actors were gathered. By considering the factor of data reliability we collect all the data from reliable threat sources as shown in Table 1.

### 4.2 Data Preparation

This section is focused on defining some functions to manipulate the document corpus. We will apply the following pre-processing operations to the corpus in order to improve the efficiency or quality of data we convert raw data into a format that is practical and easy to understand. Tokenization, stemming, and stop word elimination are all a part of the data pre-processing phase. The tokenization function is implemented by dividing the raw text into words and sentences that we can refer to as tokens because we need an individual meaning for an entity to work upon and in the text, it cannot be a character, or it cannot be a complete sentence, but each individual word holds a meaning. By importing the snowball stemmer which is actually part of the NLTK library we reduce a word to its word stem by attaching itto suffixes, prefixes, or the roots of words known as a lemma. In the third phase, we load NLTK's list of English stop words. We removed those words that occur commonly across all the documents in the corpus or words like "a", "the", or "in" which don't convey significant meaning.

### 4.3 Calculating TF-IDF and Document Similarities

To determine how relevant those words are given to the document here, we create a TF-IDF matrix using CTI records. When working with text data, it is essential to fit your text data into a TF-IDF vectorizer. For machine learning algorithms, which often require numerical data, text input in its raw form is not adequate. Text data can be transformed into numerical feature vectors using the TF-IDF vectorization technique, making it suitable for use with machine learning models. When we transform the raw text into tokens or terms in the preceding phase, we are essentially creating a list of the features that will be used in the TF-IDF matrix. First, we count word occurrences throughout the document to create a TF-IDF matrix. Using this, a document term matrix is created (DTM). Then we use a technique known as frequency-inverse document frequency weighting, where words that frequently appear ina document but not commonly in the corpus are given greater weight because they have a deeper meaning related to the content. There are a few things to be aware of regarding how the parameters are defined in our research methodology: The maximum frequency that a given feature may appear inside the documents is indicated by the parameter max- df in theTF-IDF matrix. If a phrase appears in more than 80% of the texts, its meaning is probably unimportant (in the context of CTI documents). To provide a measure of similarity betweeneach document and the other texts in the corpus, Cosine and Jaccard similarity measures are computed against the TF-IDF matrix. In n-dimensional space, Cosine similarity calculates the cosine of the angle between two non-zero vectors and by contrasting the size of two sets' intersection and union, Jaccard similarity calculates how similar they are.

In order to determine which method performs best on the data corpus, their outcomes are compared after the implementation of these two similarities.

**Table 1** Cyber Threat Sources

| SR.No | Sources | Description |
|---|---|---|
| 1 | Security Vendors | Fireeye[36],Rapid7[37],Clearskysec[38],alivenvault[39],trendmicro[40], Symantec[41],Palotonetworks[42],Arbor[43],Pwc[44],Zscalar[45],crowd strike[46],Proofpoint[47],Kaspersky[48],digital shadows[49],Novetta[50],Fox-it[51],Malware bytes[52],Group-IB[53],Cisco[54],Looking Glass Cyber solutions[55],Checkpoint[56],Inetresec[57],RiskIQ[58] |
| 2 | Security Blog | Info security magazine[59],Security affairs[60], Krebsonsecurity[61],Welivesecurity[62],Securelist[63],Microsoft security blog[64] |
| 3 | Cyber Security News and Information Platform | Threat Connect[65],ThreatPost[66] |
| 4 | US-CERT Portal | Cyber security and infrastructure security agency[67] |

## 4.4 Proposed Framework

One should be able to gather information on attack observables, indicators, patterns, incidents, threat actors, campaigns, exploit targets, and courses of action from the CTI documents that were received during the data-collecting phase. These documents cannot be easily or immediately analyzed by machines because of their unstructured nature. We suggest an automated method for categorizing cybercriminal organizations using CTI documents with unstructured text. Agglomerative hierarchical clustering, out of all the algorithms described in the literature, is the greatest choice for identifying intriguing relationships among cyber threat actors utilizing CTI data. Its main premise is that it groups objects or documents based on their differences. The type of dissimilarity can be appropriatefor the topic being examined and the type of data being used. There are some restrictions with the algorithm. It does not scale well because of its time complexity, which is at least $O(N^2 \log N)$, where N is the total number of objects. To get around this restriction, we are using the K-means clustering algorithm before hierarchical clustering because of its time complexity. The underlying rationale for applying this concept to the bottom-up approach (Agglomerative Hierarchal Clustering) is that the computational cost can be decreased if thebottom-up process begins somewhere in the middle of the hierarchy and the lower part of thehierarchy is built using a less expensive technique like partitional clustering. Divisive hierarchical clustering, a top-down approach, is renowned for its high cost, $O(2^N)$, and confirming middle-level sub-clusters by individual data points would still be expensive, therefore this notion would not work well with it. In order to successfully aggregate cyber criminals based on their common features using CTI documents, the output of the K-meansalgorithm will be used as input to agglomerative clustering. K-means clustering is applied tothe pre-processed data corpus using the different distance measures to group data points into clusters based on their similarities and dissimilarities. Distance measures like Euclidean, Manhattan, Canberra, and Minkowski are considered in this approach. In K- means clustering, we first choose the centroid points at random, and then we allocate each data object to the centroid point that is closest to it.

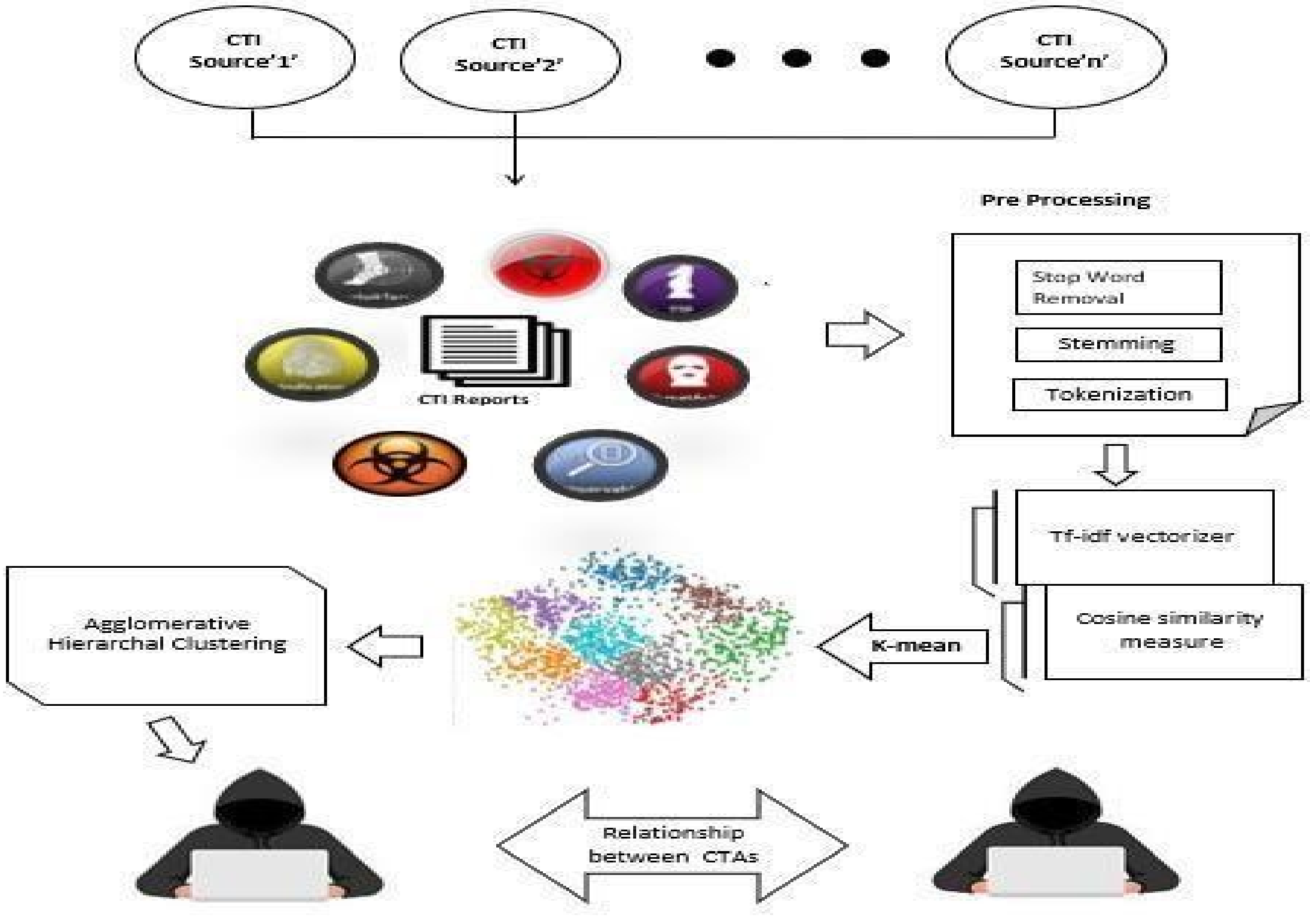

**Figure 1** Proposed Framework

Computing the central points of these recently formed clusters and revising the calculation of centroid points. This process is repeated until no data point is assigned to another cluster. For determining an optimal number of clusters, an elbow method is utilized. In this method, we continuously iterate for k=1 to 20. For every value of k, we calculate the within-cluster sum of squares (WCSS value). k values are plotted on the x-axis of Figure 2 with the appropriate WCSS on the y-axis. The elbow Method is used to locate the plot point when the rate of WCSS decline abruptly changes. This location seems to have the ideal number of clusters. After analyzing the graph, we determined that 12 is the ideal value for the K-means clustering algorithm. The method's outcome will be K clusters referred to as middle-level clusters.

### 4.5 Mathematical Prototype

The stepwise details related to the implementation of the proposed framework named as efficient agglomerative hierarchal clustering are given in algorithm1. Here pre-processing method repeats this process for each element in the D.C[$D_{i.....n}$] input document collection. In this function, D.C[$D_{i.....n}$] represents the document corpus that contains all the textual documents recognized by [$D_{i.....n}$]. To enhance the quality of data pre-processing functions; tokenization, stemming, and stop word removal are applied against all the documents present in the document corpus. In step 1 T.K denotes the tokenization function which is applied against the document corpus after breaking all the raw text into tokens. This function's output is assigned to an array entitled as P.Proc[$T_{i.....n}$]. The output of this step will be input towards another pre-processing operation, stemming symbolized by S.T.M. Step 2 will allocate each stemmed word to P.Proc[S.$T_{i.....n}$]. In step 3 S.T.R identified stop words removal operation is applied against the data stored in P.Proc[S.$T_{i.....n}$]. The outcome of this pre-processing function will be kept in P.Proc[S.$R_{i.....n}$]. The array P.Proc[S.$R_{i.....n}$] pre-processed text will be allocated to C.T[$w_{i.....n}$]. For each document and every phrase C.T[$w_{i.....n}$] in the processed collection. The value of TF-IDF (Term frequency-inverse document frequency) is calculated by the vectorizer function. After applying the TF-IDF vectorizer designated by TDIDFV to the pre-processed text in the resultant we get a two-dimensional matrix denoted by TDIDFM[T.M $_{i.....n}$][$D_{i.....n}$] that preserves all the words that occur frequently within a document but not frequently within the corpus. The compute document similarity function goes over the documents in the TF-IDF matrix repeatedly. S and R are two vectors in Cosine Similarity (C.S) and Jaccard Similarity (J.S) metrics, where it calculates document similarities and returns the distance matrix. Here dist in the K-means clustering function stores the similarity value that is calculated against the TF-IDF matrix to yield an indication of how similar each document is to the other document present in the corpus. K-means clustering is applied to the resultant of the previous step. This stage will result in K- clusters, which are represented by K.C[$K_{i.....n}$], which contain n clusters where each cluster contains all the similar documents with respect to other clusters Here AGNES represents agglomerative clustering applied on K -Clusters. F.H[$C_{i.....n}$] contains the final hierarchy of clusters. A detailed discussion of K-means and agglomerative hierarchal clustering functions are presented in algorithms 2 and 3.

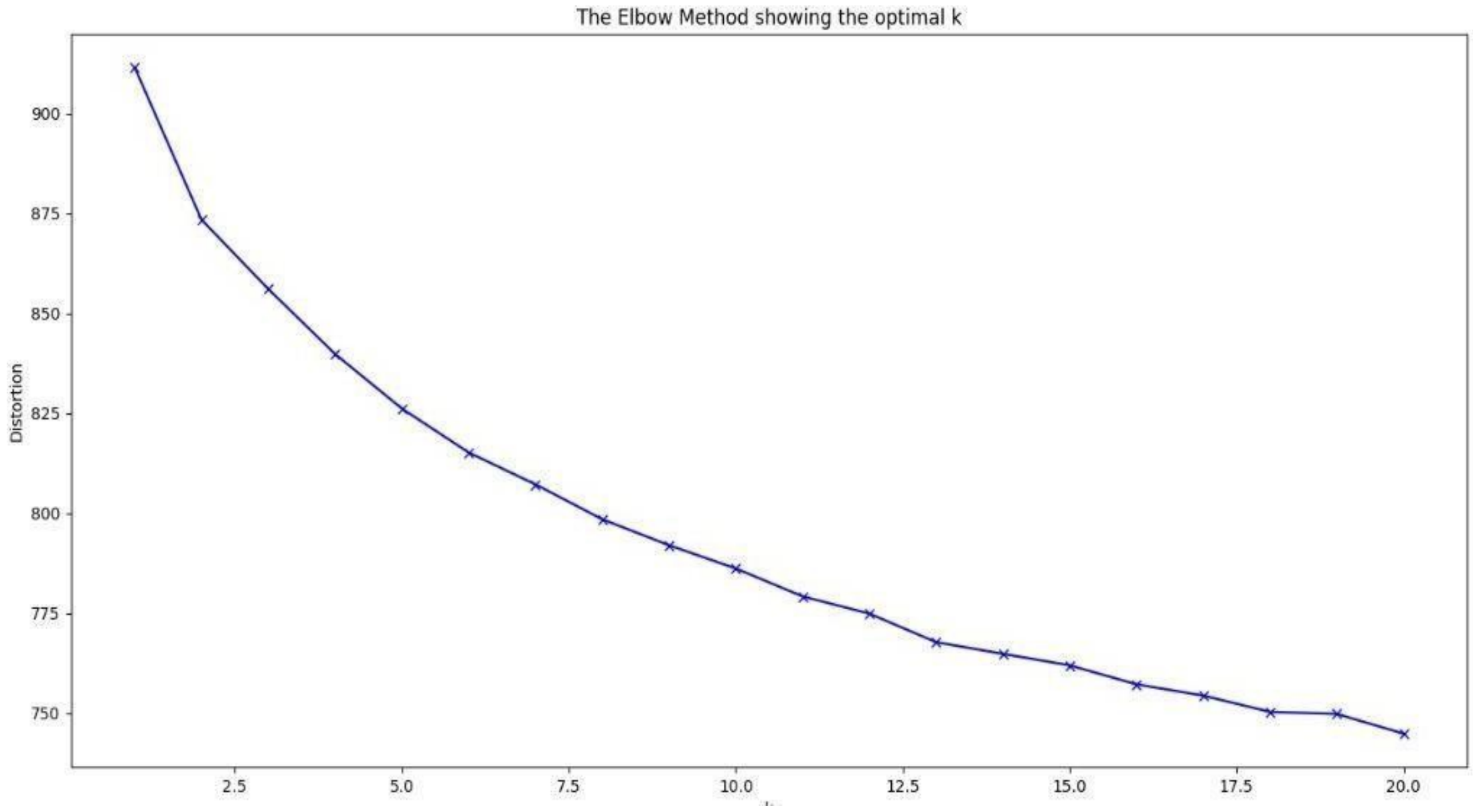

**Figure 2** Elbow Method's Result

---

**Algorithm 1** Efficient Agglomerative Hierarchical Clustering

---

**Input:** D.C[$D_{i......n}$]={d1,d2,…....,dn}

**Output:** F.C[$C_{i......n}$]={c1,c2,…,cn}

**Begin**

**function** Pre-Processing(D.C[$D_{i.......n}$])

**for** i=1 to dn **do**

P.Proc[$T_{i.....n}$]= T.K(D.C[$D_{i......n}$])

P.Proc[S.$T_{i.....n}$]= S.T.M(P.Proc[$T_{i......n}$])

P.Proc[S.$R_{i.....n}$]= S.T.R (P.Proc[S.$T_{i......n}$])

$C.T[w_{i.....n}]$= $P.Proc[S.R_{i.....n}]$

**end for**

**return** $C.T[w_{i.....n}]$

**end function**

**function** TF-IDF Vectorizer($C.T[w_{i.....n}]$, $D.C[D_{i.....n}]$)

**for** j=1 to wn **do**

TF-IDF ($C.T[w_{i.....n}]$, $D.C[D_{i.....n}]$)=TF($C.T[w_{i.....n}]$, $D.C[D_{i.....n}]$)*IDF($C.T[w_{i.....n}]$)

IDF(C.T[wi…..n])=log[n/df(C.T[wi…..n])

**end for**

**return** TDIDFM[T.M i…..n][Di…..n]

**end function**

**function** compute document similarities (TDIDFM[T.M i…..n][Di…..n])

**for** k=1 to dn do

$C.S=\frac{\sum SiRi}{\sqrt{\sum Si2}\sqrt{Ri2}}$

$J.S=\frac{|S\cap R|}{|\square\square|}$

**end for**

**return** dist

**end function**

**function** K-Means Clustering(dist)

$K.C[K_{i.....n}]$= K-Means(Dist) using algorithm2

**end function**

**function** Efficient Agglomerative Hierarchal Clustering($K.C[K_{i.....n}]$)

$F.H[C_{i.....n}]$= AGNES($K.C[K_{i.....n}]$) using algorithm 3

**end function**

**End**

---

In the K-means clustering algorithm, the clusters are initialized using the given centroids at the beginning of the procedure. Each document is then iteratively assigned to the cluster whose centroid is nearest. It adjusts the centroids in accordance with any changes to the assignments. Up until there are no more reassignments, the process is repeated while algorithm 3 focuses on the detailed description of the agglomerative hierarchal clustering function. Initializing a matrixMx and a distance matrix dist(di,dj) is the first step in the process. The distance matrix is then updated as iteratively goes over pairs of indices (di,dj) and computes the similarity between vectors v di and v dj. The least_value, least_row, and least_col variables are updated in accordance with the pair that has the smallest distance between them It sets the matrix

Depletion M.D. to 0 and the cluster size S.C. to Mx. It then goes into a loop that keeps going until thecluster S.C.'s size drops to a predetermined stop criterion.

**Algorithm 2** K-Means Clustering

**Input:**

Docs={doc1,doc2,doc3,docn}

N=No of clusters

**Output:**

K={k1,k2,k3,kn}

**Begin**

**1:** Input $N^{th}$ clusters to be allocated.

**2:** Read $N^{th}$ clusters centroid.

**3: Do**

**4:** Assign each point to its closest centroid.

**5: IF(**Pre-cluster points!=new cluster points) **THEN**

**6:** Computenewcentroid of clusters

**7: END IF**

**8: ELSE**

**9:** Centroid remains the same

**10: END ELSE**

**11: WHILE(**Re-assignment of points occur in cluster)

**Finish**

Within the loop, it increases the matrix depletion M.D., decreases the cluster size S.C., and merges the vectors with the least distance (as indicated by least_row and least_col). It ensures that the merged cluster has the proper distances from other clusters by updating the distance matrix after merging. Until the cluster size S.C. meets the stop requirement, theoperation is repeated.

**Algorithm 3** Agglomerative Hierarchal Clustering

**Begin**

**1:** Distance=dist(di , dj)

**2:** Matrix Mx={1,………..,Mx1,1, ,Mx1}

**3: for** di = 1 to Mx1 **do**

**4:** **for** dj = di to Mx1 **do**

**5:** dist(di,dj) =Compute Similarity(vdi,vdj) dist(dj, di)

```
            IF (dist(di,dj) <least_value)THEN
    {
    least_value=dist(di,dj)
    least_row=di
    least_col=dj
    }
          ENDIF
      end for
end for
Size of cluster S.C =Mx
Matrix Depletion M.D = 0
WHILE (S.C >cluster stop )
    {
        MERGE(Vleast_row, Vleast_col)
         S.C=S. C -1;
         M.D=M.D+1;
     }
    for di=1to Mx- M.D       do
            for dj=di to Mx- M.D do
                    IF di=least_row& dj=least_col THEN
                     {
                             dist(dj , di)=dist(di,dj)
                     }
                    END IF
                   IF( dist(di,dj) <least_value) THEN
                   {
                      least_value=dist(di,dj)
                      least_row= di
                       least_col =dj
                     }
                    END IF
```

```
            end for
        FOREACH dist(di,dj) do
                distance stop=Stopcriterion(dist(di,dj))
    end for
End
```

## 5. Experimental Evaluation

The evaluation of the proposed framework is presented in this section. The efficient agglomerative hierarchal clustering approach is evaluated specifically for its ability to identify a relationship between cyber threat actors based on their common features. Silhouette coefficient and Davies Bouldin index are used as evaluation parameters. The performance of the proposed approach is compared with standard K-means and agglomerative hierarchal clustering algorithms. The technique is implemented on personal computer which have Intel(R)Core(TM) m3- 7Y30 CPU @ 1.00GHz 1.61 GHz processor, 8.0 GB RAM, 64-bit operating system, x64- based processor, and Windows 10 pro. Python programming language is used to implement this research methodology. A Clustering-based proposed approach is evaluated by employing similarity and dissimilarity measures such as distance between cluster points. It involves grouping of data points unlike regression and classification we don't have the target variable in clustering. It lies under the category of an unsupervised machine learning-based approach so we can't calculate errors or accuracy or any one of these metrics. To assess the suggested framework, we employed the Silhouette coefficient and the Davies Bouldin score as evaluation measures. The proposed approach's performance is contrasted with the outcomes of standard K-means and agglomerative hierarchical clustering techniques. In the K-means clustering algorithm, we utilized different distance measures to group data points into clusters based on their similarity and dissimilarities. This method incorporates distance metrics such as Euclidean, Manhattan, Canberra, and Minkowski while in agglomerative hierarchical clustering different linkage criteria; ward, single, complete, average, and centroid are available in order to determine how clusters are merged during the hierarchical clustering process. We use all the aforementioned criteria in conjunction with several distance measures like Euclidean, Manhattan, Canberra, and Minkowski in order to achieve the results of hierarchical clustering. To identify links between different cyberthreat actors proposed technique named efficient agglomerative hierarchal clustering is implemented with different linkage criteria. In this case, valid options in our dictionary are ward, single, complete, and average while results using centroid are not applicable because we are already passing centroid as a parameter in K-means clustering when combining K-means and agglomerative hierarchal clustering approaches. From all three of these methods, we select the finest outcomes. Lastly, contrast the outcomes of the three clustering techniques to determine which one is most effective at combining and separating similar and dissimilar data points. We also implemented two similarity measures, such as Cosine and Jaccard, to offer a measure of similarity between each data point; the document and the other text in the corpus. We examine the findings of these two measurements to figure out which is most effective for our data corpus. An in-depth examination of evaluation results is presented below.

## 5.1 Silhouette Coefficient

It is a quantitative indicator that assists in determining the quality of clustering solutions. When compared to other clusters (separation), the silhouette coefficient for a single data point expresses how similar to its own cluster (cohesion) it is. It can be anywhere between -1 and +1, with +1 denoting that the data point is entirely within its own cluster and far from any nearby clusters. Data points with a value of 0 are on or very near the decision boundary between two adjacent clusters. A value of -1 suggests that a data point may belong to the incorrect cluster because it is located closer to a neighboring cluster than to its own. To get a final score for the entire clustering, the silhouette coefficient is calculated for each data point in the dataset and then averaged. The following is the formula for the silhouette coefficient of a data point 'i':

$$S(i)=\frac{a(i)-b(i)}{\max(a(i),b(i))} \qquad eq(1)$$

In eq( 1) intra-cluster distance, or a (i), is the average distance between each data point in a cluster. The shortest average inter-cluster distance, or b (i), is the distance between data points i and all other data points in all clusters of which i is not a member. Each data point is closer to its own cluster than to neighboring clusters when the silhouette coefficient is high (near +1), indicating that the data points are well-clustered. The presence of overlapping clusters or data points on or near the boundary between clusters is indicated by a silhouette coefficient that is close to 0. Data points may have been misclassified if the silhouette coefficient is negative (closer to -1). The silhouette score of all three methods is compared separately in Tables 2-5.

**Table 2** Result comparison of Clustering Algorithms using the Silhouette coefficient with Cosine Similarity

| Combinations | **Silhouette Score** | | |
|---|---|---|---|
| | Standard K-Means Clustering | Standard Agglomerative Clustering | Efficient Agglomerative Hierarchal Clustering |
| Cosine similarity, Euclidean, Ward, Silhouette coefficient | **0.055** | -0.214 | 0.09022569900359145 |
| Cosine similarity, Manhattan, Ward, Silhouette coefficient | 0.039 | -0.222 | 0.09022569900359145 |
| Cosine similarity, Canberra, Ward, Silhouette coefficient | 0.036 | **-0.173** | 0.09022569900359145 |
| Cosine similarity, Minkowski, Ward, Silhouette coefficient | **0.055** | -0.214 | 0.09022569900359145 |
| Cosine similarity, Euclidean, Single, Silhouette coefficient | **0.055** | **-0.152** | **0.095247046630965** |
| Cosine similarity, Manhattan, Single, Silhouette coefficient | 0.039 | -0.174 | **0.095247046630965** |
| Cosine similarity, Canberra, Single, Silhouette coefficient | 0.036 | -0.153 | **0.095247046630965** |
| Cosine similarity, Minkowski, Single, Silhouette coefficient | **0.055** | **-0.152** | **0.095247046630965** |
| Cosine similarity, Euclidean, Complete, Silhouette coefficient | **0.055** | -0.223 | 0.09213236026187827 |
| Cosine similarity, Manhattan, Complete, Silhouette coefficient | 0.039 | -0.233 | 0.09213236026187827 |
| Cosine similarity, Canberra, Complete, Silhouette coefficient | 0.036 | **-0.173** | 0.09213236026187827 |
| Cosine similarity, Minkowski, Complete, Silhouette coefficient | **0.055** | -0.223 | 0.09213236026187827 |
| Cosine similarity, Euclidean, Average, Silhouette coefficient | **0.055** | -0.190 | 0.09411817632067011 |
| Cosine similarity, Manhattan, Average, Silhouette coefficient | 0.039 | -0.198 | 0.09411817632067011 |
| Cosine similarity, Canberra, Average, Silhouette coefficient | 0.036 | **-0.173** | 0.09411817632067011 |
| Cosine similarity, Minkowski, Average, Silhouette coefficient | **0.055** | -0.190 | 0.09411817632067011 |
| Cosine similarity, Euclidean, Centroid, Silhouette coefficient | **0.055** | -0.104 | N.A |
| Cosine similarity, Manhattan, Centroid, Silhouette coefficient | 0.039 | -0.112 | N.A |
| Cosine similarity, Canberra, Centroid, Silhouette coefficient | 0.036 | **-0.088** | N.A |
| Cosine similarity, Minkowski, Centroid, Silhouette coefficient | **0.055** | -0.104 | N.A |

Table 2 appears to display various configurations of the K-Means clustering method and distance metrics (Cosine similarity, Euclidean, Manhattan, Canberra, and Minkowski). Each combination has a silhouette coefficient value that represents the efficiency with which the K-Means algorithm was able to produce clusters using that particular distance metric. The values of the silhouette coefficients that are given for each combination show how well the clusters were produced. In contrast, a lower value indicates that the clusters may be less distinct or poorly formed. A higher silhouette coefficient indicates that the clusters are well-separated and well-defined. In this table, two combinations (Cosine similarity with Euclidean and Cosine similarity with Minkowski) have a silhouette coefficient of 0.055, indicating that these combinations resulted in clusters with similar quality in terms of silhouette score. The other combinations (Cosine similarity with Manhattan and Cosine similarity with Canberra)have lower silhouette coefficients (0.039 and 0.036, respectively), suggesting that the clusters formed using these distance metrics may be less distinct or less cohesive compared to the first two combinations. It also illustrates various configurations of the Ward linkage distance metric (Cosine similarity with Euclidean, Manhattan, Canberra, and Minkowski) in conjunction with the Agglomerative clustering method. Each combination has a silhouette coefficient value which reflects the extent to which the Agglomerative clustering algorithm functioned to produce clusters using that particular distance metric and linking technique. In this table, the silhouette coefficient for two combinations Cosine similarity with Euclidean and Cosine similarity with Minkowski is -0.214, whereas the silhouette coefficients for Manhattan and Canberra are -0.222 and -0.173, respectively. Every value is negative. A low score implies that the clusters formed with these combinations are poorly defined, may overlap, or are not sufficiently separated from one another while in the case of single linkage silhouette coefficients for Manhattan and Canberra are -0.174 and -0.153, respectively, however, the silhouette coefficients for two combinations of Cosine similarity with Euclidean and Cosine similarity with Minkowski are -0.152 correspondingly. The negative silhouette coefficients indicate that the choice of distance metrics (Cosine similarity, Manhattan, Canberra, Minkowski) and the "Single" linkage method did not yield meaningful clusters for the given data. Manhattan and Canberra have silhouette scores of -0.233 and -0.173, respectively, while the corresponding silhouette coefficients for both instances of cosine similarity with Euclidean and Minkowski are -0.223 using complete linkage criteria. The Manhattan and Canberra have silhouette scores of -0.198 and -0.173, accordingly, whereas utilizing average linkage criteria, the corresponding silhouette coefficients for both instances of Cosine similarity with Euclidean and Minkowski are -0.190. Manhattan and Canberra have silhouette scores of -0.112 and -0.008, correspondingly, though employing the centroid linkage criterion, the associated silhouette coefficients for both instances of Cosine similarity with Euclidean and Minkowski are -0.104. In the proposed framework, referred to as efficient agglomerative hierarchical clustering, the Cosine similarity distance measure is applied with four different linkage methods: Ward, Single, Complete, and Average. The silhouette coefficient values vary slightly among the combinations, ranging from approximately 0.090 to 0.095. This suggests that all combinations are relatively similar in terms of clustering quality. This table's values are entirely positive, which is generally desirable. The positive silhouette score indicates that the proposed framework that we have implemented appears to function fairly well demonstrating that the clusters are very distinct and spaced apart.

Table 3 contains different combinations of distance metrics such as Euclidean, Manhattan, Canberra, and Minkowski with the Jaccard similarity measure used in the K-means clustering approach, along with the silhouette coefficient values associated with each combination. The findings from the analysis of Table 3 show that the silhouette scores for Manhattan and Canberra are -0.010 and the silhouette coefficients for both instances of Jaccard similarity with Euclidean and Minkowski are -0.004. The absence of recognizable clusters for the provided data is indicated by the negative silhouette coefficients for the distance metrics (Euclidean, Manhattan, Canberra, and Minkowski) and Jaccard similarity.

**Table 3** Result comparison of Clustering Algorithms using Silhouette coefficient with Jaccard Similarity

| Combinations | **Silhouette Score** | | |
|---|---|---|---|
| | Standard K-Means Clustering | Standard Agglomerative Clustering | Efficient Agglomerative Hierarchal Clustering |
| Jaccard similarity, Euclidean, Ward, Silhouette coefficient | -0.004 | -0.365 | 0.09050214016473164 |
| Jaccard similarity, Manhattan, Ward, Silhouette coefficient | **-0.010** | -0.382 | 0.09050214016473164 |
| Jaccard similarity, Canberra, Ward, Silhouette coefficient | **-0.010** | **-0.274** | 0.09050214016473164 |
| Jaccard similarity, Minkowski, Ward, Silhouette coefficient | -0.004 | -0.365 | 0.09050214016473164 |
| Jaccard similarity, Euclidean, Single, Silhouette coefficient | -0.004 | -0.463 | **0.09334084900162792** |
| Jaccard similarity, Manhattan, Single, Silhouette coefficient | **-0.010** | -0.462 | **0.09334084900162792** |
| Jaccard similarity, Canberra, Single, Silhouette coefficient | **-0.010** | **-0.430** | **0.09334084900162792** |
| Jaccard similarity, Minkowski, Single, Silhouette coefficient | -0.004 | -0.463 | **0.09334084900162792** |
| Jaccard similarity, Euclidean, Complete, Silhouette coefficient | -0.004 | -0.459 | 0.09198448728525851 |
| Jaccard similarity, Manhattan, Complete, Silhouette coefficient | **-0.010** | -0.459 | 0.09198448728525851 |
| Jaccard similarity, Canberra, Complete, Silhouette coefficient | **-0.010** | **-0.427** | 0.09198448728525851 |
| Jaccard similarity, Minkowski, Complete, Silhouette coefficient | -0.004 | -0.459 | 0.09198448728525851 |
| Jaccard similarity, Euclidean, Average, Silhouette coefficient | -0.004 | -0.286 | 0.09097884629817761 |
| Jaccard similarity, Manhattan, Average, Silhouette coefficient | **-0.010** | -0.302 | 0.09097884629817761 |
| Jaccard similarity, Canberra, Average, Silhouette coefficient | **-0.010** | **-0.254** | 0.09097884629817761 |
| Jaccard similarity, Minkowski, Average, Silhouette coefficient | -0.004 | -0.286 | 0.09097884629817761 |
| Jaccard similarity, Euclidean, Centroid, Silhouette coefficient | -0.004 | -0.462 | N.A |
| Jaccard similarity, Manhattan, Centroid, Silhouette coefficient | **-0.010** | -0.461 | N.A |
| Jaccard similarity, Canberra, Centroid, Silhouette coefficient | **-0.010** | **-0.429** | N.A |
| Jaccard similarity, Minkowski, Centroid, Silhouette coefficient | -0.004 | -0.462 | N.A |

It also exhibits four distinct groupings of distance metrics, including Euclidean, Manhattan, Canberra, and Minkowski, along with the Jaccard similarity and Ward linkage methods, which are applied in tandem with the Agglomerative clustering algorithm. The silhouette scores for Manhattan and Canberra have the values -0.382 and -0.274, and the silhouette coefficients for both instances of Jaccard similarity with Euclidean and Minkowski are, correspondingly, -0.365. The negative silhouette coefficients demonstrate that the Ward linkage method, Jaccard similarity, and distance metrics (Euclidean, Manhattan, Canberra, and Minkowski) failed to identify significant associations for the provided data. In the case of the single linkage approach, Manhattan and Canberra have silhouette scores of -0.462 and -0.430, respectively, and both cases of Jaccard similarity with Euclidean and Minkowski have silhouette coefficients of -0.463 while in the complete linkage approach, and the Euclidean, Manhattan, Canberra, and Minkowski distance metrics are clustered in four different ways. According to the results, Canberra has silhouette scores of - 0.427, but for the remaining combinations, the value is the same which is -0.459. The results also reveal that Manhattan and Canberra have silhouette scores of -0.302 and -0.254, respectively, while the corresponding silhouette coefficients for both instances of Jaccard similarity with Euclidean and Minkowski are -0.286 using average linkage criteria. In accordance with Table3's discoveries, Manhattan and Canberra have silhouette scores of -0.461and -0.429, correspondingly, though employing the centroid linkage criterion, the associated silhouette coefficients for both instances of cosine similarity with Euclidean and Minkowski are -0.462. The Jaccard similarity distance metric is used in the suggested framework, with four possible linking methods: Ward, Single, Complete, and Average. The silhouette coefficient values range from roughly 0.090 to 0.093, with some combinations showing slightly different values. This shows that, in terms of clustering quality, all combinations are quite comparable. The values in this table are all positive, which is normally a good thing. The adopted framework seems to work reasonably well, accordingto the positive silhouette score displaying how different and distant the clusters are.

**Table 4** Final Outcome of Clustering Algorithms using Silhouette Coefficient with Cosine similarity

| **Clustering Algorithms** | **Silhouette Coefficient(Cosine)** |
|---|---|
| K-Means Clustering | 0.055 |
| Agglomerative Hierarchal Clustering | -0.088 |
| Efficient Agglomerative Hierarchal Clustering | **0.095247046630965** |

After implementation of standard K-means, Agglomerative and Proposed Framework against different combinations of distance measures; Euclidean, Manhattan, Canberra, Minkowski, Linkage criteria; Ward, Single, Complete, Average, Centroid, Similarity measures; Cosine, Jaccard. By comparing the results presented in Table 2-3 we select the most suitable results; a higher silhouette score against each clustering algorithm both in the cases of cosine and Jaccard similarity measure as well. The ultimate outcomes are reported in Table 4-5 after comparison.

Table 4 presents three different clustering algorithms (K-Means, Agglomerative, Efficient Agglomerative Hierarchal Clustering ) along with their respective Silhouette Coefficient values using cosine similarity measure. The Silhouette Coefficient values vary among the algorithms. Specifically, K-Means has a Silhouette Coefficient of 0.055, indicating a moderate level of cluster quality. Agglomerative clustering has a negative Silhouette Coefficient of -0.088, which suggests that the clusters created by this algorithm have overlapping or poorly separated data points. Negative values are generally less desirable. The Proposed Approach; Efficient Agglomerative Hierarchal Clustering has a Silhouette Coefficient of 0.095247046630965, indicating relatively good cluster quality. Based on the Silhouette Coefficient values in this table, the Proposed Approach appears to perform the best among the three algorithms, as it has the highest positive Silhouette Coefficient, indicating well-defined clusters. K-Means also shows reasonable cluster quality, while Agglomerative clustering seems to produce less well-defined clusters.

Table 5 summarizes the Silhouette Coefficient values for different clustering algorithms: K-Means, Agglomerative, and a Proposed Approach, but this time using the Jaccard similarity measure. K-Means has a Silhouette Coefficient of -0.010, which is slightly negative. This suggests that the clusters created by K-Means using Jaccard similarity may not be well-separated, and data points within clusters are not significantly more similar to each other than to data points in other clusters. Agglomerative clustering has a Silhouette Coefficient of -0.254, which is also negative. This indicates that the clusters created by Agglomerative clustering using Jaccard similarity have even less separation and may exhibit significant overlap. The Proposed Approach has a Silhouette Coefficient of 0.09334084900162792, which is positive. This suggests that the clusters created by the custom approach using Jaccard similarity are relatively well-separated, and data points within clusters are more similar to each other than to datapoints in other clusters. The Proposed Approach appears to perform the best among the three algorithms when using Jaccard similarity as the similarity measure. It achieves a positive Silhouette Coefficient, suggesting better cluster quality in terms of separation. The results of clustering are substantially influenced by the similarity measure selection. We examined Tables 4-5 to show the effects of employing different similarity metrics (Cosine and Jaccard) on the values of the Silhouette Coefficient for different clustering algorithms. We concluded that the Cosine similarity measure works effectively under the suggested framework.

**Table 5** Final Outcome of Clustering Algorithms using Silhouette Coefficient with Jaccard Similarity

| Clustering Algorithms | Silhouette Coefficient(Jaccard) |
|---|---|
| K-Means Clustering | -0.010 |
| Agglomerative Hierarchal Clustering | -0.254 |
| Efficient Agglomerative Hierarchal Clustering | **0.09334084900162792** |

## 5.2 Davies Bouldin Index

The Davies-Bouldin index concentrates on the compactness and separation of clusters and is interpreted differently from the silhouette coefficient. The "average similarity" betweeneach cluster and its most comparable cluster is measured by the Davies-Bouldin Index. It measures the degree to which the clusters are distinct and well-separated. The following eq(2) is used to compute the Davies-Bouldin index of clustering algorithms. In equation 2n represents the number of clusters, Si is a metric used to assess how compact a cluster is; the average distance between data points in cluster i, Sj is a metric used to assess how compact a cluster is; the average distance between data points in cluster j, Mij is a metric used to describe how distinct clusters i and j are from one another (for example, the distance between cluster centroids or other similarity/distance metrics). Better clustering is indicatedby a lower DBI score. Smaller values imply that clusters are distinct and well-separated.

$$\text{DBI}= \frac{1}{\square}\sum_{\square=1}^{n} \max j \neq 1 \frac{Si+Sj}{(\square\square\square)} \qquad \text{eq(2)}$$

Table 6 presents five different combinations of distance metrics (Cosine similarity), linkage methods (Ward, Single, Complete, Average, Centroid), and the Davies-Bouldin Index (DBI) values associated with each combination. Each combination is associated with a DBI value, which quantifies the quality of the clusters created by the Agglomerative clustering algorithm using that specific combination of distance metric and linkage method. In this table, the DBI values vary among the combinations. The DBI values range from approximately 27.151 to 61.842. Lower DBI values are generally preferred as they indicate better-defined and well-separated clusters. Therefore, combinations with lower DBI values (such as Single linkage with DBI 27.151) are typically indicative of better clustering results. Combinations with higher DBI values (such as Average linkage with DBI 61.842) suggest that the clusters may be less distinct and might overlap more, which can negatively impact the clustering quality. In the case of an efficient agglomerative hierarchal clustering approach, all the DBI values are relatively low, which is generally desirable. Low DBI values indicate that the clusters created by the proposed approach are well-separated and distinct from each other. The DBI values in this table range from approximately 2.212 to 2.344. These values are indicative of reasonably good clustering results. The efficient agglomerative hierarchal clustering approach that is implemented appears to perform well, as indicated by the low DBI values. These values suggest that the clusters created using Cosine similarity with different linkage methods are well-defined for the given data. In order to identify which clustering algorithm performs best on the data corpus, we also compared the findings shown in Table 6 for the Davies Bouldin index assessment measure. K- Means has a DBI of 2.734, which suggests moderate cluster quality. Lower DBI values are generally preferred. Agglomerative clustering has a much higher DBI of 27.151, indicating poorer cluster quality. This high value suggests that the clusters created by Agglomerative clustering using Cosine similarity have significant issues, possibly overlapping or unclear separation. The Proposed Approach has a DBI of 2.212146287008467, which is lower than both K-Means and Agglomerative. This suggests that the Proposed Approach performs better in terms of cluster quality when using the Cosine similarity measure.

**Table 6** Result comparison of Clustering Algorithms using Davies Bouldin Index with Cosine similarity

| Combinations | **Davies Bouldin Index** | | |
|---|---|---|---|
| | Standard K-Means Clustering | Standard Agglomerative Hierarchal Clustering | Efficient Agglomerative Hierarchical Clustering |
| Cosine similarity, Ward, Davies-Bouldin Index | **2.734** | 46.569 | 2.286823755134645 |
| Cosine similarity, Single, Davies-Bouldin Index | **2.734** | **27.151** | 2.2639113000167033 |
| Cosine similarity, Complete, Davies-Bouldin Index | **2.734** | 44.757 | 2.344123476806612 |
| Cosine similarity, Average, Davies-Bouldin Index | **2.734** | 61.842 | **2.212146287008467** |
| Cosine similarity, Centroid, Davies-Bouldin Index | **2.734** | 38.321 | N.A |

Table 7 lists the five possible linking methods (Ward, Single, Complete, Average, Centroid), similarity measurements (Jaccard similarity), and corresponding Davies-Bouldin Index (DBI)values for each combination. The quality of the clusters produced by the Agglomerative clustering algorithm employing a particular combination of similarity measure and linking mechanism is quantified for each combination using a DBI score. The DBI values in this table show a big difference between the combinations. The DBI values are somewhere between 12.137 and 73.721. Lower DBI values are desired since they signify clusters that are more clearly defined and well-separated. Inferring superior clustering outcomes are combinations with lower DBI values, such as Average linkage with DBI 12.137 and Ward linkage with DBI12.804. Higher DBI combinations(such as Single linkage with DBI 73.721) imply that the clusters may be less distinct and may overlap more, which may have a detrimental effect on theclustering quality. In the case of an efficient agglomerative hierarchal clustering approach, all of the DBI numbers in Table 7 are low, which is normally a good thing. Low DBI values showthat the clusters produced by the suggested method are clearly segregated from one another. In this table, the DBI values fall between around 2.274 and 2.389. These numbers show that the clustering findings were fairly good. The low DBI values suggest that the implemented efficientagglomerative hierarchical clustering method performs well. The Proposed Approach has the lowest DBI of2.274033025778767 among the three algorithms. Davies Bouldin's score of the proposed framework using Cosine similarity is lower as compared to Jaccard similarity which indicates that Cosine similarity is a good choice for assessing interactions between cyber threatactors.

**Table 7** Result comparison of Clustering Algorithms using Davies Bouldin Index with Jaccard Similarity

| Combinations | **Davies Bouldin Index** | | |
|---|---|---|---|
| | Standard K-Means Clustering | Standard Agglomerative Clustering | Efficient Agglomerative Hierarchical Clustering |
| Cosine similarity, Ward, Davies-Bouldin Index | **13.884** | 12.804 | 2.3333191295704205 |
| Cosine similarity, Single, Davies-Bouldin Index | **13.884** | 73.721 | 2.3333191295704205 |
| Cosine similarity, Complete, Davies-Bouldin Index | **13.884** | 27.713 | 2.38953195138396 |
| Cosine similarity, Average, Davies-Bouldin Index | **13.884** | **12.137** | **2.27403302577876** |
| Cosine similarity, Centroid, Davies-Bouldin Index | **13.884** | 64.287 | N.A |

# 6. Cyber Criminals Profiling

In order to estimate the impact of oncoming assaults on the system, understand the characteristics associated with these attacks, and quickly respond to cybersecurity threats, it is essential to observe the attacker's knowledge, skills, and behaviors. Profiling, in accordance with [16], comprises assessing character traits or behavioral tendencies that permit an investigator to draw generalizations about a subject or a crime scene. It is possibleto predict the likely course of action and targets of cyber threat actors by using profiles based on the whole scope of their attack. Then, security personnel can take steps to preventthese anticipated acts. Additionally, it aids businesses in determining the exact risk profiles most relevant to the cybercrime most likely to target them. A more focused allocation of resources for cybersecurity measures is made possible by this assessment. Cybercrime patterns and trends can be identified using this information, making it faster to identify and stop upcoming attacks. In current literature, cyber security experts and the cyber community identify some sort of cyber threat actors based on their own perceptions although we cannot depend on this type of information. They also profile them on the basis of limited features such as TTPs(Attack patterns), experience level, and risk averseness there is some sort of link that exists between cyber threat actors but there are no traces of identifying such a relationship in literature. We identify an interesting association between cyber threat actors based on their shared traits and aggregate them using the unsupervised clustering approach covered in section 4.4. This section talks about the profile of these cybercriminal groups as illustrated in Figure 3 based on comprehensive threat information, considering elements like malware, vulnerability, victims who were specifically targeted, motive, attacker's country, attack IDs, and targeted entities. In total 12 groups of cybercriminals are identified. Table 8 lists the cyber threat actors that are part of each group, and profiles of individual groups are depicted in Figure 4-27. Cybercriminals commence by acquiring details about concerning possible victims. Based on their goals and reasons, this may entail conducting research on groups, people, or entities in certain nations. They select or create malware that is suited to their goals. This malware can take the form of trojans, ransomware, spyware, or specially created tools, afterwards they find weaknesses that can be exploited within the target organizations. This might involve investigating known vulnerabilities or possibly finding zero-day flaws; vulnerabilities that were previously unidentified and unpatched. The most effective attackpaths are selected by cybercriminals based on the vulnerabilities found. To install malware on the target's systems, cybercriminals create alluring phishing emails, malicious attachments, and other delivery methods. The malware then takes advantage of the weaknesses in the targeted systems after the victim interacts with the adverse payload. It carries out its nefarious tasks, which may involve data theft, file encryption, network penetration, or the exfiltration of private data.

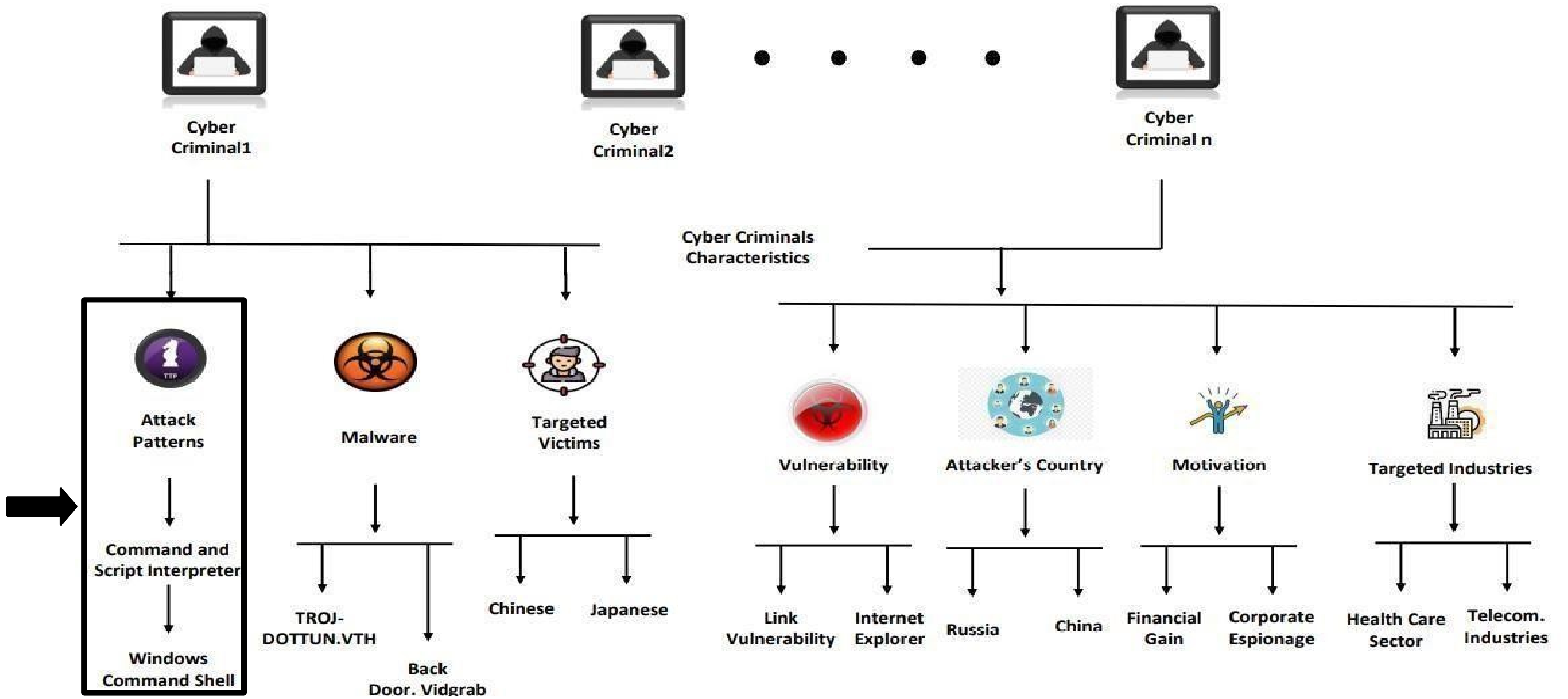

**Figure 3** Features for cyber criminals profiling

Cybercriminals may steal important information, influence systems, or launch other assaults from the infected systems, depending on their objectives.

## 6.1 Cyber Criminal Group1's Profile

Figure 4 demonstrates the cyber threat Group 1's profile, including the malware, target victims, industries, vulnerabilities, motives, and attack methods used. The group is believed to have affiliations with several countries, including China, North Korea, Russia, Iran, and Lebanon. It has a wide range of targeted victims across various countries such as the US, Thailand, Vietnam, Malaysia, Singapore, Philippines, Indonesia, Georgia, Saudi Arabia, South Korea, Bangladesh, Taiwan, Mexico, India, Africa, Japan, Hong Kong, UAE, Lebanon, Kuwait, Yemen,

Oman,Canada, Switzerland, Australia, UK, Israel, Germany, Netherlands, Afghanistan, Pakistan, Ukraine, Italy, Belgium, Poland, Turkey, andFrance.

**Table 8** Overview of Cyber Threat Actors Groups

| Groups | Cyber Threat Actors |
|---|---|
| Group 1 | Admin@338, Andariel, APT1, APT17, APT28, APT29, APT30, APT32, APT33, APT38, APT41, Backdoor Diplomacy, Blacktech, Carbanak, Cleaver, Cobalt Group, Copy Kittens, Dark Caracal, Dark Hotel, Dragonfly 2.0, Dragonok, Dust Storm, Equation, Fox Kitten, Gall Maker, Gama Redon, GCMAN, HAFNIUM, Higaisa, Inception, Ke3chang, Kimsuky, Lazarus, Leaf Miner, Lotus_Blossom, Magic_Hound, Moafee, Molerats, Muddy Water, Mustang Panda, Naikon, NEODYMIUM, Nomadic Octopus, Oilrig, Orangeworm, Pitty_Tiger, Poseidon Group, PROMETHIUM, Putter_Panda, Rancor, Sandworm, Sharpshooter, Silent_Librarian, Sowbug, Strider, Suckfly, TA459, Threat Group-3390, Thrip, Tropic_Trooper, Turla, Volatile Cedar, Windigo, Winnti_Group. |
| Group 2 | Andariel, APT29, APT33, APT37, APT41, Carbanak, Cobalt Group, Costaricto, DarkCaracal, Darkhotel, Darkhydrus, Deep Panda, Dragonfly, Evilnum, Ferocious_Kitten, FIN6, FIN7, FIN8, Frankenstein, Gamaredon, GCMAN, Inception, Ke3chang, Lazarus, Machete, Molerats, Muddywater, Mustang_Panda, Naikon, Oilrig, Rocke, Sharpshooter, Silence, StealthFalcon, TA505, TA551, TeamTNT, TransparentTribe, Tropic_Trooper, Turla,Volatile Cedar, Whitefly, Windigo, Winnti_Group, WIRTE. |
| Group 3 | APT1, APT28, APT29, APT33, Black Oasis, Carbanak, Chimera, Cobalt Group, Darkhotel, Dragonfly 2.0, Equation, FIN4, FIN6, FIN7, Frankenstein, Gallium, Gamaredon, GCMAN, GOLD SOUTHFIELD, Kimsuky, Lazarus, Lotus_Blossom, Machete, Muddy Water, Mustang_Panda, Naikon, Night_Dragon, Oilrig, Orangeworm, Putter_Panda, Rancor, Sandworm, Sharpshooter, Silent_Librarian, Silver Terrier, TeamTNT, Thrip, Tropic_Trooper, Whitefly, Winnti_Group, Zirconium. |
| Group 4 | Apt28,Apt29 , Apt33, Carbanak, Cobalt Group, Kimsuky,Mustang_Panda,Putter_Panda,Rocke,Sandworm,Silent_Librarian,TransparentTribe, Windshift, Winnti_Group. |
| Group 5 | APT1, APT3, APT12, APT17, APT28, APT29, APT33, APT41, Bouncing Golf, Carbanak, Chimera, Dark Caracal, Equation, Evilnum, FIN7, FIN8, Gamaredon, HAFNIUM, Higaisa, Ke3chang, Kimsuky, Lazarus, Leviathan, Machete, Menupass, Muddy water, Mustang_Panda, Naikon, Oilrig, PLATINUM, Sandworm, Sidewinder, Silence, Stealth Falcon, TA505, Threat Group-3390, Tropic_Trooper, Turla, Windigo, Winnti_Group. |
| Group 6 | Ajax_Security_Team, Andariel, APT1, APT3, APT17, APT18, APT28, APT29, APT30, APT33, APT38, APT41, Carbanak, Cobalt Group, Dark Caracal, Darkhotel, Deep Panda, Dragonfly 2.0, Elderwood, Equation, Ferocious_Kitten, FIN4, FIN5, FIN7, FIN8, Fox Kitten, Gamaredon, GCMAN, GOLD SOUTHFIELD, Gorgon, Higaisa, IndrikSpider, Kimsuky, Lazarus, Lotus_Blossom, Menupass, Moafee, Mustang_Panda, Naikon, Night_Dragon, Oilrig, Orangeworm, Pitty_Tiger, Putter_Panda, RTM, Sandworm, Scarlet Mimic, Silent_Librarian, Silverterrier, Stealth Falcon,TA459, TA505, TeamTNT, TEMP.Veles, Threat Group-1314, Transparent Tribe, Turla, Windigo, Winnti_Group, Zirconium. |
| Group 7 | admin@338, APT3, APT16, APT28, APT37, APT38, APT41, Axiom, Black oasis, Bouncing Golf, Bronze Butler, Carbanak, Cobalt Group, Darkhotel, Darkhydrus, Dragonfly2.0, Dragonok, Evilnum, FIN7, Fox Kitten, Frankenstein, Gallium, Gamaredon, Indigo zebra, Indrik Spider, Ke3chang, Kimsuky, Lazarus, Leviathan, Magic_Hound, Menupass, Moafee, Molerats, Muddy water, Mustang_Panda, Naikon, Oilrig, Patchwork, Poseidon Group, PROMETHIUM, Rocke, Sandworm, Scarlet Mimic, Sidewinder, Silverterrier, Strider, TA459, TA505, TeamTNT,Threat Group-3390, TontoTeam, Transparent Tribe, Turla, Volatile Cedar, WIRTE, Wizard Spider. |
| **Group 8** | Andariel, APT12, APT29, APT33, APT37, APT-C-36, Carbanak, Chimera, Cobalt Group, Darkhotel, Darkhydrus, Deep Panda, Equation, FIN6, FIN7, FIN8, Frankenstein, Gamaredon, GCMAN, GOLD SOUTHFIELD, HAFNIUM, Indrik Spider, Kimsuky, Lazarus, Magic_Hound, Muddy water, Mustang_Panda, Naikon, Oilrig, Rocke, Sandworm, Sidewinder, Silent_Librarian, Stealth Falcon, TA505, TeamTNT, Transparent Tribe, Tropic_Trooper, Turla, Winnti_Group, Zirconium. |
| **Group 9** | Admin@338, Andariel, APT1, APT3, APT12, APT29, APT30, APT32, APT37, Blacktech, Carbanak, Cobalt Group, Darkhotel, Darkhydrus, Dragonok, Evilnum, FIN8, Ferocious_Kitten, Gamaredon, HAFNIUM, Inception, Indrik Spider, Ke3chang, Kimsuky, Lazarus, Lotus_Blossom, Machete, Menupass, Moafee, Molerats, Muddy water, Mustang_Panda, Nomadic Octopus, Oilrig, Patchwork, Rancor, Rocke, RTM, Sandworm, Scarlet Mimic, Sidewinder, Silence, TA459, TA505, TA551, TeamTNT, Tonto Team, Transparent Tribe, Tropic_Trooper, Turla, Windigo, Winnti_Group, Wizard Spider. |
| **Group 10** | Admin@338, Ajax_Security_Team, Apt3, Apt12, Apt28, Apt29, Apt32, Apt33, Apt41, Blue Mockingbird, Bronze Butler, Carbanak, Cleaver, Cobalt Group, Costaricto, Dark Caracal, Darkhotel, Deep Panda, Dragonfly 2.0, Dust Storm, Elderwood, Equation, Evilnum, Fin6, Fin7, Fin8, Fin10, Fox Kitten, Gamaredon, GCMAN, Hafnium, Indrik Spider, Ke3chang, Kimsuky, Lazarus, Molerats, Mustang_Panda, Naikon, Nomadic Octopus, Oilrig, Orangeworm, Poseidon Group, Promethium, Rocke, RTM, Sandworm, Scarlet Mimic, Sharpshooter, Silent_Librarian, TA459, TA505, TA551, TeamTNT, ThreatGroup-3390, Tropic_Trooper, Turla, Volatile Cedar, Winnti_Group. |
| **Group 11** | Admin@338, Andariel, Apt3, Apt12, Apt16, Apt18, Apt28, Apt41, Apt-C-36, BlackOasis, Blue Mockingbird, Bronze Butler, Carbanak, Cobalt Group, Darkhotel, Elderwood, Equation, Fin6, Fin7, Hafnium, Inception, Lazarus, Lotus_Blossom, Muddy Water, Orangeworm, Pitty_Tiger, Sandworm, TA505, Windigo, Zirconium. |
| **Group 12** | Admin@338, Ajax_Security_Team, APT1, APT18 , APT28, APT29, APT32, APT33, APT39, APT41, Axiom, Bronze Butler, Carbanak, Cleaver, Cobalt Group, Copy kittens, Darkhotel, Darkhydrus, Dragonok, Evilnum, Ferocious_Kitten, FIN4, FIN6, FIN7, FIN8 ,Fin10, Fox Kitten, Gallium, Gamaredon, Group5, HAFNIUM, Higaisa, Indigo zebra, Indrik Spider, Lazarus, Leafminer, Lotus_Blossom, Magic_Hound, Menupass, Mofang, Molerats, Muddy water, Mustang_Panda, Operation Wocao, Sandworm, Silence, Silent_Librarian, Silverterrier, TA505, TA551, TeamTNT, TEMP.Veles, Threat Group-3390, TransparentTribe, Tropic_Trooper, Winnti_Group. |

Cybercriminals present in this group are likely involved in activities aimed at stealing sensitive information for intelligence purposes or advancing political agendas. This may take place through a variety of attack techniques depicted in Figure 5, including brute force assaults, social engineering attacks, and user-keystroke recording on PCs. Emails with spear phishing are among the most frequently encountered instances of social engineering. Hackers send emails from accounts that have been compromised that seem authentic and pretend to be reputable organizations like banks, governments, or well-known businesses. The intention is to mislead victims into exposing confidential details such as credit card numbers, login passwords, or personal information. As opposed to many other cyberattacks that focus on technological flaws, social engineering takes advantage of human psychology and belief via the use of techniques involving baiting, quid pro quo, and pretexting. In pretexting, an attacker fabricates a situation or pretext in order to gather information from a target. Under the pretense of resolving a technical issue, an attacker may pretend to be an IT support technician and ask for login credentials. Baiting entails providing something alluring, such as free software, movie downloads, or USB drives, and the Quid Pro Quo tactic involves the attacker promising something in exchange for information or access. In return for a user's login information, they provide free software. They also infiltrate websites that their intended targets visit on a regular basis.

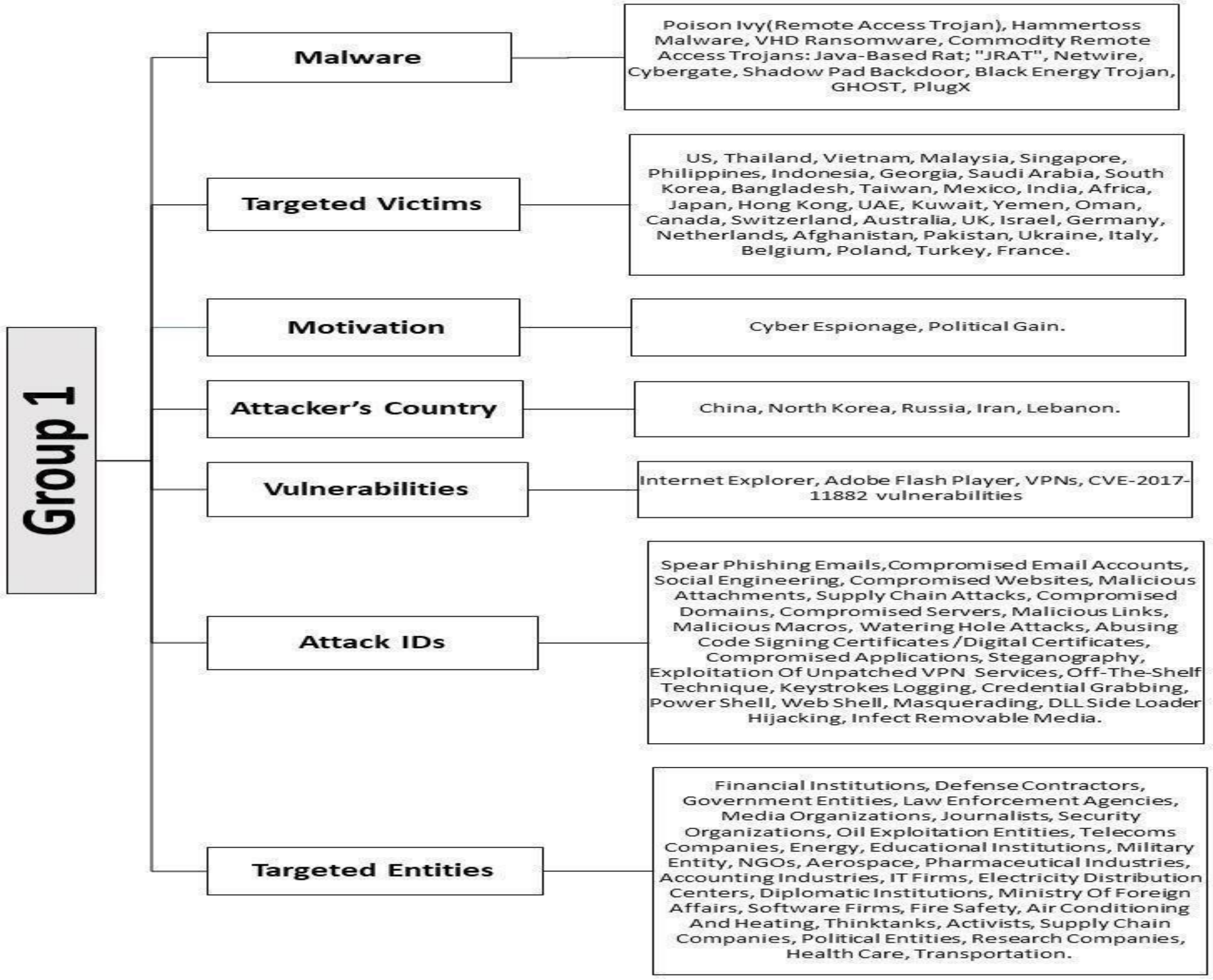

**Figure 4** Group1's Threat Profile

These "watering hole" websites are deliberately chosen by the attackers because they are aware that their victims—typically certain people or organizations—are likely to visit them furtherly legitimate website domains have been hacked. After getting inside, attackers use backdoors, malicious scripts, and content modification to stay in control of the hacked domain. Attackers load and run malicious DLL files during the program execution by taking advantage of applications that are vulnerable or incorrectly configured. They insert a malicious DLL file at a spot that will be tried to be loaded by the susceptible program. This might be network sharing or a local directory. The malicious DLL is unintentionally loaded by the susceptible program upon launch, providing the attacker access to or control over the machine. If the group fails to exploit websites and applications, they will use code-signing techniques to make their own malicious websites and applications and make them legitimate and trustworthy by signing them with a valid digital certificate. Software integrity and authenticity are confirmed by digital certificates that are issued by reputable certificate authorities (CAs). Attackers, however, may take advantage of code signing in the following ways: hackers might use stolen credentials or false identities to get digital certificates through fraudulent means. Either way, they started certificate expiry attacks, in which the attacker targets certificates that are about to expire since they know that consumers could become less careful when they see a notice about an expired certificate. Additionally, they engaged in certificate spoofing, which entails creating phoney certificates that closely resemble real ones in order to deceive consumers into believing they are connected to a reliable website. Man-in-the-middle (MITM) attacks may result from this, in which the attacker intercepts data being sent back and forth between the user and the trustworthy server. This group exploits Microsoft Office, Internet Explorer, Adobe Flash Player, VPNs, and CVE-2017-11882 vulnerabilities in order to install a wide range of malware by using different attack techniques. According to the attacker’s motivation, malware is deployed in malicious attachments, links, macros, and powershell scripts and delivered to the victim’s machine via spear-phishing emails, compromised legitimate websites or applications, or by creating their own malicious websites and applications. Additionally, they contaminated detachable media like SD cards, external hard discs, and USB devices. The virus can propagate to and possibly compromise a new device if the compromised media is attached to it. They often used readily available, pre-built tools, or malware to launch cyberattacks. The Poison Ivy virus offers a remote-control interface that permits attackers to remotely infect the victim's computer without being present on the victim's premises. A web shell is uploaded by attackers to a web server that is vulnerable. After installation, it offers backdoor access and remote control over the hacked remote server. During remote control communication, cybercriminals concealed malware using steganography in seemingly innocuous files like pictures, music files, or other digital media. They also leveraged social media sites for C2 communications, suchas GitHub and Twitter. It uses these trustworthy channels to transmit and receive data and orders that are encoded, making it more difficult to identify as malicious activity. Group1 targets a wide range of entities, including financial institutions, defence contractors, government entities, law enforcement agencies, media organizations, journalists, security organizations, oil exploitation entities, telecoms companies, energy, educational institutions, military entities ,NGOs ,aerospace, pharmaceutical industries, accounting industries, , IT firms, electricity distribution centers, diplomatic institutions, ministry of foreign affairs, software firms, fire safety, air conditioning and heating, thinktanks, activists, supply chain companies, political entities, research companies, health care, transportation by using Hammertoss, VHD, JRAT, Netwire, Cybergate, Shadow Pad, Black Energy PCRAT/GHOST, and PLUGX malware.

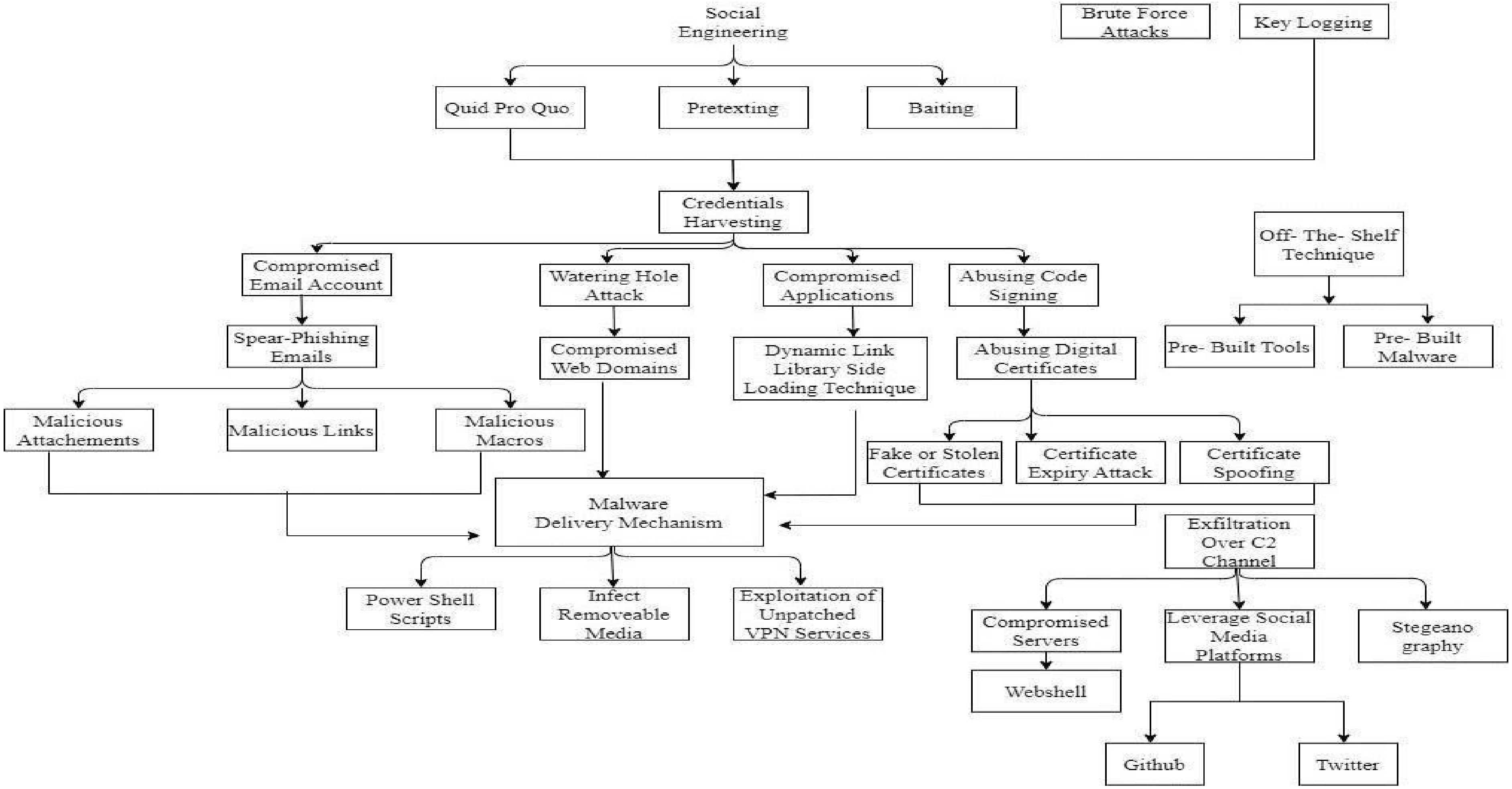

**Figure 5** Digital Intrusion Pathway - Group 1

## 6.2 Cyber Criminal Group2's Profile

The profile of the cyber threat Group 2 is presented in Figure 6. This group is thought to have ties to a number of nations, including North Korea, Russia, China, Lebanon, and Iran.It targets a wide variety of individuals in numerous nations, such as South Korea, the US, the UK, Russia, China, and many others. Installing a variety of malware like BIOLAND, Carbanak, More- Eggs"(a.k.a Terra Loader or Spicy Omelette), Cobalt, Royal DNS, Win32/Stealth Falcon, PlugX, Trickbot by taking advantage of flaws in Microsoft Word, Windows, and servers. This diverse set of targets suggests a broad and sophisticated cyber operation for cyber espionage and financial gain. This indicates that they are interested in both stealing sensitive information and making money through their cyber activities. Different attack methods as shown in Figure7 are used in order to achieve this goal. These tactics include social engineering assaults, keylogging, and hijacking a real Internet email account; which is the process by which an unauthorized person accesses and controls a validemail account that belongs to someone else. To keep control of the account and stop the rightful owner from getting access again, the attacker modifies all account settings, including the password and recovery details. Emails with spear phishing are among the mostprevalent instances of social engineering. Hackers send emails from hacked accounts that seem authentic and pretend to be reputable organizations like banks, governments, or well- known businesses. The intention is to deceive the receivers into disclosing private information, such as credit card numbers, login passwords, or personal information. In contrast to many other cyberattacks that focus on technological flaws, social engineering abuses human psychology and conviction via the use of strategies including baiting, quid pro quo, and pretexting. They also penetrate websites that their intended targets visit on an ongoing basis. These "watering hole" websites are deliberately chosen by the attackers because they are aware that their victims—typically certain people or organizations—are likely to visit them while other authentic website domains have been compromised. Afterward, they compromise a specific website, hackers insert malicious Visual Basic scripts into websites that eventually get utilized by web browsers. Attackers load and run malicious DLL files during program execution by taking advantage of applications that are vulnerable or incorrectly configured. They insert a malicious DLL file at a spot that will be tried to be loaded by the susceptible program. This might be network sharing or a local directory. If the group fails to exploit websites and applications they will use code-signing techniques to make their own malicious websites and applications and make them legitimate and trustworthy by signing them with a valid digital certificate. Cybercriminals have the ability to get digital certificates in deceitful ways, such as creating false identities or launching certificate expiry attacks. Additionally, they engage in certificate spoofing, which entails creating phoney certificates that closely resemble real ones in order to deceive consumers into believing they are connected to a reliable website. Man-in-the-middle (MITM) attacks may result from this, in which the attacker intercepts data being sent back and forth between the user and the trustworthy server.

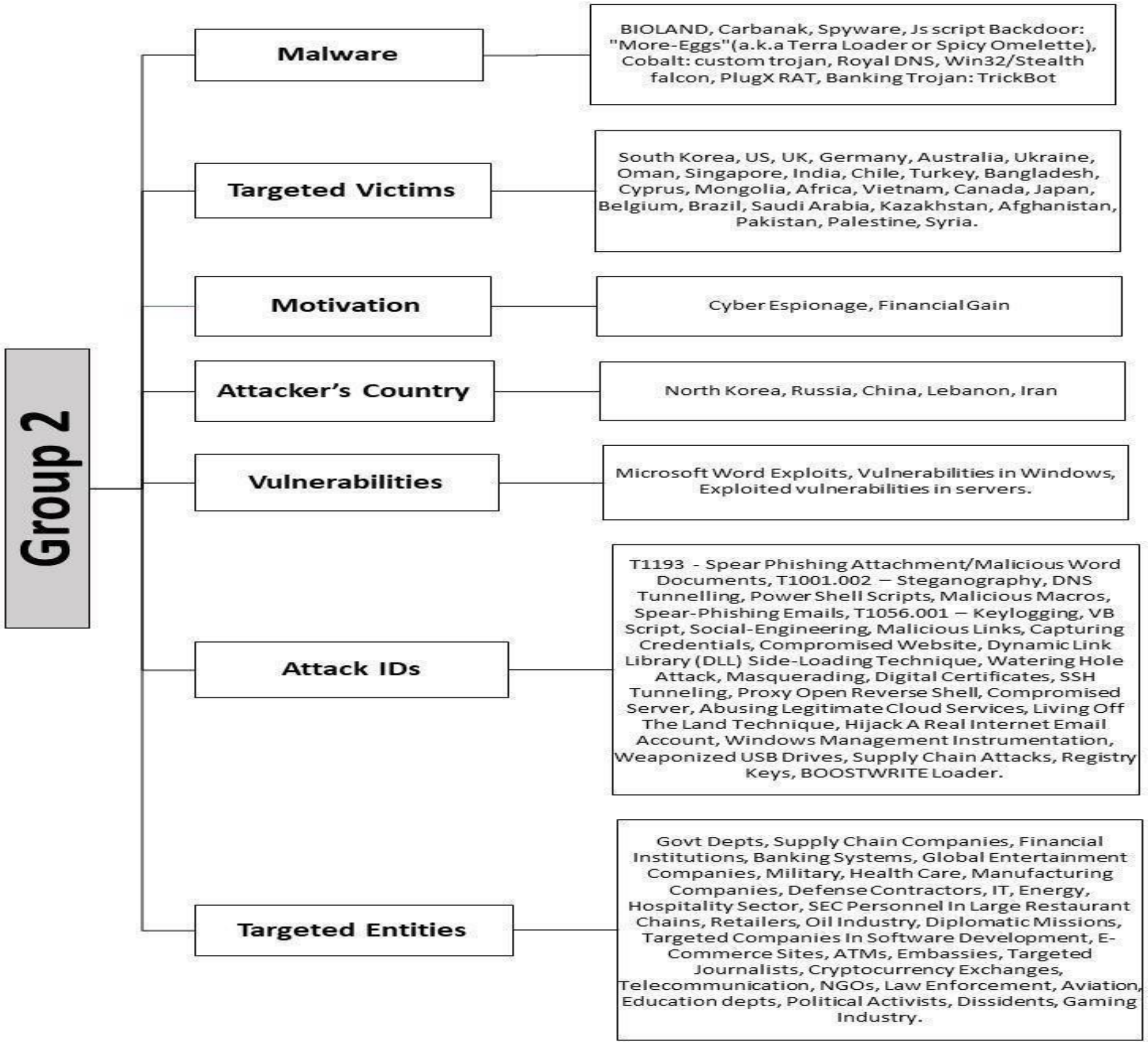

**Figure 6** Group2's Threat Profile

They hosted, distributed, or had control over adverse infrastructure or material via using popular and reliable cloud platforms and services. On cloud computing platforms such as AWS(Amazon Web Services), Azure, Google Cloud, and Dropbox, they set up or breach authentic accounts. Group2 utilized different malware delivery mechanisms which included malicious attachments, links, macros, and supply chain attacks; aimed to subtly undermine an organization by compromising its partners, suppliers, or service providers, and Boost write loader. Malware is deployed in malicious attachments, links, and macros that are delivered to the victim's machine via spear-phishing emails, compromised legitimate websites, applications, and cloud platforms, or creating their own malicious websites and applications.

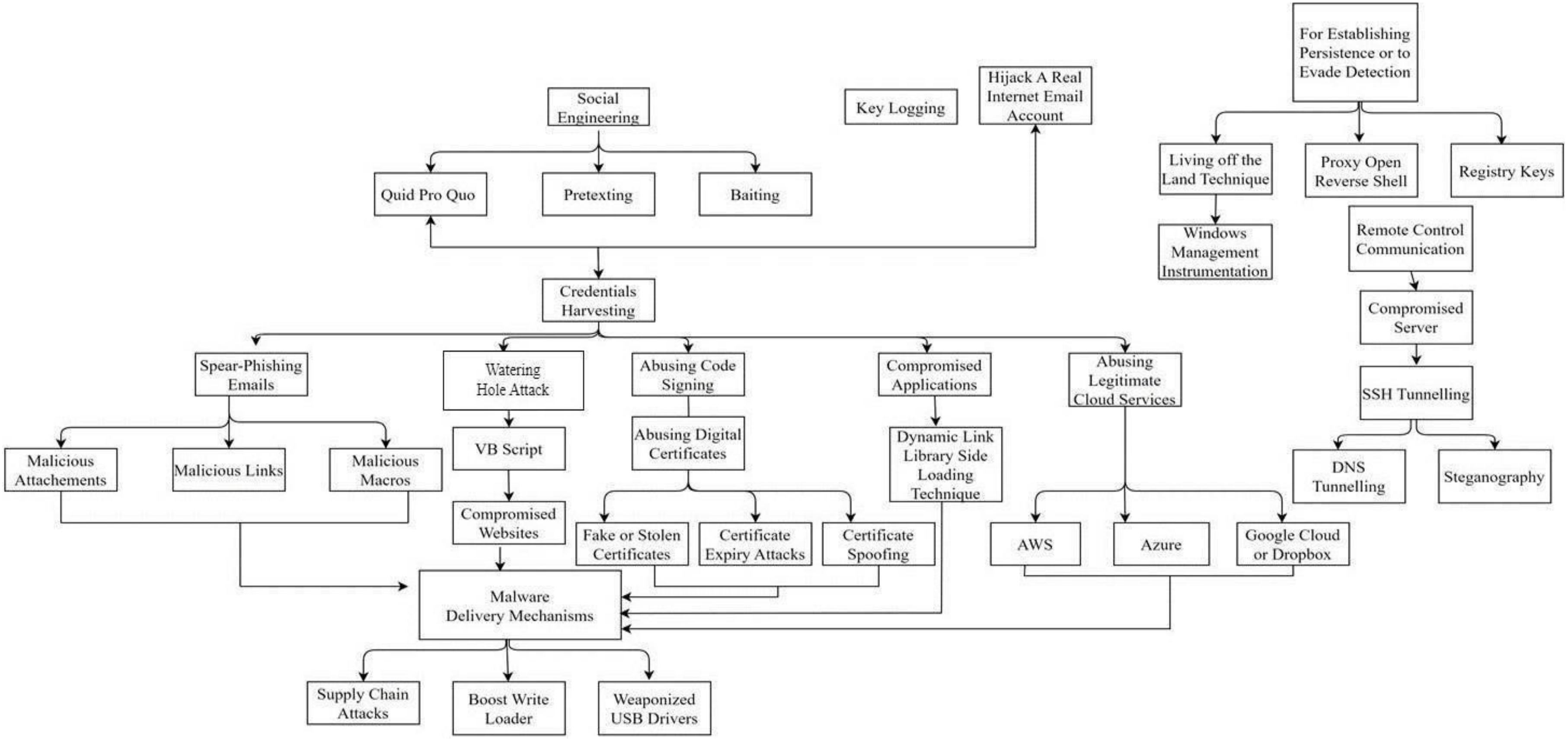

**Figure 7** Digital Intrusion Pathway - Group 2

Moreover, they contaminated SD cards, external hard drives, and USB devices. Once a new device is linked to infected media, the malware can propagate to it and possibly compromise it. They employed living off the land, proxy reverse shell, and registry key strategies for persistence or evading detection in a targeted environment. Threat actors leverage authentic, built-in tools, utilities, or processes on a target system or network, such as Windows Management Instrumentation (WMI), in the Living Off the Land (LOL) approach. WMI is abused by attackers to create persistence on a hacked system. While the Proxy Open Reverse Shell technique uses a compromised system (the "proxy") to establish an unauthorized reverse shell connection from a target system to an attacker-controlled server, they also create scheduled tasks or event triggers that regularly execute malicious scripts using WMI, to make sure their access remains undetected. The attacker may essentially take control of the target machine remotely after it creates a reverse shell connection with the compromised proxy. On the other hand, attackers can also create or modify registry keys to make sure that their malicious code or malware runs every time the system boots up or a specific user logs in to maintain persistence on compromised systems. This connection can be used to execute commands, exfiltrate data, or establish a persistent backdoor for future access. They established secure communications using encryption between a nearby PC and a distant server. SSH tunnels have been employed by attackers to provide a safe path for transferring confidential data from a compromised machine to a distant server. In order to get around network security measures, they either use a steganographic technique or encode the data they want to send into DNS queries which are then forwarded to a DNS server under the attacker's control. During remote control communication, it entails concealing malware within seemingly innocuous items like pictures, music files, or other digital media. They target a wide range of entities, including government departments, financial institutions, supply chain companies, military organizations, healthcare, manufacturing, defence contractors, and many others.

## 6.3 Cyber Criminal Group3's Profile

As shown in Figure 8 the attacks were launched by a group of cybercriminals referred to as Group 3 affiliated with a variety of states like Russia, Iran, China, and North Korea. Theytargeted organizations in the energy sector, companies in other ICS sectors such as industrial/machinery, manufacturing companies, private companies, defence industry, government entities, financial institutions, pharmaceutical, military, NGOs, banks, intelligence agencies, telecommunications, education depts, civil society organizations, oil companies, aviation industry, chemical companies, transportation sector, healthcare, high-tech industries, hospitality sector of Russia, U.S., UK, Switzerland, Japan, South Korea, India, Taiwan, Belgium, Brazil, Germany, Malaysia, Spain, Ukraine, Turkey, Saudi Arabia, China, Hong Kong, Myanmar, Thailand, Vietnam, Philippines, Singapore for financial profit and cyber espionage by taking advantage of the flash exploit. This indicates that they are interested in both stealing sensitive information and making money through their cyber activities. Different attack methods presented in Figure 9 are used in order to achieve this goal. Among these attack techniques are social engineering attacks. Emails with spear phishing are among the most frequently encountered instances of social engineering.In contrast to many other cyberattacks that focus on technological flaws, social engineering takes advantage of human psychology and belief via the use of strategies including baiting, quid pro quo, and pretexting. Cyber adversaries manipulate or redirect the normal executionflow of a legitimate website to an attacker 's-controlled website.

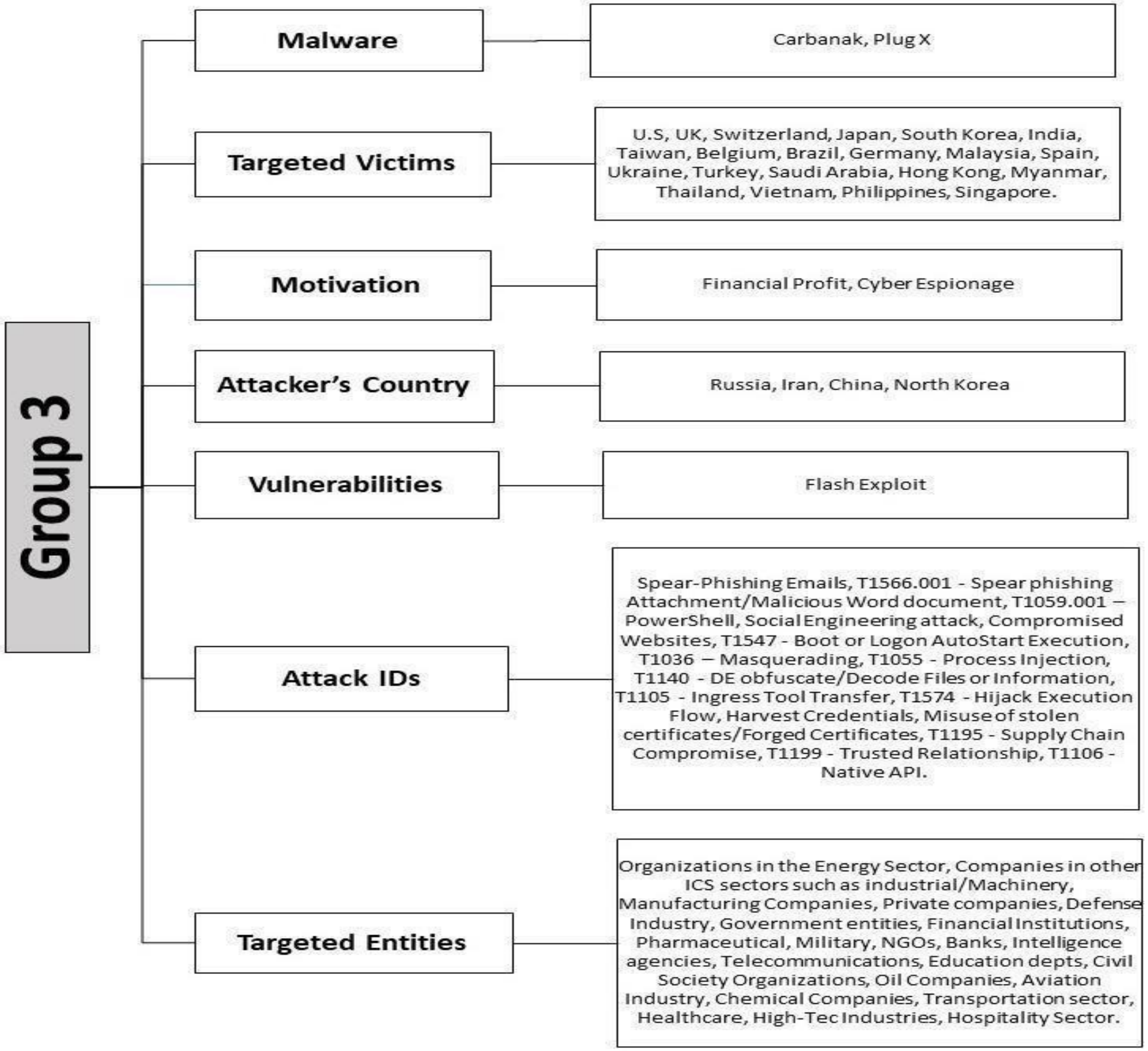

**Figure 8** Group3's Threat Profile

Attackers employed a variety of methods for distributing malware, including supply chain attacks, process injection, ingress tool transfer, malicious word documents,and interpreters or languages for command and scripting. Spear- phishing emails or legitimate websites are used to infect victims' computers with malware that is installed in malicious word attachments, command and scripting interpreters, or languages such as Python, JavaScript, Windows Command Prompt, PowerShell, and Unix- based systems.

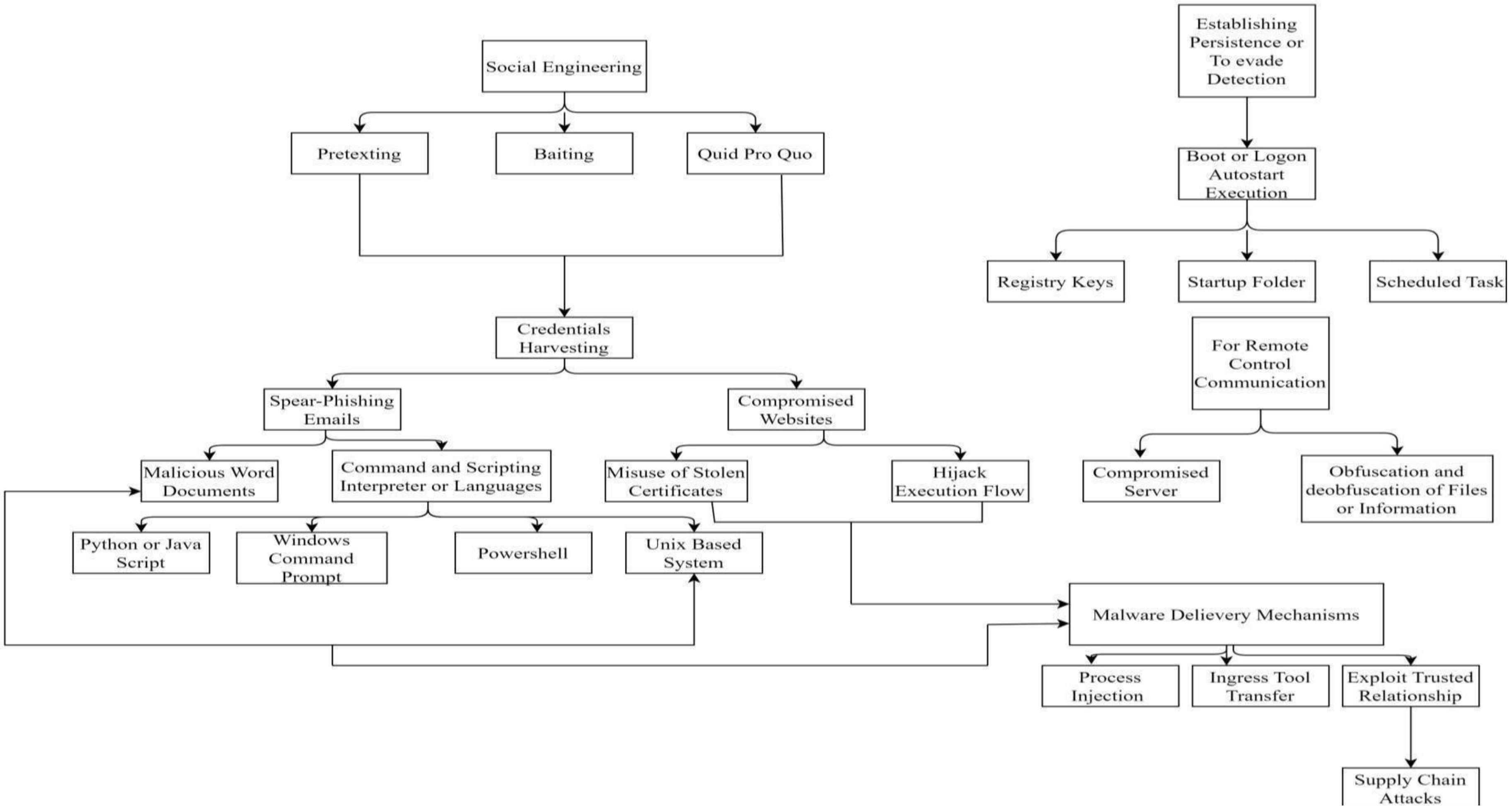

**Figure 9** Digital Intrusion Pathway-Group3

Using the process injection technique, a hacker accesses a legitimate functioning process on a compromised system and injects malicious code. As a result, the attacker's code can run in the legitimate process's context and memory. Cyber adversaries use malicious equipment like USB drives or DVDs to move sensitive data or stolen data from the victim's network to an external place. The supply chain of the target, which consists of vendors, subcontractors, outside service providers, and other partners, is investigated by the attackers. They search for holes or weak points in this chain. Once the supply chain entity is breached, the attackers take advantage of the supplier's and target organization's mutual trust. They approach the ultimate target by using the resources or access granted to them by the compromised entity. To get access to the target's systems and accomplish their goals, they could employ malware, stolen credentials, or other techniques. Attackers used boot or auto logon start execution tactics, which include configuring or manipulating system components to launch malicious code or programs automatically at system startup or user logon, in order to persist or elude detection. On the other hand, they also create scheduled tasks that execute malicious scripts or code at specific times, such as during system startup or user logon. They create or modify registry keys to ensure that their malicious code or malware runs every time the system boots or a specific user logs in. They also place malicious shortcuts or executable files in the system's startup folders, such as the "Startup" folder within the "Start" menu or the "Startup" folder in the user's profile directory. These three techniques were used for performing boot or logon auto-start execution technique. Communication between the compromised system and the attacker's controller server is encrypted by using an effective encryption algorithm like AES.

### 6.4 Cyber Criminal Group4's Profile

Figure 11 depicts that Group 4 belongs to Russia. The main goals of Russian hackers are cyber espionage and financial gain, and they target government entities, military, defence personnel/contractors, and telecommunication companies in the US, Europe, Afghanistan, Hong Kong, Asia, and Middle Eastern nations by using different attack techniques like spear-phishing emails, abusing digital certificates, and social engineering attacks. To evade detection, they utilized shortened URL services in which they convert long and complex URLs into shorter, more manageable links as shown in Figure 10.

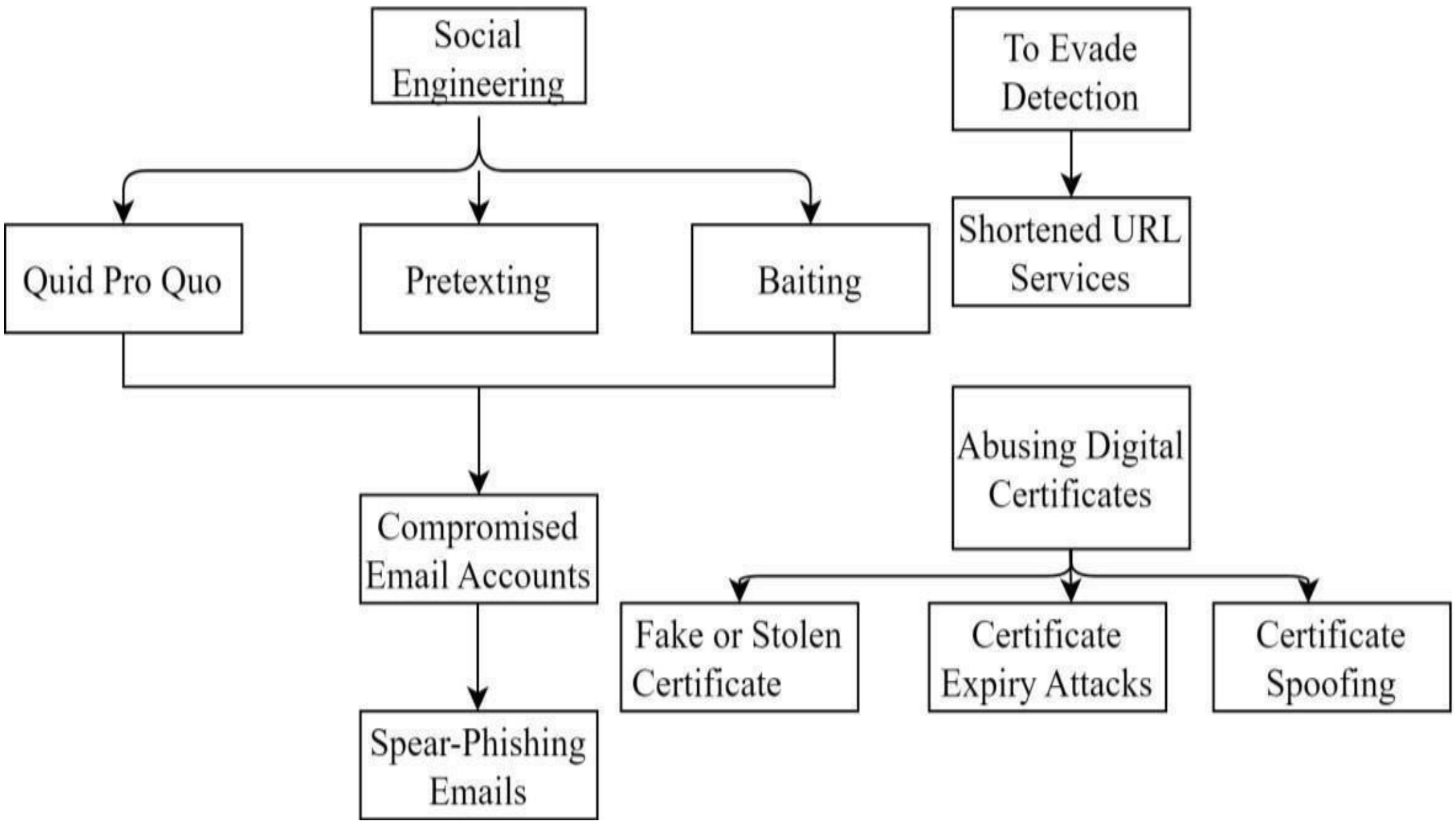

**Figure 10** Digital Intrusion Pathway-Group 4

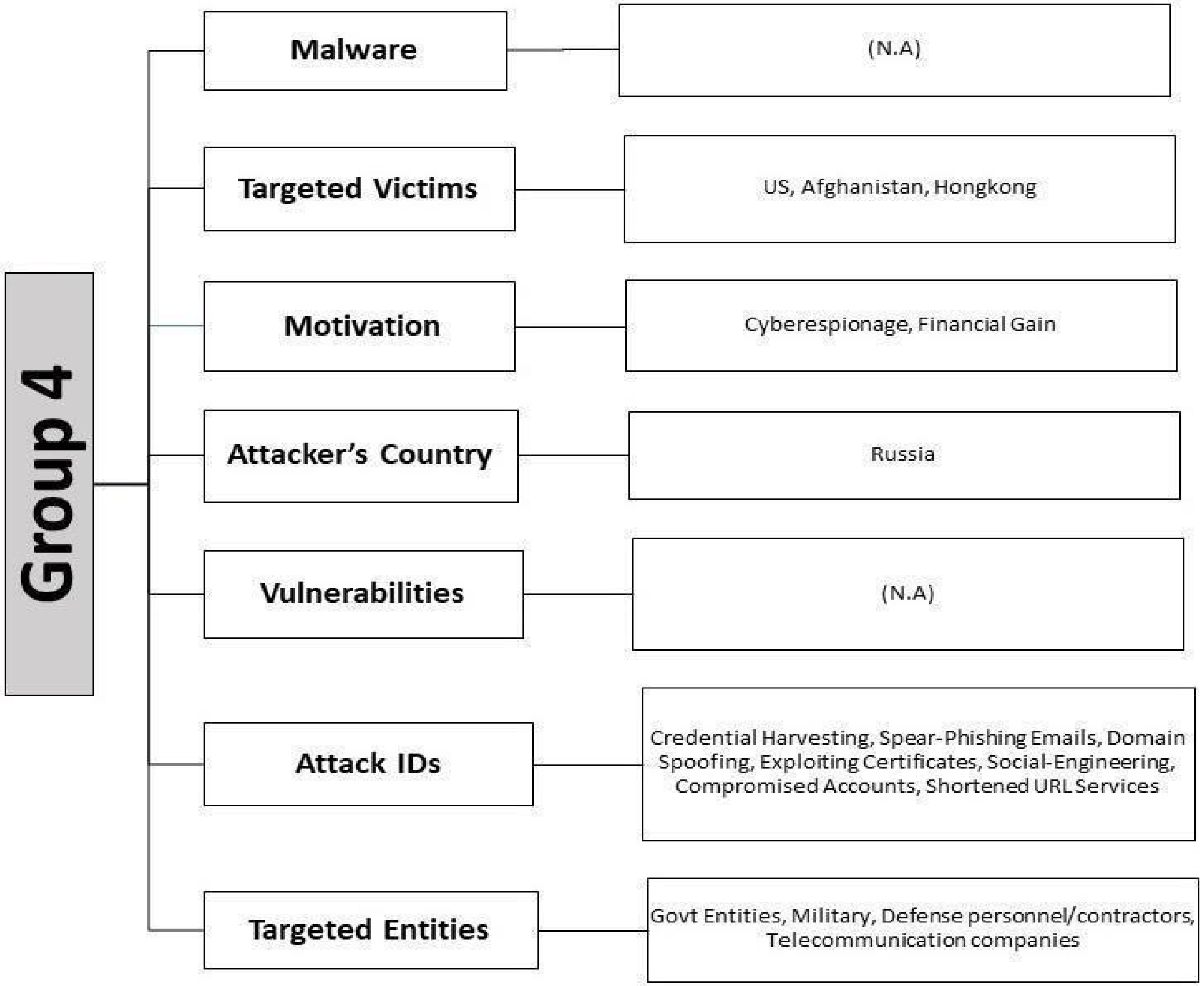

**Figure 11** Group4's Threat Profile

## 6.5 Cyber Criminal Group5's Profile

Figure 12 and 15 refers to the profiling of Group 5 and Group 6. Group 5 is linked to China, Russia, and Korea. They utilize Windows, CVE-2017-0199, exchange server, and browser-based zero-day exploits to install Bateleur, Royal DNS, Poison Ivy, PlugX, and Win32/StealthFalcon as malicious payloads. Their targets encompass telecommunication companies, government entities, defence contractors, transportation, restaurant chains, hospitality organizations, retailers, merchant services, suppliers, academic institutions, etc. As shown in Figure 13 Group5 implements the reconnaissance phase of the cyber kill chain before delivering the weaponized payloads to the target by using system information discovery, file and directory discovery, system connections discovery, system network configuration discovery, system owner/user discovery, system time discovery, process discovery, application window discovery techniques. They start by collecting comprehensive data on the targeted system, including its configurations, user accounts, hardware, and software. After that, they look for files and folders on a compromised system. At different phases of an assault, such as locating sensitive data or locating possible targets, this information will be helpful. They get details about active network connections and a targeted system's network setup in the third and fourth steps of this phase. In addition, they find out which processes are active on a hacked system and details about the owner or users of the machine. Discovering the system's date and time settings in the fifth and sixth steps of this phase will help to understand the system's time configuration, which is crucial for coordinating attacks and evasion. This information will also aid in understanding user roles and privileges to target specific individuals or groups. The final step involves the discovery of program window information on a compromised system by the attackers, who then target certain application windows to steal sensitive data. Attackers captured the information entered through input devices such as a keyboard and mouse as well as they captured a screenshot of the victim's desktop or specific application windows. Attackers collect information or tools from network-shared drives that are accessible within a compromised environment with the intention of finding the intended victim's login credentials and other private data. Emails with spear phishing are among the most prominent instances of social engineering. In addition, they breach websites that their intended targets visit regularly, and they have also successfully hacked legitimate web servers. Once inside, attackers use backdoors, malicious scripts, and website content modification to stay in control of the compromised systems. On that particular website, they additionally incorporate malicious javascript code that creates user profiles, gathers information on users' browsing preferences, and may even steal private data. Users' web browsers immediately run the embedded JavaScript code when they visit the hacked web page. After that, the attacker uses the obtained data to infiltrate a distant server under his control. Attackers deliver malicious or misleading adverts byusing this data for targeted advertising. Attackers used different malware delivery mechanisms which include malicious attachments, macros, command and scripting interpreters or languages; Python or JavaScript, windows command prompt, powershell, Unix-based systems, exploitationfor client execution, shortcut modification, hidden windows, and infect removable media. Malware embedded in malicious attachments, macros, command and scripting interpreters, andlanguages is delivered via spear-phishing emails and compromised websites. They also run malicious malware on a victim's computer or device by taking advantage of flaws in client-side software or apps. They target the victim's system's applications, including media players, document readers, and web browsers. Additionally, they alter shortcut files and produce invisible windows that are not visible to users but may still interact with the system and exe cute commands, enabling the virus to do harmful tasks covertly. Additionally, they contaminated detachable media like SD cards, external hard discs, and USB devices. Attackers used boot or auto logon auto-start execution tactics to remain persistent, manipulating or configuring systemcomponents such that malicious code or programs would run automatically when the victim's machine started up. To evade detection or to bypass security firewalls Group 5 mainly utilizedthree main techniques which include indicator removal from tools, virtualization sandbox evasion, and time stamp manipulation. In addition to deleting or altering event records, logs, and other artifacts produced by their tools or malware during an attack, attackers also destroy indications of compromise (IOCs). Additionally, they put code in their malware to look for indications of virtualization or sandbox environments, such as certain registry entries or file locations, and they also modify the timestamps on files and artifacts. In order to evade detection or forensic examination, malware that identifies a virtualized or sandbox environment may function differently or may not execute malicious actions. The Poison Ivy malware offers a remote-control interface that lets attackers remotely infect the victim's computer without being present on the victim's premises. Attackers transfer files from a remote system to a local systemas well as exfiltrate data from a compromised machine to a server or external location. During remote control connection, data is compressed and encrypted using a powerful encryption method, such as AES. A steganographic approach is then used to conceal highly confidential data inside seemingly innocent files, such as music files, photographs, or other digital media. They move data from compromised systems to their attacker's control servers and vice versa via a variety of application, non-application layers, and cryptographic protocols, or they employ automated tools or scripts to exfiltrate data automatically to a predefined location. Theyalso established alternative communication channels or methods that can be used by malware or attackers if the primary communication channel is compromised or blocked. They target the US, UK, Venezuela, Pakistan, Russia, Austria, Vietnam, Indonesia, Myanmar, China, Jordan, South Africa, Ukraine, Azerbaijan, Germany, Czech Republic, Israel, Cyprus, Greece, Turkey, Taiwan, Malaysia, UAE, Saudi Arabia, Thailand, Georgia, Singapore, Canada, South-Korea, Hongkong, Philippines, Spain, Belarus, Brazil countries for the purpose of cyber-espionage, and financial gain.

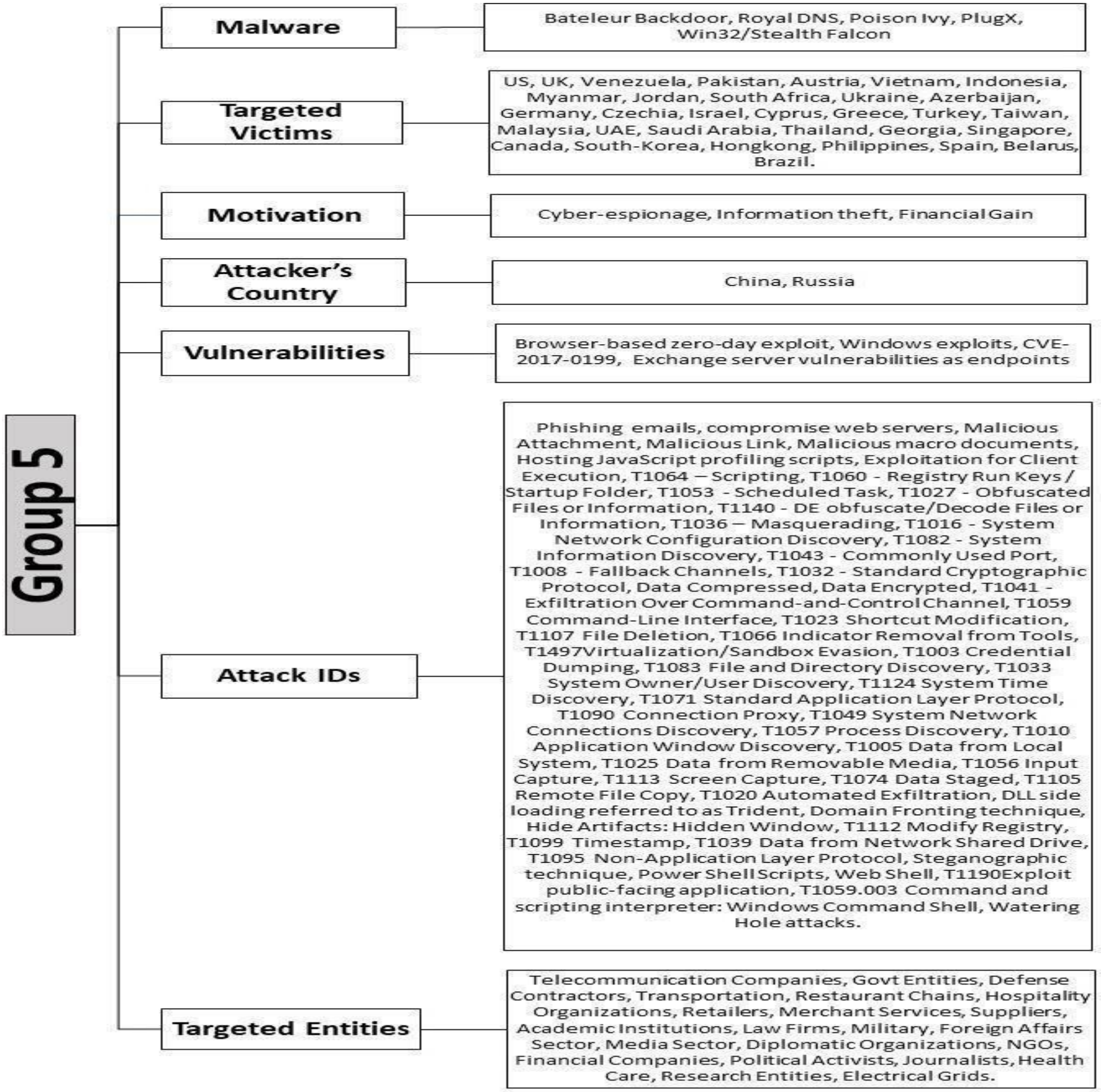

**Figure 12** Group5's Threat Profile

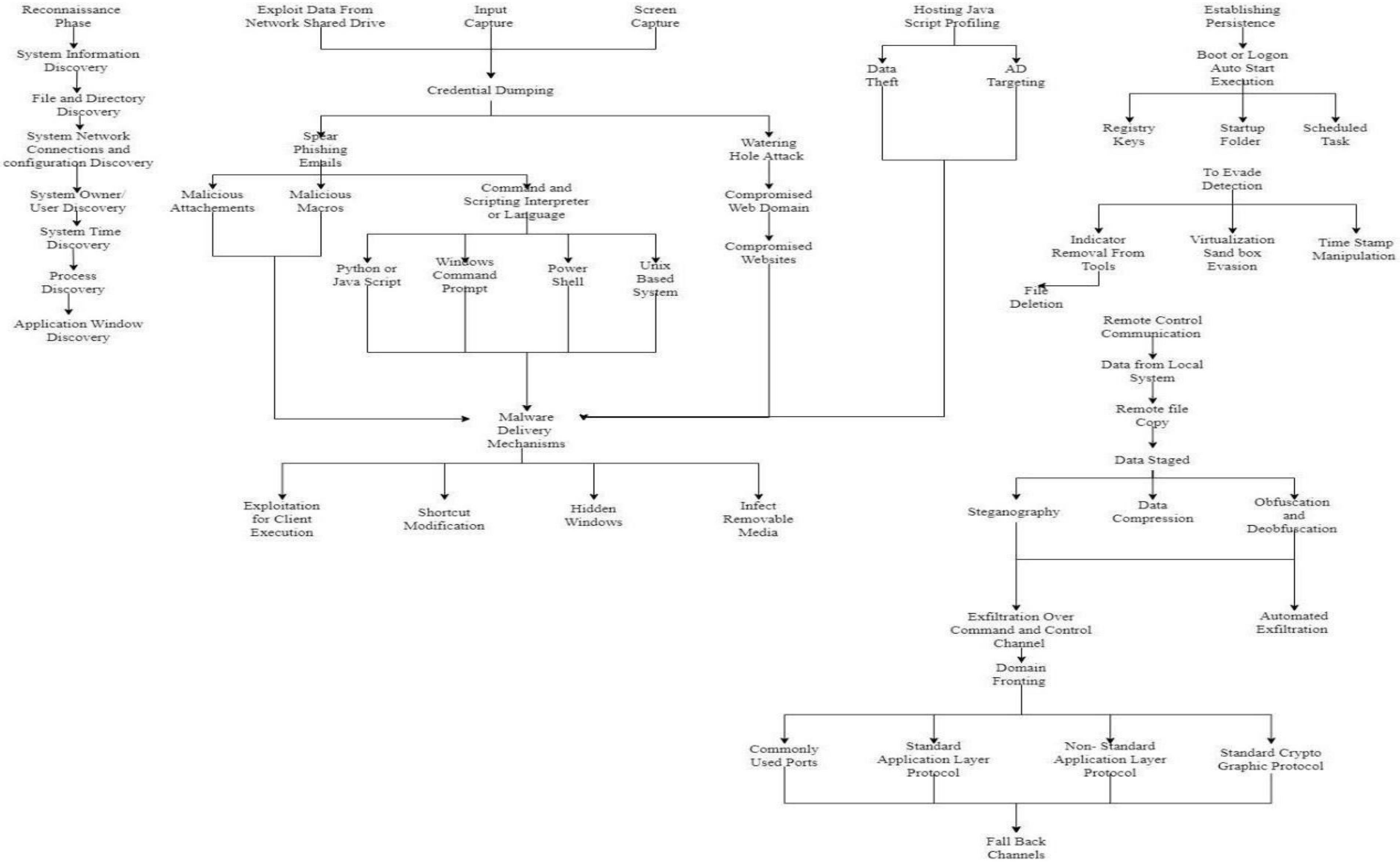

**Figure 13** Digital Intrusion Pathway-Group 5

## 6.6 Cyber Criminal Group6's Profile

Conversely, Group 6 belongs to Iran, North Korea, China, Russia, and Ukrainian nations. They leverage VPN networks, Windows, Adobe Reader, and Microsoft office weaknesses to deploy noxious payloads including Raw POS Malware, Carbanak, ZeroCleare, Dustman, Gand Grab, Revil/Sodinokibi, PlugX, and Triton malware to interact with victims' machines of US, Southeast Asia, South Korea, Japan, Hongkong, Malaysia, Germany, UK, Netherlands, Taiwan, Russia, Australia, Switzerland, China, Belarus, Brazil, Bulgaria, Spain, Afghanistan, Pakistan, Iraq, Columbia, Canada, France, Israel, Ukraine, UAE, Belgium, Canada, France, Israel, Ukraine, UAE, Belgium, Vietnam, Mexico, South Africa, India, Indonesia, Philippines, Myanmar, Singapore, Thailand, Georgia, Chile, Saudi Arabian nations. They concentrate on government departments, defence companies, aerospace, banks, IT, financial institutions, telecommunication, supply chain organizations, media organizations, universities, NGOs, the hospitality industry, hotels, energy sector, oil refinery, military, health care, pharmaceutical companies, casinos, security sectors, insurance companies, law enforcement, thinktanks, diplomats, chemical industry, transportation, electrical utilities, retailers for achieving their objectives of cyber espionage, financial gain, and economic gain. Different attack methods are used in order

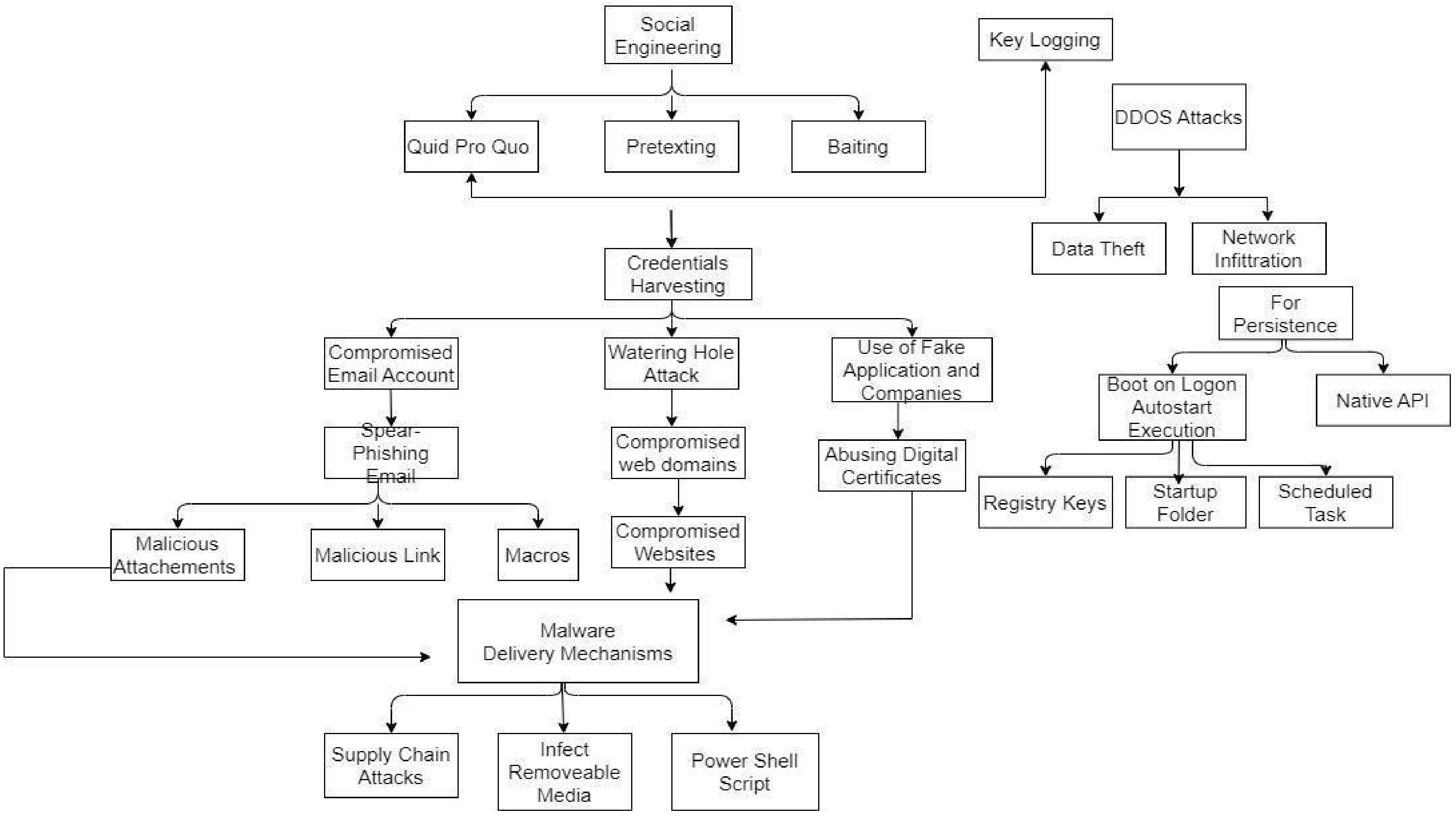

**Figure 14** Digital Intrusion Pathway -Group 6

to achieve this goal presented in Figure 14.

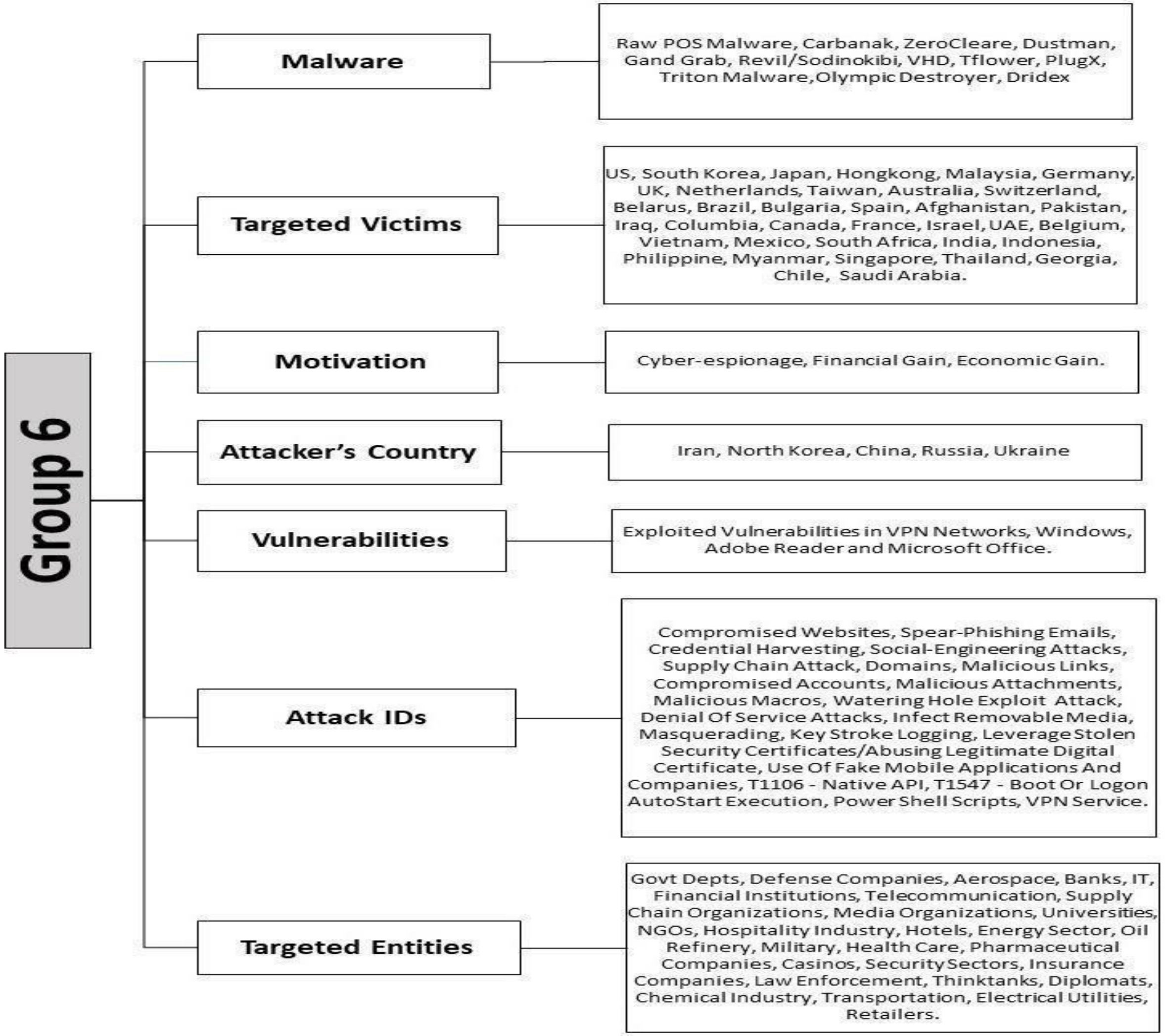

**Figure 15** Group6's Threat Profile

These attack techniques are social engineering attacks and keyboard captures from users' PCs.Hackers send emails from compromised accounts that seem authentic and pretend to be reputable organizations like banks, governments, or well-known businesses. As opposed to many other cyberattacks that focus on technological flaws, social engineering takes advantage of human psychology and confidence via the use of strategies including baiting, quid pro quo,and pretexting. They also infiltrate websites that their intended targets visit on a daily or weekly basis. Attackers load and run malicious DLL files during program execution by taking advantage of applications that are vulnerable or erroneously configured. If the group fails to exploit websites and applications, they will use code-signing techniques to make their own malicious websites and applications and make them legitimate and trustworthy by signing them with a valid digital certificate. In order to subtly infiltrate an organization, Group 6 used a variety of malware distribution techniques, including malicious attachments, URLs, macros, powershell scripts, and supply chain attacks. Malicious attachments, URLs, macros, and powershell scripts are used by malware to infect a victim's computer. These can be distributed by spear-phishing emails, or compromised legitimate websites, applications, and companies. Additionally, they poisoned detachable media like SD cards, external hard discs, and USB devices. They also used denial-of-service (DDOS) attacks for data theft and network infiltration. Attackers used native API methods and boot or logon auto-start execution for persistence. By using this method, the attacker's code will be able to run continuously on the computer that was compromised.

## 6.7 Cyber Criminal Group7's Profile

The cyber threat Group 7's profile is shown in Figure 16, along with the malware, target victims, industries, weaknesses, motivations, and attack techniques utilized. The group is thought to have ties to a number of nations, including China, Russia, North Korea, and Iran. It targets a wide variety of individuals in several nations, notably the US, Japan, South Korea, Taiwan, Vietnam, Malaysia, Philippines, Indonesia, India, Nigeria, Jordan, UK, Russia, Singapore, Germany, China, Hongkong, Iran, Mexico, Saudi Arabia, Kuwait, UAE, Australia, France, Poland, Hungary, Italy, Africa, Ukraine, Afghanistan, Cambodia, Israel, Turkey and many more. Exploited certain weaknesses such as zero-day vulnerability in Internet Explorer, Adobe Flash zero-day vulnerability, windows vulnerability, CVE-2017- 11882, exploited internet-facing services in order to deploy a variety of malware: Poison Ivy RAT, Carbanak RAT, POS Malware, Lizar Malware, Tirion RAT, Ghost RAT, Shammon malware, PlugX RAT, Conti Ransomware, Trickbot banking malware by utilizing different attack methods. Group7 implements the reconnaissance phase of the cyber kill chain before delivering the weaponized payloads to the target by using system information discovery, file and directory discovery, system network connection discovery, system network configuration discovery, network share discovery, system owner/user discovery, process discovery, windows component object model and distributed component object model and obfuscated meterpreter stager techniques. They start by collectingcomprehensive data on the targeted system, including its configurations, user accounts, hardware, and software. After that, they look for files and folders on a compromised system. At different phases of an assault, such as locating sensitive data or locating possible targets, this information will be beneficial. They get details about active network connections and atargeted system's network setup in the third and fourth steps of this phase. In order to find available network shares, such as file servers, shared drives, and other network-attached storage, attackers also enumerate the network. Attackers obtain data on the content, permissions and vulnerabilities linked to the shares after they have been recognized as accessible network shares. An attacker's ability to acknowledge possible sources of confidential information for espionage is enhanced by their understanding of network sharing. In addition, they find out which processes are active on an attacked system and details about the owner or users of the machine. Understanding user roles and privileges to target particular people or groups will be made easier with the use of this knowledge. In order to execute malicious code, they also used the Windows component object model andthe distributed component object model, which enable software components to communicate with one another on Windows systems. Additionally, they implemented simulated cyberattacks using the meterpreter technique against targeted systems in order tocheck for vulnerabilities. Employing different techniques including credential harvesting and hash method passing, attackers avail the use of legitimate user accounts, both with andwithout related permissions and rights, to perform malicious operations within a compromised environment. By using the hash approach, attackers can authenticate on othercomputers, usually those connected to a network, by stealing hashed user credentials from one machine. Instead of using actual passwords in plaintext, it makes use of password hashes. Once an attacker has access to a system, they can take credentials for hashed passwords from memory or storage. While credential harvesting happens through a variety of techniques such as social engineering attacks, keylogging, and brute force attacks, attackers utilize these hashes directly to authenticate on other systems or services where thesame password is used rather than breaking the hashes. Emails with spear phishing are among the most prevalent instances of social engineering. These "watering hole" websites are deliberately chosen by the attackers because they know that their victims—typically certain people or organizations—are likely to visit them and take advantage of browser or plugin vulnerabilities. The malicious code is automatically run in the browser context of the user when they visit the infected website. They mostly affect websites that their intended audience trusts, including news organizations and official government websites. Via the use of visual basic scripts included in online forms that collect sensitive data or simply redirect to malicious websites, they take advantage of it. Attackers load and run malicious DLL files during program execution by taking advantage of applications that arevulnerable or incorrectly configured. The malicious DLL is unintentionally loaded by the susceptible program upon launch, providing the attacker access to or control over the machine. Additionally, they abuse Windows data exchange features while compromising web applications by injecting malicious code into templates or documents that are subsequently processed by web application software. These forms of malicious code injection are used by web applications to interact with their database. Additionally, they hosted, distributed, or had control over harmful infrastructure or material via using popularand reliable cloud platforms and services. On cloud computing platforms such as AWS (Amazon Web Services), Azure, Google Cloud, and Drop box, they set up or hack authenticaccounts. As shown in Figure 17 Group 7 utilized different malware delivery mechanismswhich included malicious attachments, links, macros, and supply chain attacks. Malware isdeployed in malicious attachments, links, macros, and command and scripting interpreters or languages that are delivered to the victim’s machine via spear-phishing emails, compromised legitimate websites, applications, and cloud platforms.

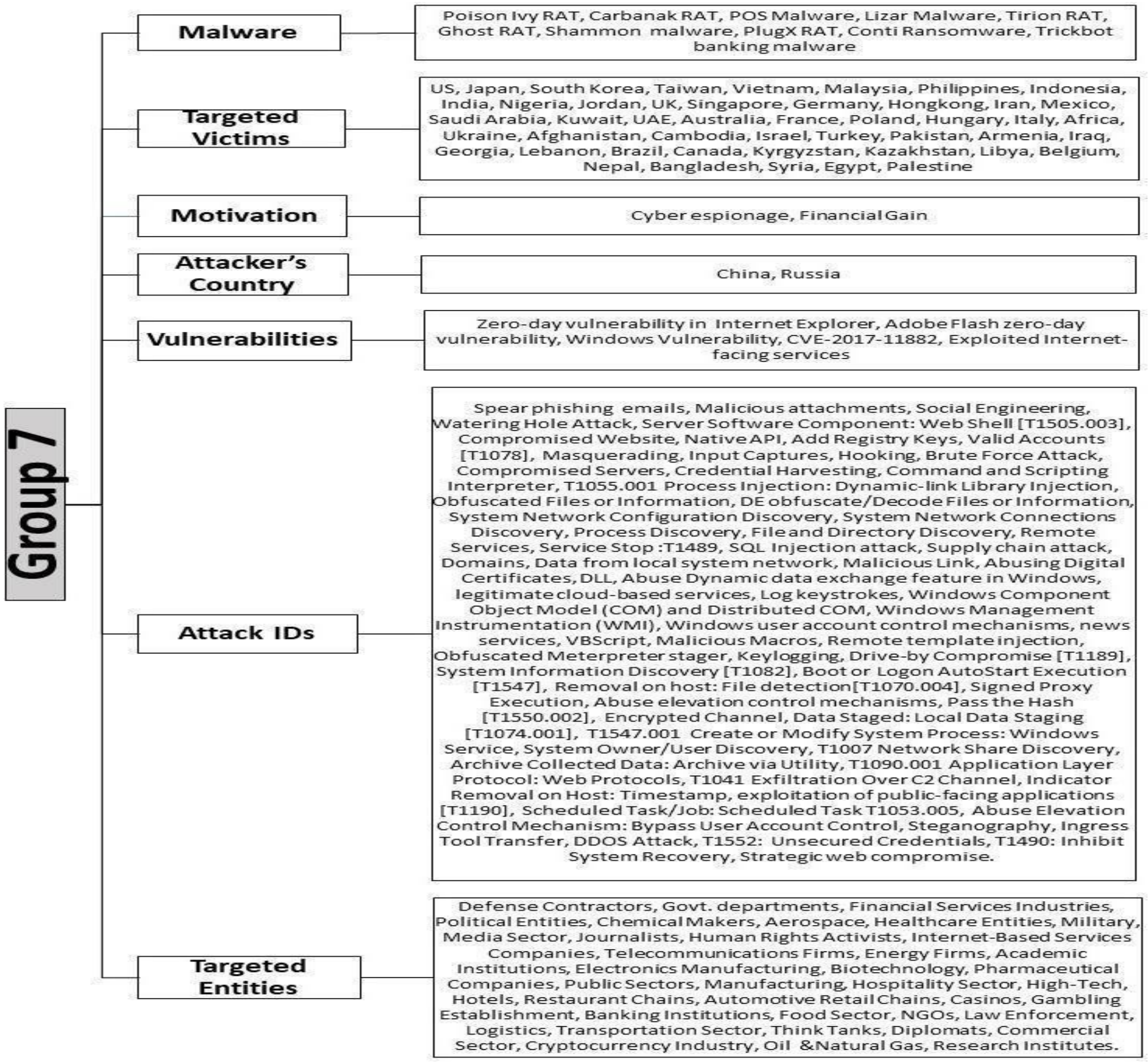

**Figure 16** Group7's Threat Profile

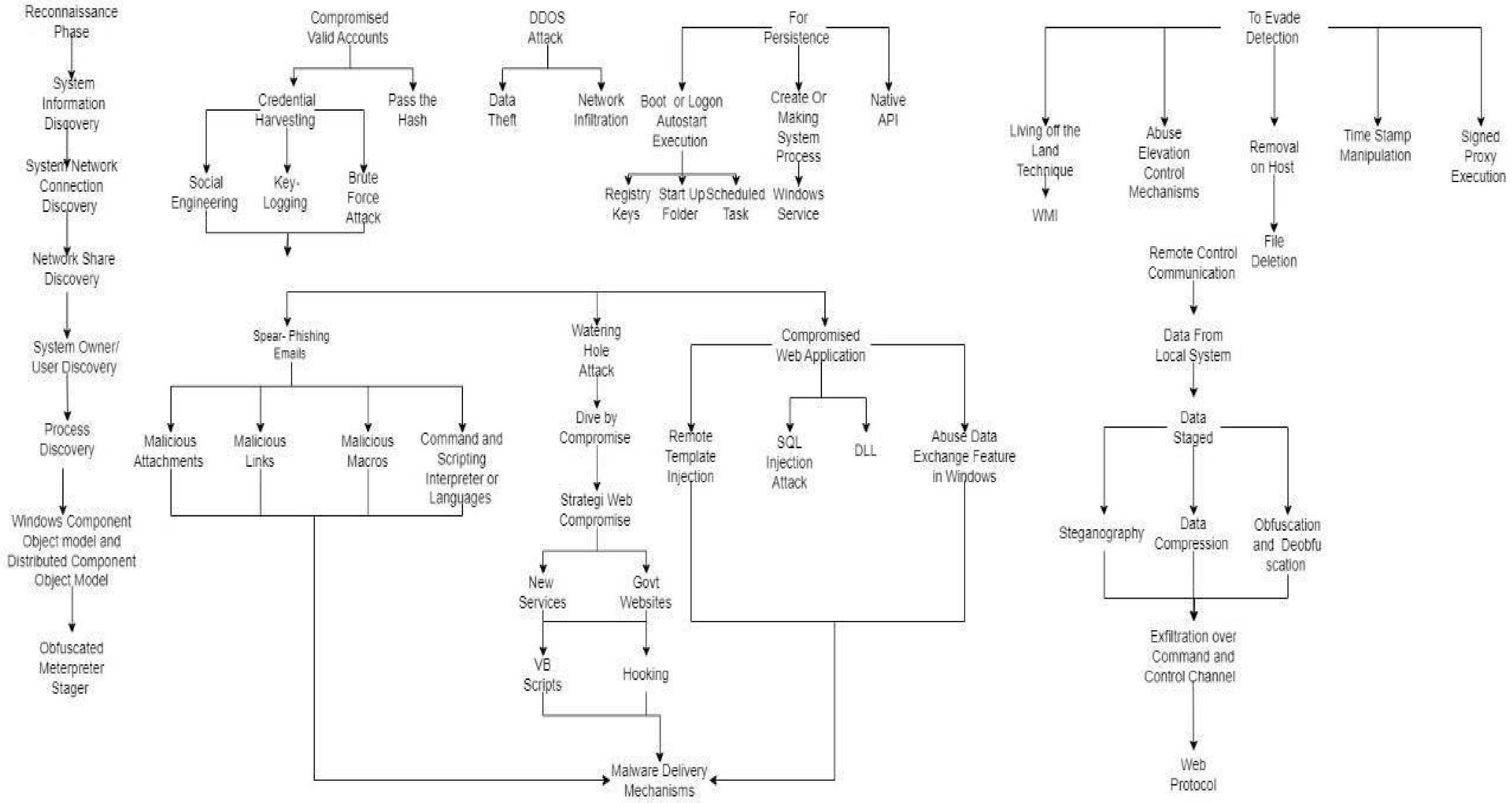

**Figure 17** Digital Intrusion Pathway-Group7

They also deployed DDOS attacks for data theft and network infiltration. For persistence, attackers utilized boot or logon auto-start execution, modified system Windows services, and NativeAPI techniques. To evade detection or to bypass security firewalls Group 7 mainly utilized five main techniques which include indicator removal from tools, abuse elevation control mechanisms, timestamp manipulation, signed proxy execution, and living off the land. In addition to deleting or altering event records, logs, and other artifacts produced by their tools or malware during an attack, attackers also eliminate indications of compromise (IOCs). Additionally, they tamper with the timestamps of files and artifacts, and on Windows systems, they take advantage of the mechanisms that give users or processes enhanced rights. They try to get around security safeguards, such as user account control procedures that serve to prevent unauthorized modifications to the computer, in order to access sensitive resources or carry out malevolent operations. Threat actors use legitimate, built-in tools, utilities, or processes on a target machine or network, such as Windows Management Instrumentation (WMI), in the Living Off the Land (LOL) approach. WMI isabused by attackers to create persistence on a hacked system. While the signed proxy execution technique exploits signed executable files with valid digital signatures to get around security controls and execute unauthorized commands, they create scheduled tasks or event triggers that use Windows Management Interface (WMI) to launch malicious scripts on a regular basis, guaranteeing that their access stays were undetected. With the use of the remote-control interface offered by the Poison Ivy virus, attackers may remotely infect a victim's computer without having to personally visit them. Data from a compromisedsystem is exfiltrated by attackers to a server or external location. Using a powerful encryption method like AES, data transferred during remote control contact is compressed and encrypted. Malicious material is concealed inside seemingly innocuous items like photos, music files, or other digital media using a steganographic technique. They use various web protocols to transfer data from comprised systems to their attacker's control servers and vice versa. They target defence contractors, government departments, financial services industries, political entities, chemical makers, aerospace, healthcare entities, military, media sector, journalists, human rights activists, internet-based services companies, telecommunications firms, energy firms, academic institutions, electronics manufacturing, biotechnology, pharmaceutical companies, public sectors, manufacturing, hospitality sector, high-tech, hotels, restaurant chains, automotive retail chains, casinos, gambling establishment, banking institutions, food sector, NGOs, law enforcement, logistics, transportation sector, thinktanks, diplomats, the commercial sector,cryptocurrency industry, oil and natural gas and research institutes. This wide range of targets points to a complex and extensive cyber operation for financial gain and cyber espionage.

## 6.8 Cyber Criminal Group8's Profile

Figure 19 exhibits the profile of cyber threat Group 8. This group is said to have connections to several countries, including Russia, China, North Korea, and Iran. It addresses a broad spectrum of individuals in many different countries such as the US, Germany, Japan, Colombia, UAE, Ukraine, Saudi Arabia, China, India, Pakistan, Taiwan, Russia, and Mongolia. They use bugs in Adobe Acrobat Reader or Flash Player, CVE-2017-0199 to target government entities, manufacturing firms, financial institutions, the banking sector, aerospace, energy vehicles, health care, defence industrial base, educational institutions, casual-dining restaurants, casinos, hotels, law firms, policy think tanks, NGOs, retail sector, technology, oil and natural gas, chemical, telecommunication, gaming industry,military, IT industry, and cloud-based providers by employing different attack techniques for financial gain and cyber espionage. This suggests that they are motivated by both the desire to steal private data and the want to profit from their digital actions. As depicted in Figure 18 using techniques such as system information discovery, file and directory discovery, system service discovery, system network connection discovery, system network configuration discovery, network service scanning, system owner/user discovery, system time discovery, and process discovery, Group 8 implements the reconnaissance phase of the cyber kill chain prior to delivering the weaponized payloads to the target. They commence by acquiring comprehensive data on the targeted system, including its configurations, user accounts, hardware, and software. After that, they look for files and folders on a compromised system. At different phases of an assault, such as locating sensitive data or locating possible targets, this information will be essential. On a target system, they also locate and list all of the services that are active. Programs or processes known as services operate in the background and perform a variety of tasks, including network communication, file sharing, or remote administration. In the fourth and fifth stages of this phase, attackers seek out established network connections and the network configuration of a targeted machine. They can also look for known vulnerabilities linked with those services by finding active services and their versions. In a network, adversariesalso find and recognize the ports and services that are accessible on target computers. Usually included in the reconnaissance stage, network service scanning enables attackers to map the network and obtain data about possible targets. Furthermore, they find out whichprocesses are active on a hacked system and details about the owner or users of the machine.Understanding user roles and privileges to target particular people or groups will be made easier with the use of this knowledge. During the reconnaissance stage, they also collect data on the date and time settings of an attacked system. This knowledge can be useful forplanning assaults or avoiding discovery, among other things. Credential harvesting happensvia a number of techniques, including brute force assaults, social engineering attacks, logging user keystrokes on computers, taking pictures or recording the user's screen, and exploiting browser extensions. Hackers send emails from hacked accounts that seem authentic and claim to be credible organizations like banks, governments, or well-known businesses. Cybercriminals gather email addresses following credential dumping and breach legitimate user accounts, both with and without related privileges and permissions. These compromised accounts can be used in spear-phishing attempts to target particular people or businesses. They also infiltrate websites that their intended targets visit on a regular basis. The malicious code is automatically run in the browser context of the user when they visit the infected website. By either rerouting users to malicious websites or putting visual basic scripts in online forms that collect sensitive data, they take advantage of those websites. Attackers load and run malicious DLL files during program execution by taking advantage of programs that are vulnerable or incorrectly configured. In order to conceal their actions or execute malicious code without generating new processes, attackersmight either establish a new process in a suspended state, replace its code with malicious code, and then resume it or inject malicious code into the address space of an already-running process—often an authentic one. Malware is deployed in malicious attachments, links, macros, and command and scripting interpreters or languages that are delivered to the victim’s machine via spear-phishing emails, compromised legitimate websites, and applications. For persistence, attackers utilized the boot or logon auto-start execution technique. To evade detection in a targeted environment they used living off the land, Indicator removal on the host, and dynamic domain name system services techniques. With dynamic domain name services, criminal activity is concealed or obfuscated by the use of dynamic, or continuously changing, DNS services. Adversaries employ dynamic DNS in place of static IP addresses to make it more difficult for defenders to locate and disable theirinfrastructure. Along with wiping event logs and other artifacts produced by their tools or malware utilized in an attack, they also remove indications of compromise (IOCs). They initially collect data about the systems in a remote environment or network in order to facilitate remote control communication. They can use it to map the network, find servicesthat are available, and find possible targets to further exploit. In order to obtain access, attackers look for known flaws in server software and take advantage of them. Additionally,they used remote desktop exploit assaults, which entailed taking advantage of holes in remote desktop services to obtain unauthorized access to a network or machine. Through anetwork, users may access and manage a distant computer or server with the help of remotedesktop services. A web shell is uploaded by attackers to a web server that is weak. After installation, it offers backdoor access and remote control over the hacked remote server. All data communication from the local computer or through removable media is encrypted at the sender’s side and further decrypted at the receiver’s side. They use automated tools or scripts to automatically exfiltrate data to a predetermined location or use variousapplication, and non-application layers protocols or encapsulate one network protocol within another to evade network security measures.

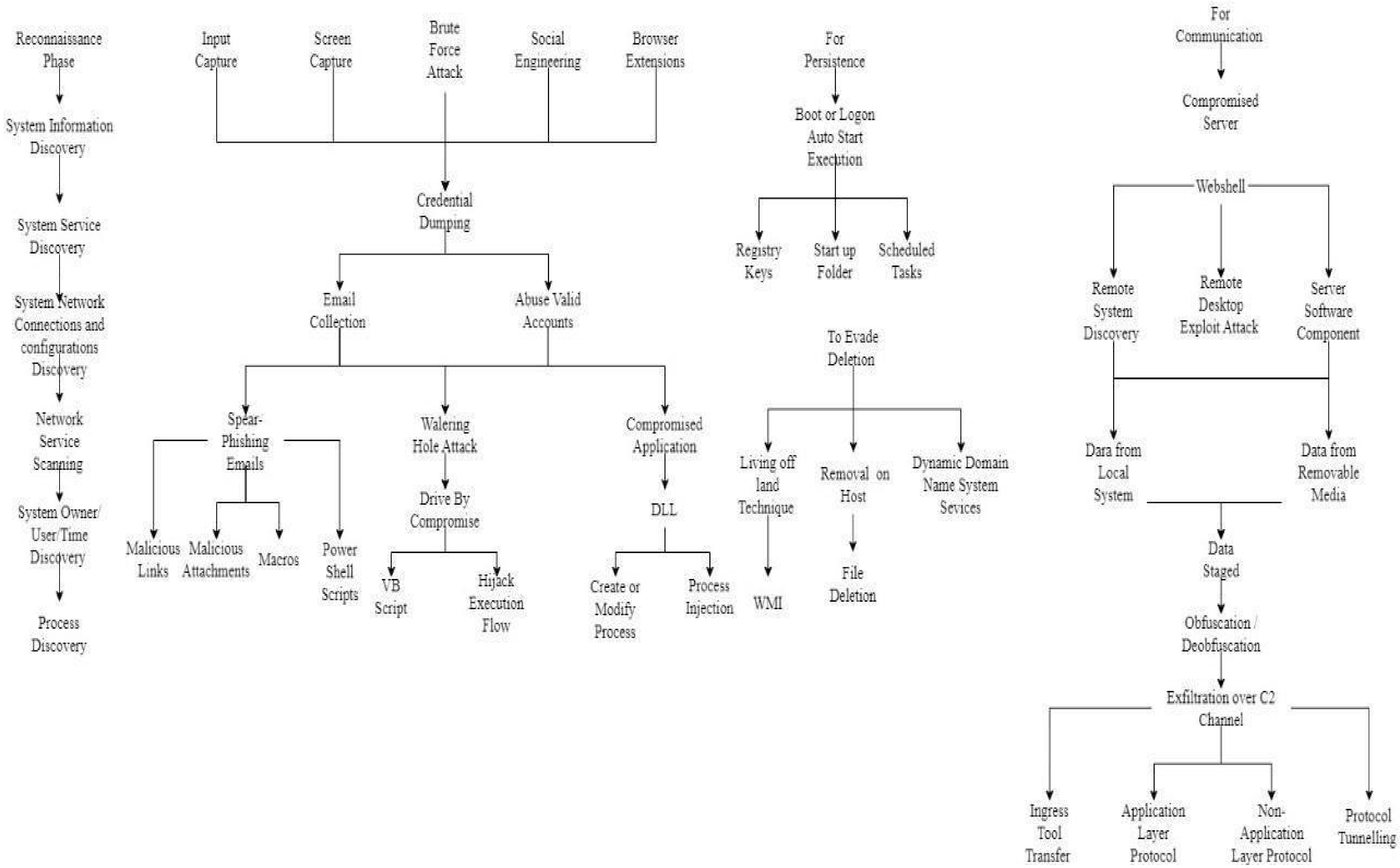

**Figure 18** Digital Intrusion Pathway-Group 8

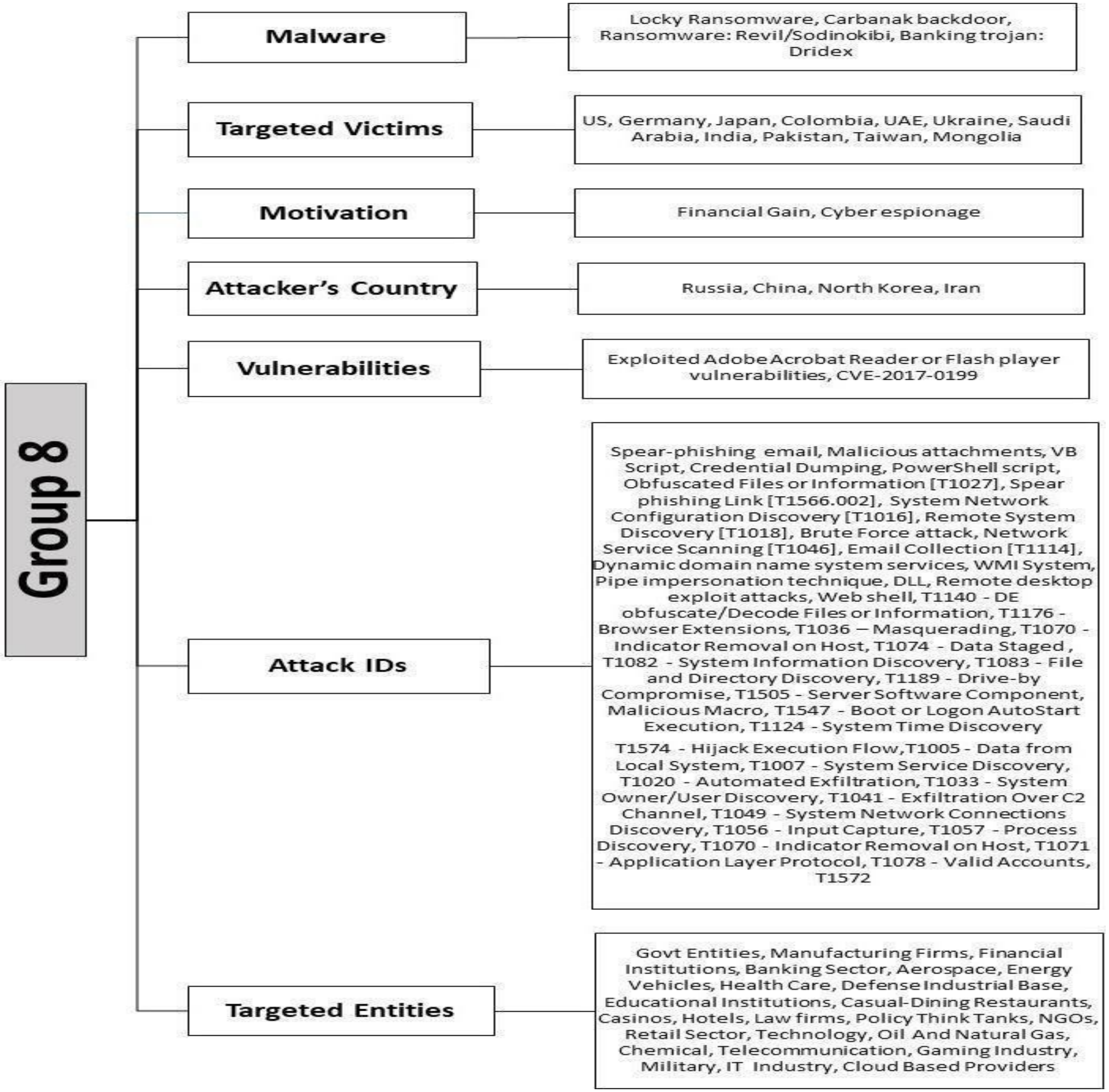

**Figure 19** Group8's Threat Profile

## 6.9 Cyber Criminal Group9’s Profile

As shown in Figure 21 the attacks were launched by a group of cybercriminals referred to as Group 9 affiliated with a variety of states like China, Russia, North Korea, and Iran. They targeted government entities, commercial entities, human rights activists, financial verticals, technology companies, military, manufacturing firms, banks, journalists, ministry of foreign affairs, high-tech companies, diplomat, education, US defence contractors, aerospace firm,telecommunication, thinktanks, political entities of US, South Korea, Japan, Vietnam, Hongkong, Thailand, India, Germany, Taiwan, Ukraine, Russia, China, UK, Saudi Arabia, Turkey, Lebanon, Israel, UAE, Myanmar, Kuwait, Cambodia, Nepal, Afghanistan, Pakistan, Bangladesh, China, Kazakhstan, Africa, Georgia, Czechia, Malaysia, Uzbekistan, Chile, Belarus, Mongolia, Austria, Mexico, Romania, Canada, Italy, France, Brazil for cyber espionage, and financial gain. The group employs a wide array of attack techniques, as indicated by the provided Attack IDs. Firstly they collect extensive information on the targeted system, including its configurations, user accounts, hardware, and software. After that, they attempt to locate files and folders on a compromised system. Credential harvesting can take place in a number of ways, such as through social engineering attacks, screen recording, taking screenshots, or collecting a user's keystrokes on their computer. Emails with spear phishing are among the most common examples of social engineering. They also infiltrate websites that their intended targets visits on an ongoing basis. Attackers load and run malicious DLL files during program execution by taking advantage of programs that are vulnerable or incorrectly configured. Additionally, they insert maliciouscode into a document or template that web application software would subsequently handle. Additionally, they hosted, distributed, or had control over harmful infrastructure or material via using popular and reliable cloud platforms and services. Group9 utilized different malware delivery mechanisms shown in Figure 20 which included malicious attachments, links, macros, and supply chain attacks. Malware is deployed in malicious attachments, links, macros, and command and scripting interpreters or languages that are delivered to the victim's machine via spear-phishing emails, compromised legitimate websites, applications, and cloud platforms. Attackers can execute arbitrary code by dynamically loading malicious programs ormodules into the memory of an active process through the use of legal application programming interfaces (APIs). Attackers used the boot or logon auto-start execution approach to be persistent. In addition to deleting or altering event records, logs, and other artifacts produced by their tools or malware during an attack, attackers also eliminate indications of compromise (IOCs) via software packing or binary obfuscation tools, and data transferred during remote control contact is compressed and encrypted. This detrimental data is then hidden via steganography with in seemingly benign files like photos, music files, or other digital media. They use various application layers, cryptographic protocols, and malicious tools or software to transfer data from compromised systems to their attacker's control servers and vice versa. These methods are employed to obtain access to target systems, compromise them, and accomplish their goals.

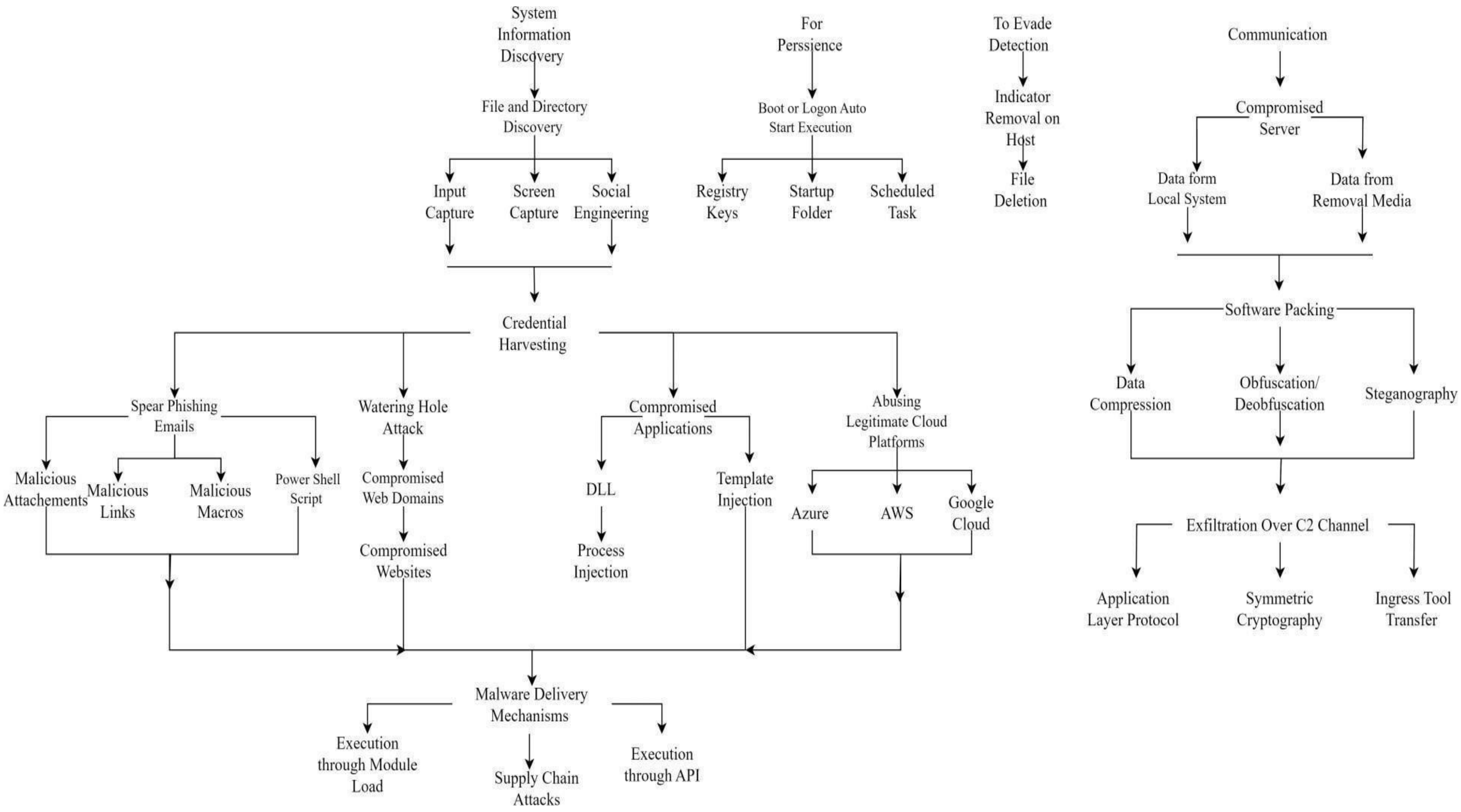

**Figure 20** Digital Intrusion Pathway-Group 9

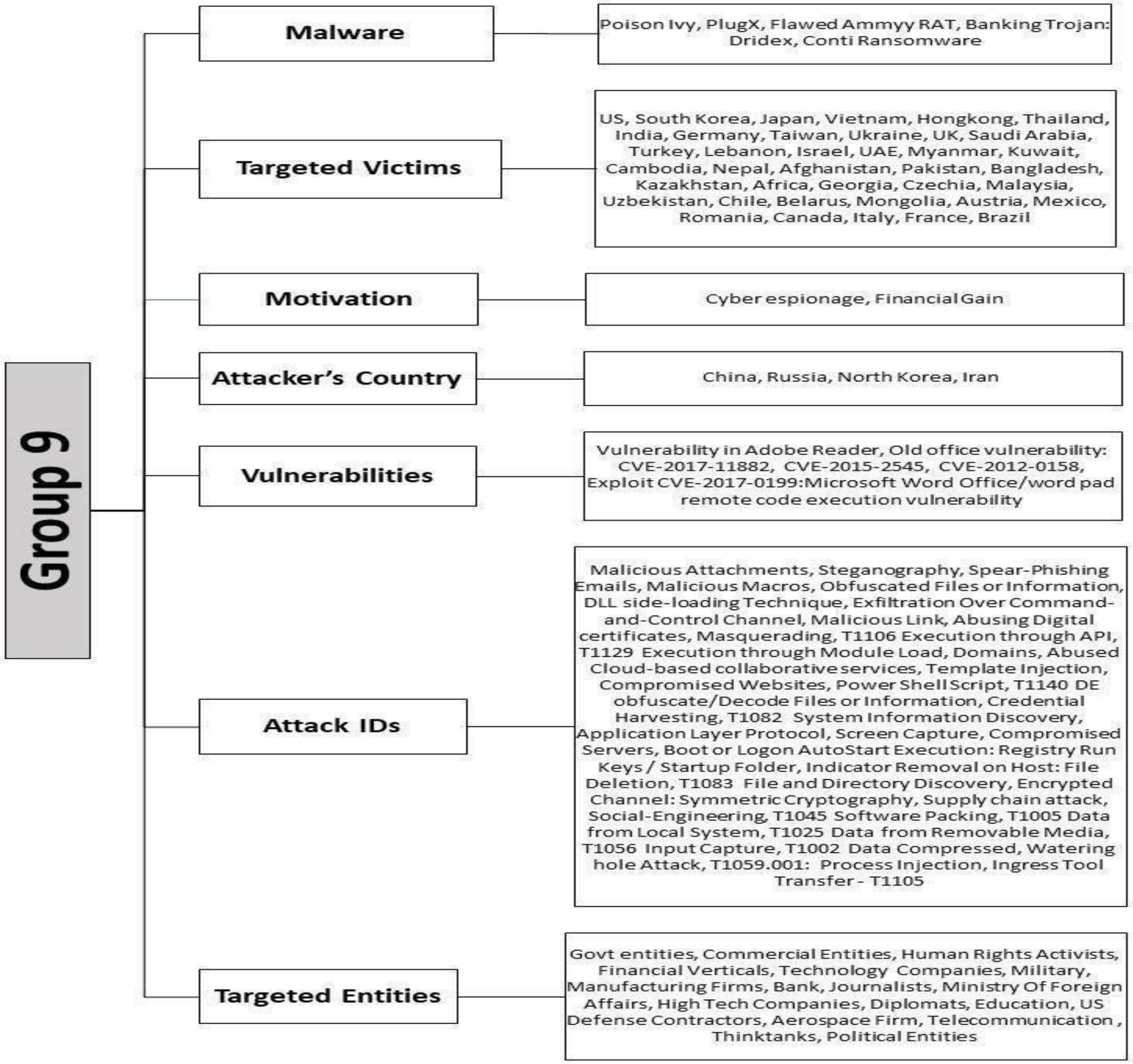

**Figure 21** Group9's Threat Profile

## 6.10 Cyber Criminal Group10's Profile

The profile of Group 10 is shown in Figure 23. They use vulnerabilities in Microsoft Exchange servers, Adobe Flash Player, popular regional IT Products, Microsoft Office, Internet Explorer, VPN servers, Microsoft Exchange proxy login flaws, Windows, Adobe Cold Fusion servers, Oracle WebLogic servers for distributing malicious payloads including Carbanak, Trick bot, PupyRAT, Cyclops Blink, PlugX, and DEPLOYLOG. Their targets encompass journalists, think tanks, government departments, human rights activists, NGOs, academic organizations, aerospace, finance sectors, politicians, retail markets, telecommunications companies, manufacturing, media, military, energy sector, oil and gas industry, chemical industry, defence companies, hospitality, restaurant, hotel networks, healthcare, diplomatic institutions, software development companies, electric utilities, transportation sector, insurance, technology, casinos, automotive, supply chain, researchers, law firms, mining companies, pharmaceutical companies. They employed various attack techniques in order to target the US, Hongkong, South Korea, Taiwan, Russia, Malaysia, North Africa, China, Singapore, Germany, Japan, Canada, Israel, Ukraine, the UK, France, Mexico, Australia, India, Bangladesh, Turkey, Thailand, Myanmar, Saudi Arabia, Kuwait, UAE, Poland, Spain, Romania, Palestine, Lebanon, Iranian nations for the purpose of cyber- espionage, financial and political gain. To extract plain text passwords, hashes, pins, and other sensitive authentication credentials from memory or files on infected Windows computers, attackers used well-known post-exploitation tools in conjunction with simulated cyberattacks. They frequently fabricate or use phoney online personas on social media platforms, taking advantage of people's trust and psychology using social engineering techniques. Attackers imitate reliable organizations and send emails from compromised accounts that seem authentic. They also infiltrate websites that their intended targets visit frequently. They primarily impact websites that their intended audience trusts as well, including news organizations and official government websites. If the group fails to exploit websites, they will use code code-signingtechniques to make their own malicious websites and applications and make them legitimate and trustworthy by signing them with a valid digital certificate. On cloud computing platforms such as AWS (Amazon Web Services), Azure, Google Cloud, and Dropbox, they set up or breach into authentic accounts. Group10 utilized different malwaredelivery mechanisms shown in Figure 22 which included malicious attachments, links, macros, supply chain attacks, and malicious HTML applications. Malware is deployed in malicious attachments, links, macros, and command and scripting interpreters or languages that are delivered to the victim's machine via spear- phishing emails, compromised legitimate websites, and cloud platforms. They utilized the Living Off the Land (LOL) technique on a target system or network such as Windows Management Instrumentation (WMI). WMI is abused by attackers to create persistence on an accessed system. To keep their access hidden, they design scheduled tasks or event triggers that use Windows Management Instruments (WMI) to execute malicious scripts periodically.

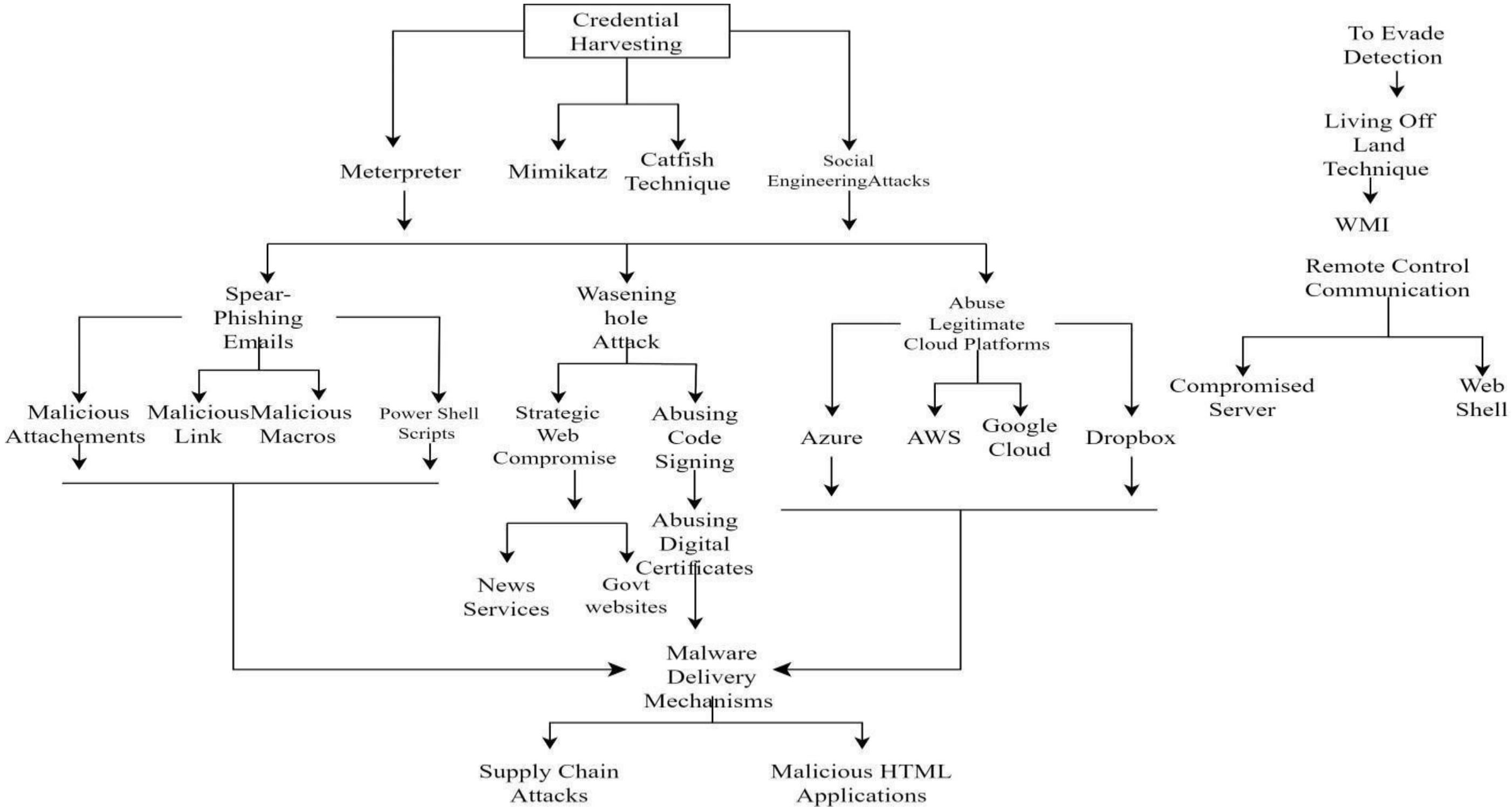

**Figure 22** Digital Intrusion Pathway-Group 10

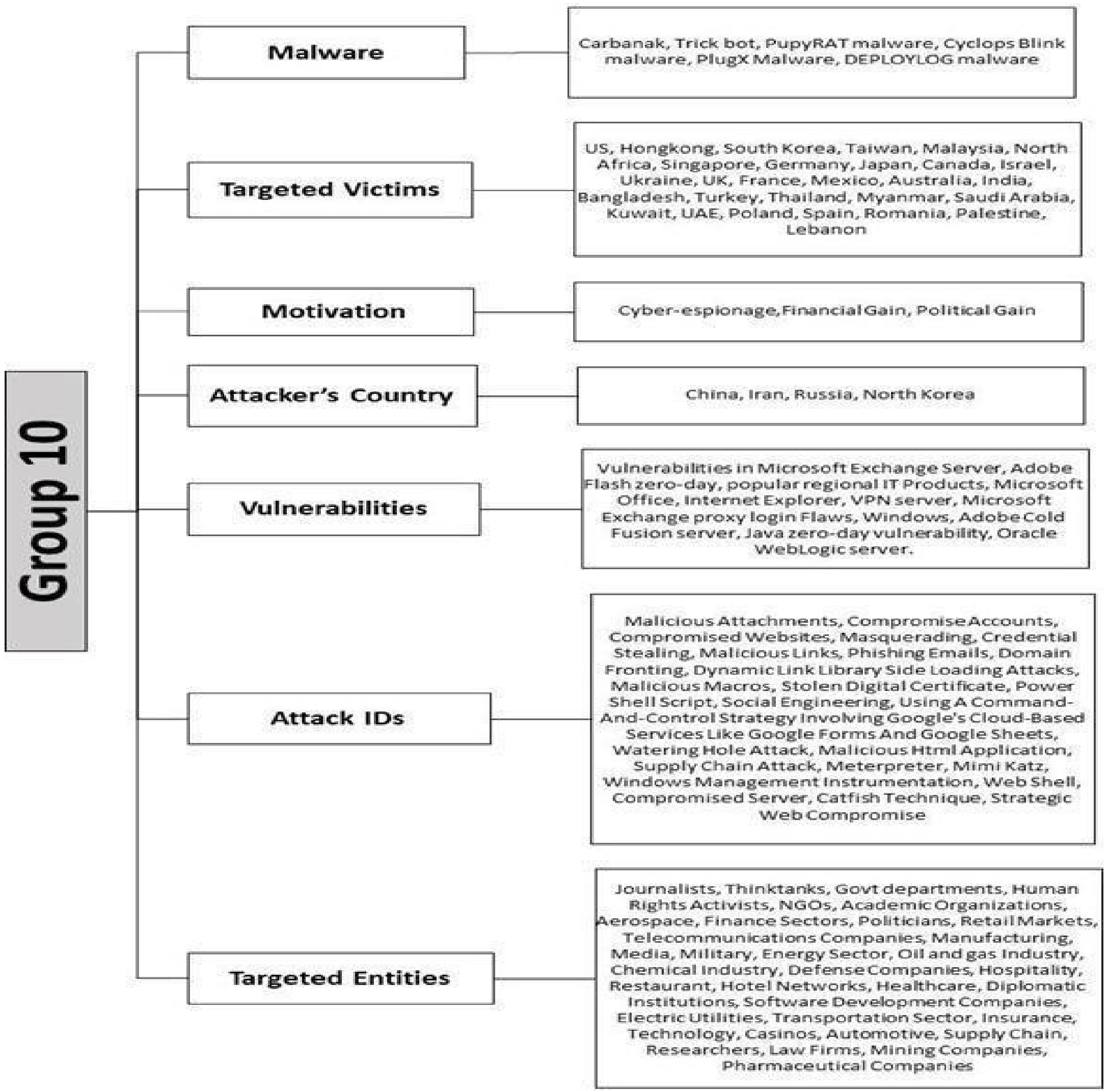

**Figure 23** Group10's Threat Profile

## 6.11 Cyber Criminal Group11's Profile

Group 11 belongs to China, North Korea, and Russian nations as illustrated in Figure 24. They leverage Internet Explorer, Microsoft Office, Exploits for vulnerabilities in Cisco products, CVE-2021-44228;Log4 shell vulnerability, CVE-2014-4114(windows OLE remote code execution flaw), windows kernel vulnerability(CVE-2017-0005). They interact with victims' computers in South Korea, the US, the Philippines, Denmark, India, the UK, Colombia, Switzerland, Saudi Arabia, Bahrain, Russia, Japan, Singapore, Myanmar, Iraq, UAE, Ukraine, Poland, and Canada by using various attack methods.

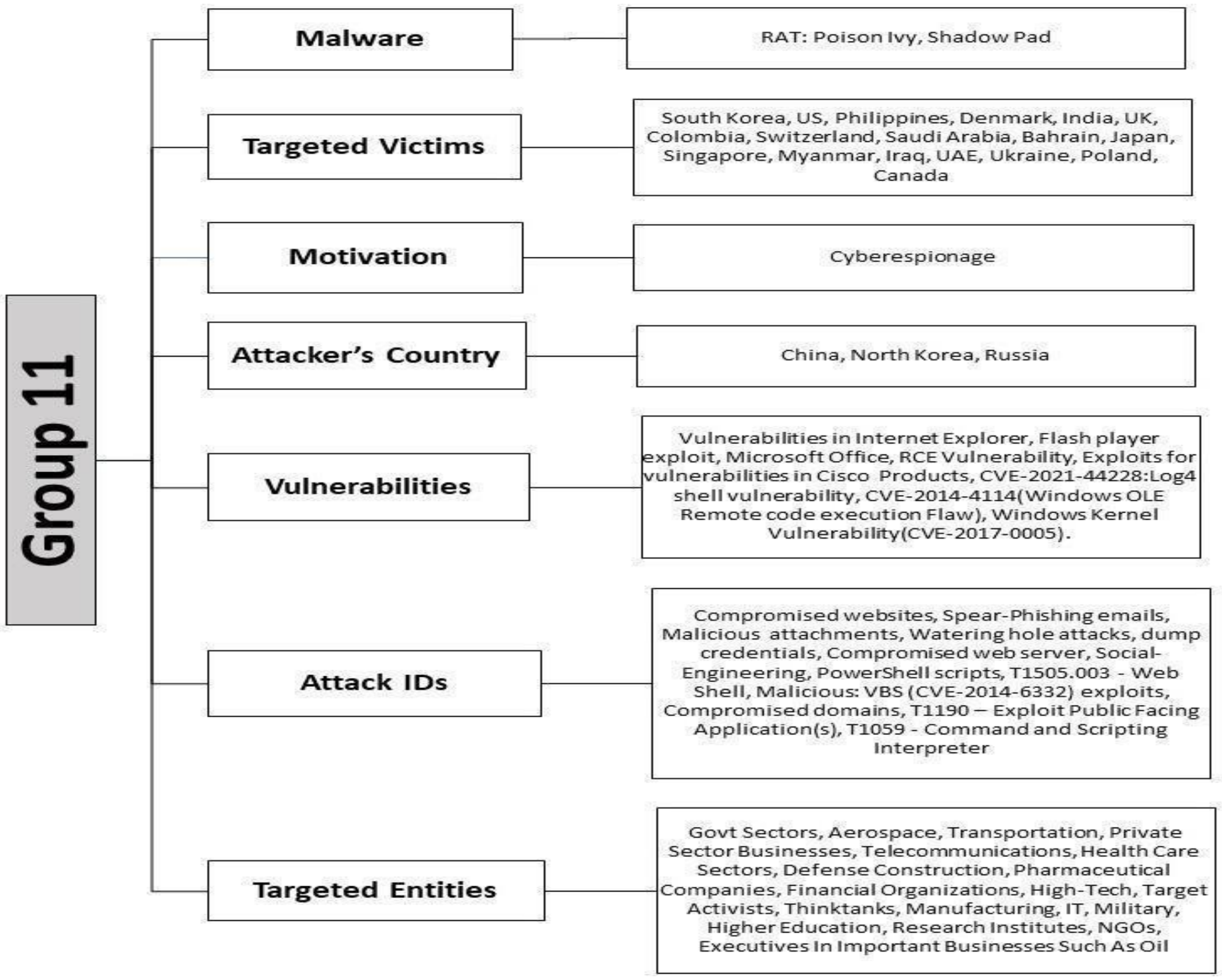

**Figure 24** Group11's Threat Profile

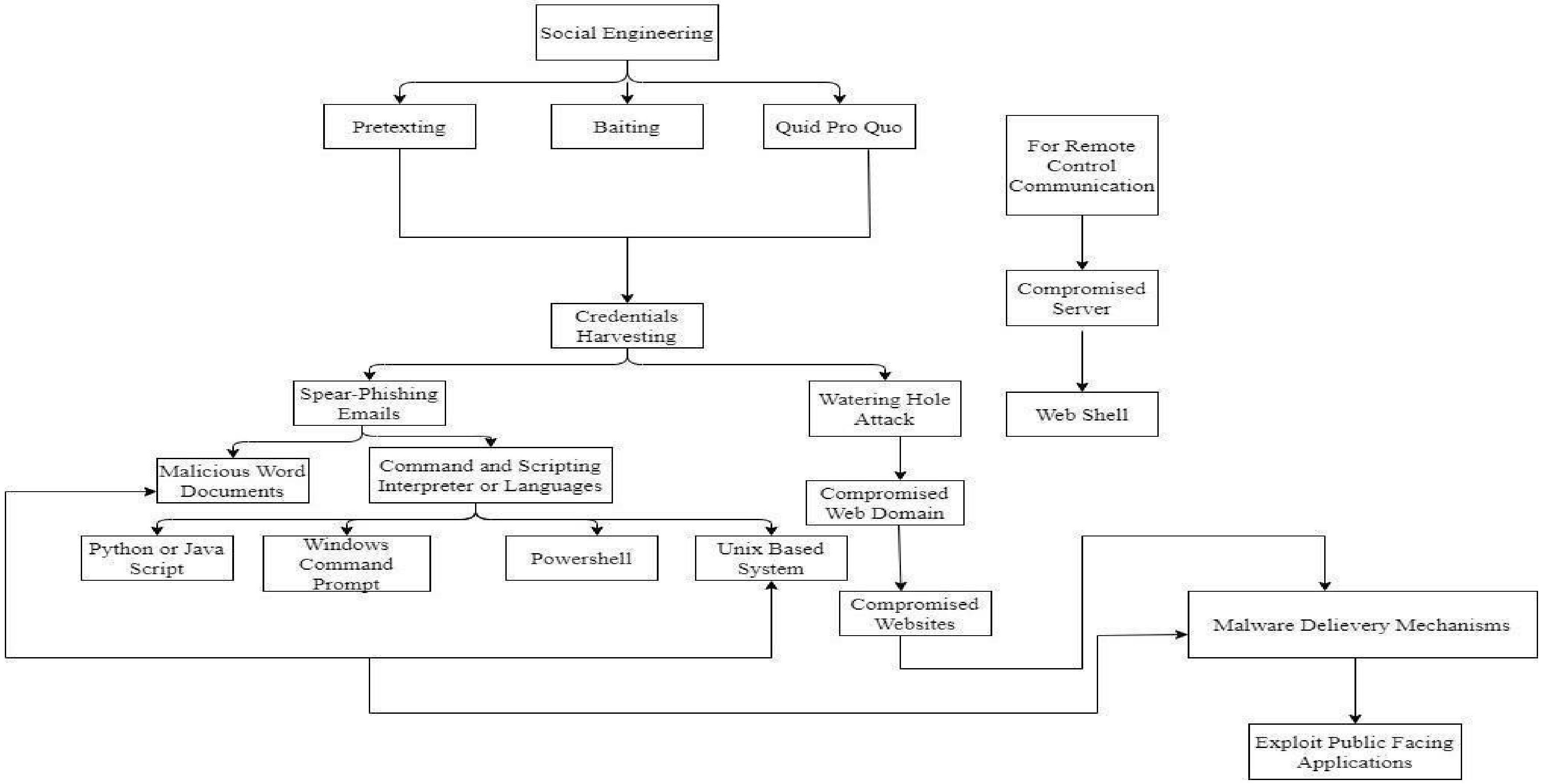

**Figure 25** Digital Intrusion Pathway-Group 11

To accomplish their goal of cyber espionage, they concentrate on targeting govt sectors, aerospace, transportation, private sector businesses, telecommunications, health care sectors, defence construction, pharmaceutical companies, financial organizations, high-tech, target activists, thinktanks,manufacturing, IT, military, higher education, research institutes, NGOs, executives in important businesses such as oil.

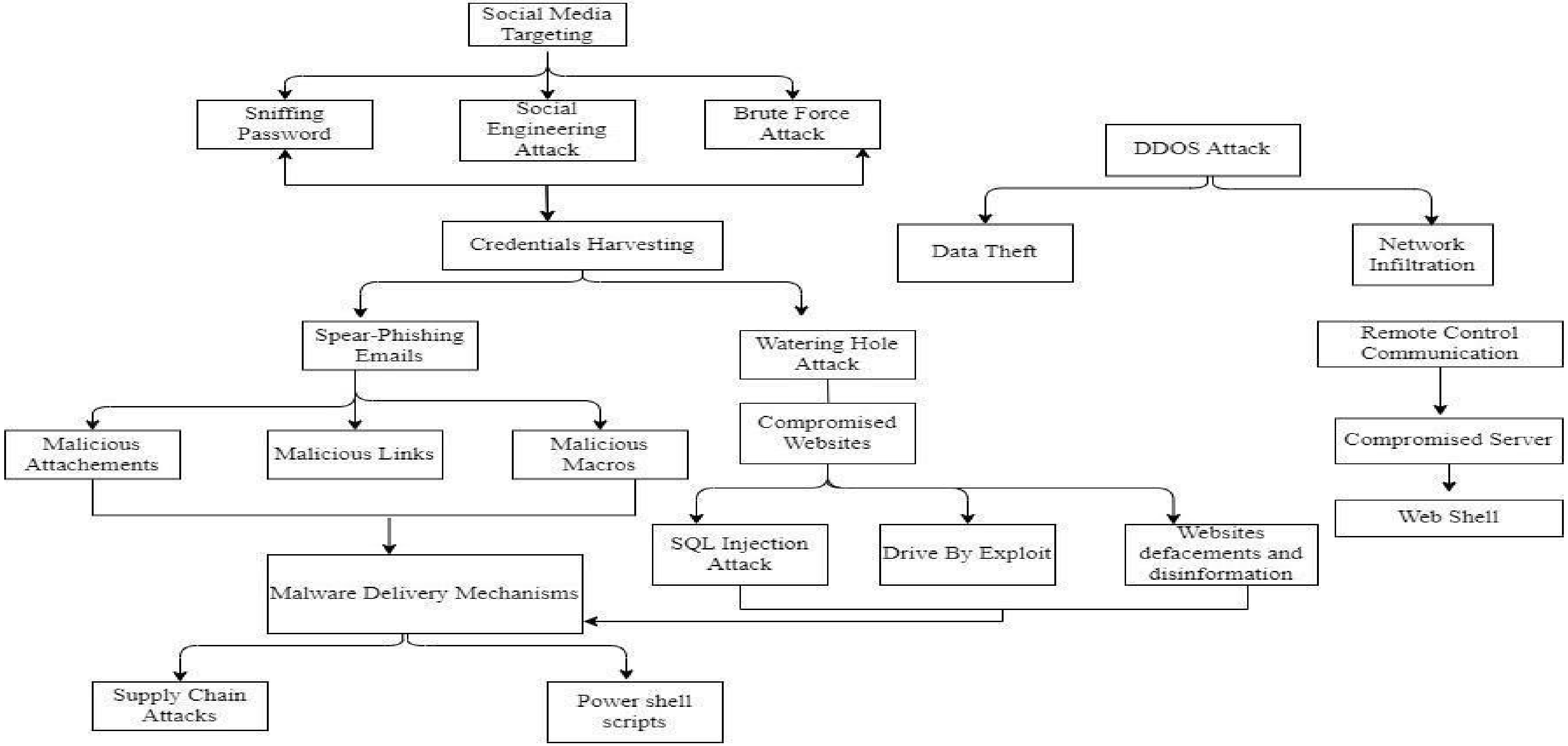

**Figure 26** Digital Intrusion Pathway-Group 12

As presented in Figure 25 among these attack techniques are social engineering assaults. Hackers send emails from hacked accounts that seem authentic and pretend to be reputable organizations like banks, governments, or well-known businesses. These "watering hole" websites are deliberately chosen by the attackers because they are aware that their victims— typically certain people or organizations—are likely to visit their legitimate website domains that have been compromised. Malicious word documents, command and scripting interpreters or languages are used by attackers to target weaknesses in software programs that are accessible via the internet and distribute malware to specific victims. Malware is transferred to the victim's computer by spear-phishing emails or hacked official websites. It is installed in malicious word attachments, command and scripting interpreters, or languages such as Python, javascript, windows command prompt, powershell, and Unix-based systems. With the use of the remote-control interface offered by the Poison Ivy virus, attackers may remotely infect a victim's computer without having to personally visit them. A web shell is uploaded by attackers to a web server that is weak. After installation, it offers backdoor access and remote control over the hacked remote server.

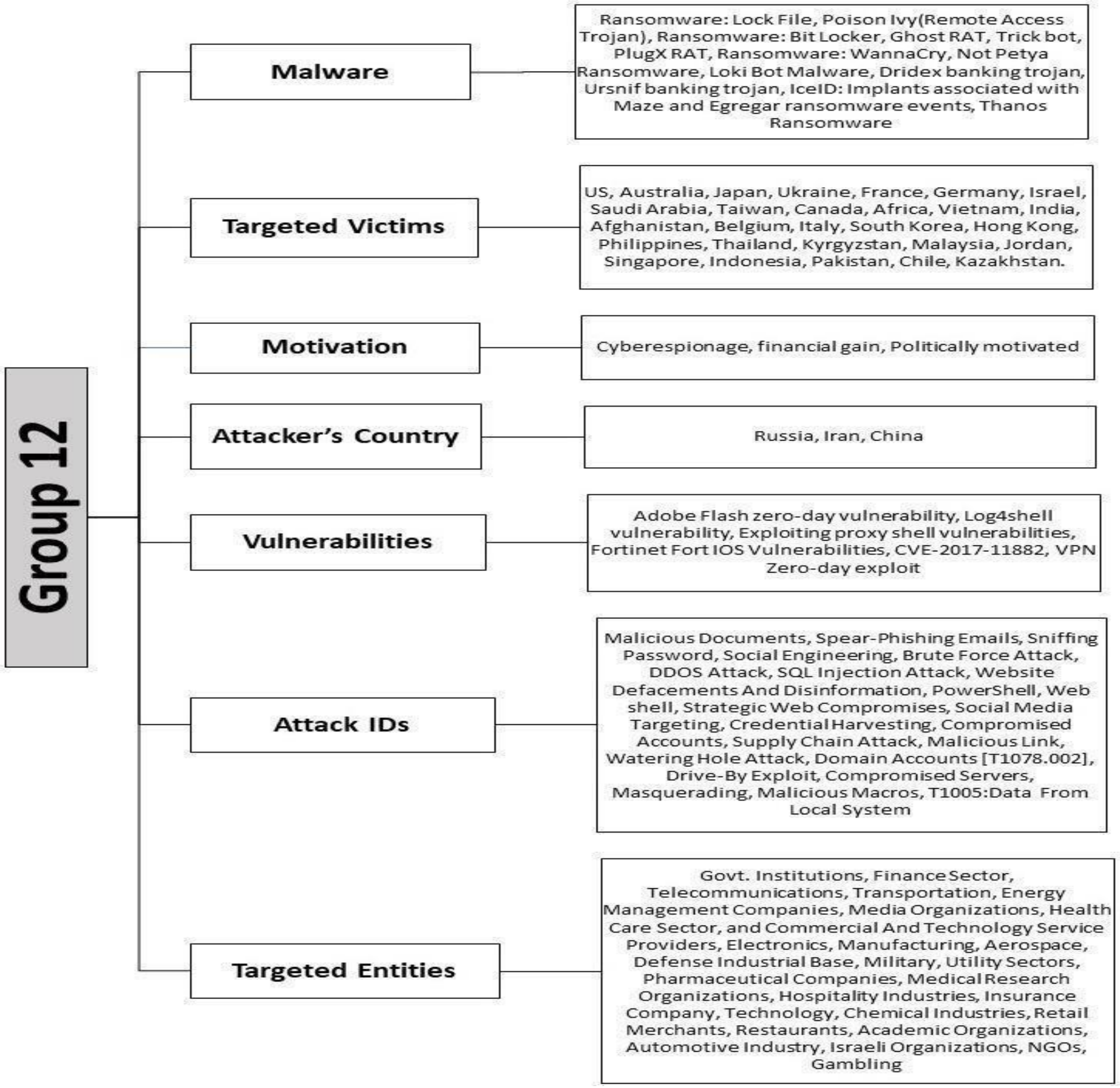

**Figure 27** Group12's Threat Profile

## 6.12 Cyber Criminal Group12's Profile

The cyber threat Group 12's profile is shown in Figure 27, along with the malware, target victims, industries, weaknesses, motivations, and attack techniques utilized. The group is thought to have ties to a number of nations, including Russia, Iran, China, and Korea. It targets a wide variety of individuals in several nations, notably the US, Australia, Japan, Ukraine, France, Germany, Russia, Israel, Saudi Arabia, China, Taiwan, Canada, Africa, Vietnam, India, Afghanistan, Belgium, Italy, South Korea, Hong Kong, Philippines, Thailand, Kyrgyzstan, Malaysia, Jordan, Singapore, Indonesia, Pakistan, Chile, Kazakhstan, Iran. Exploited certain weaknesses such as Adobe Flash zero-day, Log4shell, proxy shell, Fortinet Fort IOS, CVE-2017-11882, VPN in order to deploy a variety of malware such as Lock File, Poison Ivy(Remote Access Trojan), Bit Locker, Ghost RAT, Trick bot, PlugX RAT, WannaCry, Not Petya, Loki Bot Malware, Dridex banking trojan, Ursnif banking trojan, IceID; Implants associated with Maze and Egregar ransomware events, Thanos Ransomware by utilizing different attack methods. They target government institutions, the finance sector, telecommunications, transportation, energy management companies, media organizations, sectors, pharmaceutical companies, medical research organizations, the healthcare sector, commercial and technology service providers, electronics, manufacturing, aerospace, defence industrial base, military, utility, hospitality industries, insurance companies, technology, chemical industries, retail merchants, restaurants, academic organizations, automotive industry, Israeli organizations, and NGOs. This wide range of targets points to a complex and extensive cyber operation for cyberespionage, financial and political gain. Social media platforms are used by attackers to get personal or organizational information in preparation for cyberattacks. To engage with their targets, they fabricate social media profiles or assume the identity of real people. Through social engineering, they learn about the relationships, interests, and affiliations of their targets. Attackers get victims' usernames and passwords. Numerous techniques, such as network sniffing tools and brute force attacks, might cause this. Emails with spear phishing are among the most common examples of social engineering. Hackers send emails from compromised accounts that seem authentic and pretend to be reputable organizations like banks, governments, or well-known businesses. These "watering hole" websites are deliberately chosen by the attackers because they are aware that their victims typically certain people or organizations are likely to visit them after other genuine website domains have been compromised. The three methods they use to strategically breach websites are drive-by exploits, Website defacements, and disinformation. With these, they may serve malware, spread detrimental material, or steal sensitive information from users as shown in Figure 26. By using a technique known as SQL injection technique, attackers can extract, modify, or delete data from a database while in a drive-by exploit, the victim's device by merely visiting a malicious website or clicking on malicious advertisements. In Website defacements and disinformation techniques, they obtain unauthorized access to the web server or content management system (CMS) and replace the original text with their own words, pictures, or misinformation. They also used denial-of-service (DDOS) operations to breach networks and steal data. Malicious Word documents, hyperlinks, PowerShell scripts, macros, and supply chain assaults are some of themethods used by attackers to infect their intended targets with malware. Attackers that target a company's partners, suppliers, or service providers aim to subtly undermine the company through supply chain assaults. Spear-phishing emails or hacked official websites, malicious word attachments, links, macros, and powershell scripts are used to install malware on a victim's computer. With the use of the remote-control interface offered by the Poison Ivy virus, attackers may remotely infect a victim’s computer without having to personally visit them. A web shell is uploaded by attackers to a web server that is vulnerable. After installation, it offers backdoor access and remote control over the hacked remote server.

## 7. Research Findings and Discussion

An overview of attack techniques used against each category of cybercriminal is presented in Table 9. The total of attack techniques is 108 out of which we have determined what makes each cybercriminal gang unique as well as common points between them. After analysis, we have observed that most cyber threat actor groups steal sensitive information by exploiting human psychology and belief via the use of social engineering attacks. The intention is to mislead victims into exposing confidential details such as credit card numbers, login passwords, or personal information. Emails with spear phishing are among the most frequently encountered instances of social engineering. Hackers send emails from accounts that have been compromised that seem authentic and pretend to be reputable organizations like banks, governments, or well-known businesses. They also infiltrate websites that their intended targets visit on a regular basis. These "watering hole" websites are deliberately chosen by the attackers because they are aware that their victims—typically certain people or organizations— are likely to visit them. Additionally, they also engaged in abusing digital certificates in the following ways. Hackers might use stolen credentials or false identities to get digital certificates through fraudulent means. Either way, they started certificate expiry attacks, in which the attacker targets certificates that are about to expire since they know that consumers could become less careful when they see a notice about an expired certificate. They are also involved in certificate spoofing, which entails creating phoney certificates that closely resemble real ones in order to deceive consumers into believing they are connected to a reliable website. Man-in-the-middle (MITM) attacks may result from this, in which the attacker intercepts data being sent back and forth between the user and the trustworthy server. They frequently deployed the malware in malicious attachments, links, macros, and powershell scripts and delivered it to the victim’s machine via spear-phishing emails or compromised legitimate websites. The attacker may essentially take control of the target machine remotely and maintain persistence by utilizing the boot or logon AutoStart execution technique. They also create scheduled tasks that execute malicious scripts or code at specific times, such as during system startup or user logon or they create registry keys to ensure that their malicious code or malware runs every time the system boots or a specific user logs in. They also place malicious shortcuts or executable files in the system's startup folders, such as the "Startup" folder within the "Start" menu or the "Startup" folder in the user's profile directory. These three techniques were used for performing boot or logon auto-start execution. We also draw attention to offensive strategies that set one group apart from another. These include Off The Shelf Technique, Exploitation of unpatched VPN Services, Leveraged Social Media Platforms, hijacking a Real Internet Email Account, Boost Write Loader, Proxy Open Reverse Shell, SSH Tunnelling, DNS Tunnelling, Shortened URL Services, Application Window Discovery, Exploit Data From Network Shared Drive, Exploitation for Client Execution, Shortcut Modification, Hidden Windows, Hosting Java Script Profiling, Virtualization Sandbox Evasion, Commonly Used Ports, Standard Cryptographic Protocol, Fall Back Channels, Automated Exfiltration, Use of Fake Applications and Companies, Network Share Discovery, Windows Component Object Model And Distributed, Pass the hash, Hooking, Abuse Data Exchange Feature In Windows, Abuse Elevation Control Mechanisms, Signed Proxy Execution, System Service Discovery, NetworkService Scanning, Browser Extensions, Dynamic Domain Name System Services, Remote System Discovery, Remote Desktop Exploit Attack, Server Software Component, Protocol Tunnelling, Execution through module load, Execution through API, Mimi Katz, Catfish Technique, Malicious HTML Applications, Exploit Public Facing Applications, Website defacements and disinformation.

**Table 9** An In-Depth Analysis of Attack Techniques Against Cyber-Criminal Groups

| Attack IDs | Profile 1 | Profile 2 | Profile 3 | Profile 4 | Profile 5 | Profile 6 | Profile 7 | Profile 8 | Profile 9 | Profile 10 | Profile 11 | Profile 12 |
|---|---|---|---|---|---|---|---|---|---|---|---|---|
| Social Engineering Attack | ✓ | ✓ | ✓ | ✓ | | ✓ | ✓ | ✓ | ✓ | ✓ | ✓ | ✓ |
| Brute Force Attack | ✓ | | | | | | ✓ | ✓ | | | | ✓ |
| Keylogging | ✓ | ✓ | | | | ✓ | ✓ | | | | | |
| Compromised Email Account | ✓ | | | ✓ | | ✓ | ✓ | ✓ | | | | |
| Spear-Phishing emails | ✓ | ✓ | ✓ | ✓ | ✓ | ✓ | ✓ | ✓ | ✓ | ✓ | ✓ | ✓ |
| Malicious Attachments | ✓ | ✓ | ✓ | | ✓ | ✓ | ✓ | ✓ | ✓ | ✓ | ✓ | ✓ |
| Malicious Links | ✓ | ✓ | | | | ✓ | ✓ | ✓ | ✓ | ✓ | | ✓ |
| Malicious Macros | ✓ | ✓ | | | ✓ | ✓ | ✓ | ✓ | ✓ | ✓ | | ✓ |
| Watering Hole Attack | ✓ | ✓ | | | ✓ | ✓ | ✓ | ✓ | ✓ | ✓ | ✓ | ✓ |
| Compromised Web domains | ✓ | | | | ✓ | ✓ | | | ✓ | | ✓ | |
| Compromised applications | ✓ | ✓ | | | | | ✓ | ✓ | ✓ | | | |
| Dynamic link library side loading technique | ✓ | ✓ | | | | | | ✓ | ✓ | | | |
| Abusing code signing | ✓ | ✓ | | | | | | | | ✓ | | |
| Abusing Digital Certificates | ✓ | ✓ | ✓ | ✓ | | ✓ | | | | ✓ | | |
| Off The Shelf Technique | ✓ | | | | | | | | | | | |
| Powershell Scripts | ✓ | | ✓ | | ✓ | ✓ | ✓ | ✓ | ✓ | ✓ | ✓ | ✓ |
| Infect Removable Media | ✓ | ✓ | | | ✓ | ✓ | | | | | | |
| Exploitation of unpatched VPN Services | ✓ | | | | | | | | | | | |
| Remote Control Communication | ✓ | ✓ | | | ✓ | | ✓ | | | | ✓ | ✓ |
| Compromised servers | ✓ | ✓ | ✓ | | | | | ✓ | ✓ | ✓ | ✓ | ✓ |
| Web shell | ✓ | | | | | | | ✓ | | ✓ | ✓ | ✓ |
| Leveraged social Media Platforms | ✓ | | | | | | | | | | | |
| Steganography | ✓ | ✓ | | | ✓ | | ✓ | | ✓ | | | |
| Hijack A Real Internet Email Account | | ✓ | | | | | | | | | | |
| VB Script | | ✓ | | | | | | | | | | |
| Compromised Websites | ✓ | ✓ | ✓ | | ✓ | ✓ | | | ✓ | | ✓ | ✓ |
| Abusing Legitimate Cloud Services | | ✓ | | | | | | | ✓ | ✓ | | |
| Supply Chain Attack | | ✓ | ✓ | | | ✓ | | | ✓ | ✓ | | ✓ |
| Boost Write Loader | | ✓ | | | | | | | | | | |
| Living Off the Land Technique: WMI | | ✓ | | | | | ✓ | ✓ | | ✓ | | |
| Proxy Open Reverse Shell | | ✓ | | | | | | | | | | |
| Registry Keys | | ✓ | | | ✓ | ✓ | ✓ | ✓ | ✓ | | | |
| SSH Tunnelling | | ✓ | | | | | | | | | | |
| DNS Tunnelling | | ✓ | | | | | | | | | | |
| Shortened URL Services | | | | ✓ | | | | | | | | |
| Windows Command Prompt | | | ✓ | | ✓ | | ✓ | | | | ✓ | |
| Python Or Java Script | | | ✓ | | ✓ | | ✓ | | | | ✓ | |
| Unix Based System | | | ✓ | | ✓ | | ✓ | | | | ✓ | |
| Hijack Execution Flow | | | ✓ | | | | | ✓ | | | | |
| Startup Folder | | | ✓ | | ✓ | ✓ | ✓ | ✓ | ✓ | | | |
| Scheduled Task | | | ✓ | | ✓ | ✓ | ✓ | ✓ | ✓ | | | |
| Obfuscation and Deobfuscation of files and Information | | | ✓ | | ✓ | | ✓ | ✓ | ✓ | | | |
| Process Injection | | | ✓ | | | | | ✓ | ✓ | | | |
| Ingress Tool Transfer | | | ✓ | | | | | ✓ | ✓ | | | |
| Exploit Trusted Relationship | | | ✓ | | | | | | | | | |
| System Information Discovery | | | | | ✓ | | ✓ | ✓ | ✓ | | | |
| File And Directory Discovery | | | | | ✓ | | | | ✓ | | | |
| System Network Connections And Configuration Discovery | | | | | ✓ | | ✓ | ✓ | | | | |
| System Owner/User Discovery | | | | | ✓ | | ✓ | ✓ | | | | |
| System Time Discovery | | | | | ✓ | | | ✓ | | | | |
| Process Discovery | | | | | ✓ | | ✓ | ✓ | | | | |
| Application Window Discovery | | | | | ✓ | | | | | | | |
| Exploit Data From Network Shared Drive | | | | | ✓ | | | | | | | |
| Input Capture | | | | | ✓ | | | ✓ | ✓ | | | |
| Screen Capture | | | | | ✓ | | | ✓ | ✓ | | | |
| Exploitation for Client Execution | | | | | ✓ | | | | | | | |
| Shortcut Modification | | | | | ✓ | | | | | | | |
| Hidden Windows | | | | | ✓ | | | | | | | |
| Hosting Java Script Profiling | | | | | ✓ | | | | | | | |
| File Deletion | | | | | ✓ | | ✓ | ✓ | ✓ | | | |
| Virtualization Sandbox Evasion | | | | | ✓ | | | | | | | |
| Time Stamp Manipulation | | | | | ✓ | | ✓ | | | | | |
| Data From the Local System | | | | | ✓ | | ✓ | ✓ | ✓ | | | |
| Remote File Copy | | | | | ✓ | | | | | | | |
| Data Compression | | | | | ✓ | | ✓ | | ✓ | | | |
| Domain Fronting | | | | | ✓ | | | | | | | |
| Commonly Used Ports | | | | | ✓ | | | | | | | |
| Standard Application Layer Protocol | | | | | ✓ | | ✓ | ✓ | ✓ | | | |
| Non-Standard Application Layer Protocol | | | | | ✓ | | | ✓ | | | | |
| Standard Cryptographic Protocol | | | | | ✓ | | | | | | | |
| Fall Back Channels | | | | | ✓ | | | | | | | |
| Automated Exfiltration | | | | | ✓ | | | | | | | |
| Use of Fake Applications and Companies | | | | | | ✓ | | | | | | |
| DDOS Attacks | | | | | | ✓ | ✓ | | | | | ✓ |
| Native API | | | | | | ✓ | ✓ | | | | | |
| Network Share Discovery | | | | | | | ✓ | | | | | |
| Windows Component Object Model And Distributed Component Object Model | | | | | | | ✓ | | | | | |
| Obfuscated Meterpreter Stager | | | | | | | ✓ | | | ✓ | | |
| Pass the hash | | | | | | | ✓ | | | | | |
| Windows Service | | | | | | | ✓ | ✓ | | | | |
| Drive-By Compromise | | | | | | | ✓ | ✓ | | | | ✓ |
| Strategic Web Compromise | | | | | | | ✓ | | | ✓ | | |
| VB Script | | | | | | | ✓ | ✓ | | | | |
| Hooking | | | | | | | ✓ | | | | | |
| Remote Template Injection | | | | | | | ✓ | | ✓ | | | |
| SQL Injection Attack | | | | | | | ✓ | | | | | ✓ |
| DLL | | | | | | | ✓ | | | | | |
| Abuse Data Exchange Feature In Windows | | | | | | | ✓ | | | | | |
| Abuse Elevation Control Mechanisms | | | | | | | ✓ | | | | | |
| Signed Proxy Execution | | | | | | | ✓ | | | | | |
| System Service Discovery | | | | | | | | ✓ | | | | |
| Network Service Scanning | | | | | | | | ✓ | | | | |
| Browser Extensions | | | | | | | | ✓ | | | | |
| Dynamic Domain Name System Services | | | | | | | | ✓ | | | | |
| Remote System Discovery | | | | | | | | ✓ | | | | |
| Remote Desktop Exploit Attack | | | | | | | | ✓ | | | | |
| Server Software Component | | | | | | | | ✓ | | | | |
| Data From Removable Media | | | | | | | | ✓ | ✓ | | | |
| Protocol Tunnelling | | | | | | | | ✓ | | | | |
| Execution through module load | | | | | | | | | ✓ | | | |
| Execution through API | | | | | | | | | ✓ | | | |
| Mimi Katz | | | | | | | | | | ✓ | | |
| Catfish Technique | | | | | | | | | | ✓ | | |
| Malicious HTML Applications | | | | | | | | | | ✓ | | |
| Exploit Public Facing Applications | | | | | | | | | | | ✓ | |
| Sniffing Password | | | | | | | | | | | | ✓ |
| Website defacements and disinformation | | | | | | | | | | | | ✓ |

## 8. Conclusion And Future Work

A rising number of cyberattacks are becoming a major concern in the modern world. Experts in the cyber security world will be able to predict upcoming assaults and take preventative actions to protect against them by profiling cybercriminal groups. In the existing literature, multiple machine learning-based methods are used for attacker profiling, but they are all restricted to supervised approaches and consider a small numberof features. These approaches are costly in terms of structuring as well as labeling data. Therefore, we have proposed an unsupervised clustering-based approach for profiling cyber criminals based on complete contextual information on cyber-attacks available in textual cyber threat incident documents. One publicly accessible threat source alone has one billion threat occurrences, and the amount of cyber threat information is expanding dramatically. These reports contain valuable information, but the size of the data set makes it challenging for cyber security analysts to gather the information and then analyze it to develop effective defence strategies, which is a time-consuming process. The final outcome of the proposed framework is theprofile of cyber-criminal groups that experts can utilize to create mitigation strategies or defence mechanisms to get rid of these cyber-attacks. The performance of this approach is evaluated using multiple parameters such as the silhouette coefficient and Davies Bouldin index. A proposed framework achieves a 0.095 silhouette score and a 2.212 Davies Bouldin index. The efficient agglomerative hierarchical clustering technique achievesa **4%** increase in Silhouette score and a **5%** increase in the Davies Bouldin Index as compared to standard k means and the agglomerative hierarchal clustering approach. However, this proposed technique is not without limitations. There is still a need to enhance the framework because it is dependent on the data of threats. In the future, we will work on defence mechanisms or mitigation strategies against the profiles of cybercriminal groups.

## 9. References

[1] "How Many Cyber Attacks Happen Per Day in 2023?" Accessed: Nov. 15, 2023. [Online]. Available: https://techjury.net/blog/how-many-cyber-attacks-per-day/

[2] "An Overview of the Recent Cyber-Attacks 2022." Accessed: Nov. 16, 2023. [Online]. Available: https://www.sangfor.com/blog/cybersecurity/recent-cyber-attacks-2022

[3] "An update on our security incident." Accessed: Nov. 15, 2023. [Online]. Available: https://blog.twitter.com/en_us/topics/company/2020/an-update-on-our-security-incident

[4] "The biggest cyber attacks of 2022 | BCS." Accessed: Nov. 15, 2023. [Online]. Available: https://www.bcs.org/articles-opinion-and-research/the-biggest-cyber-attacks-of-2022

[5] Nunes, E., Shakarian, P., & Simari, G. (2016, March). Toward argumentation-based cyber attribution. In *Workshops at the Thirtieth AAAI Conference on Artificial Intelligence*.

[6] Nunes, E., Shakarian, P., Simari, G. I., & Ruef, A. (2016, August). Argumentation models for cyber attribution. In *2016 IEEE/ACM International Conference on Advances in Social Networks Analysis and Mining (ASONAM)* (pp. 837-844). IEEE.

[7] Maglaras, L., Ferrag, M., Derhab, A., Mukherjee, M., Janicke, H., & Rallis, S. (2018). Threats, countermeasures and attribution of cyber attacks on critical infrastructures. *EAI Endorsed Transactions on Security and Safety*, *5*(16).

[8] Noor, U., Anwar, Z., Amjad, T., & Choo, K. K. R. (2019). A machine learning-based FinTech cyber threat attribution framework using high-level indicators of compromise. *Future Generation Computer Systems*, *96*, 227-242.

[9] Naveen, S., Puzis, R., & Angappan, K. (2020, September). Deep learning for threat actor attribution from threat reports. In *2020 4th International Conference on Computer, Communication and Signal Processing (ICCCSP)* (pp. 1-6). IEEE.

[10] Jaafar, F., Avellaneda, F., & Alikacem, E. H. (2020, August). Demystifying the cyber attribution: An exploratory study. In *2020 IEEE Intl Conf on Dependable, Autonomic and Secure Computing, Intl Conf on Pervasive Intelligence and Computing, Intl Conf on Cloud and Big Data Computing, Intl Conf on Cyber Science and Technology Congress (DASC/PiCom/CBDCom/CyberSciTech)* (pp. 35-40). IEEE.
[11] Pitropakis, N., Panaousis, E., Giannakoulias, A., Kalpakis, G., Rodriguez, R. D., & Sarigiannidis, P. (2018). An enhanced cyber attack attribution framework. In *Trust, Privacy and Security in Digital Business: 15th International Conference, TrustBus 2018, Regensburg, Germany, September 5–6, 2018, Proceedings 15* (pp. 213-228). Springer International Publishing.
[12] Haddadpajouh, H., Azmoodeh, A., Dehghantanha, A., & Parizi, R. M. (2020). MVFCC: A multi-view fuzzy consensus clustering model for malware threat attribution. *IEEE Access*, *8*, 139188-139198.
[13] Sentuna, A., Alsadoon, A., Prasad, P. W. C., Saadeh, M., & Alsadoon, O. H. (2021). A novel Enhanced Naïve Bayes Posterior Probability (ENBPP) using machine learning: Cyber threat analysis. *Neural Processing Letters*, *53*, 177-209.
[14] Warikoo, A. (2021). The triangle model for cyber threat attribution. *Journal of Cyber Security Technology*, *5*(3-4), 191-208.
[15] "What is Cyber Threat Intelligence? | NETSCOUT." Accessed: Nov. 15, 2023. [Online]. Available: https://www.netscout.com/what-is/cyber-threat-intelligence
[16] Yepes, R. "The Art of Profiling in a Digital World." Police Chief (2016).
[17] Wagner, T. D., Mahbub, K., Palomar, E., & Abdallah, A. E. (2019). Cyber threat intelligence sharing: Survey and research directions. *Computers & Security*, *87*, 101589.
[18] Wang, T., & Chow, K. P. (2019, July). Automatic tagging of cyber threat intelligence unstructured data using semantics extraction. In *2019 IEEE International Conference on Intelligence and Security Informatics (ISI)* (pp. 197-199). IEEE.
[19] Barnum, S. (2012). Standardizing cyber threat intelligence information with the structured threat information expression (stix). *Mitre Corporation*, *11*, 1-22.
[20] Bouguettaya, A., Yu, Q., Liu, X., Zhou, X., & Song, A. (2015). Efficient agglomerative hierarchical clustering. *Expert Systems with Applications*, *42*(5), 2785-2797.
[21] Federal Privacy Council (FPC), Cybersecurity Information Sharing Act of 2015 (CISA),https://www.fpc.gov/19081/ (2015).

[22] U. S. Congress, S. 754 cybersecurity information sharing act of 2015 (2015).
[23] Noor, U., Anwar, Z., & Rashid, Z. (2018, July). An association rule mining-based framework for profiling regularities in tactics techniques and procedures of cyber threat actors. In *2018 International Conference on Smart Computing and Electronic Enterprise (ICSCEE)* (pp. 1-6). IEEE.
[24] Liao, X., Yuan, K., Wang, X., Li, Z., Xing, L., & Beyah, R. (2016, October). Acing the ioc game: Toward automatic discovery and analysis of open-source cyber threat intelligence. In *Proceedings of the 2016 ACM SIGSAC conference on computer and communications security* (pp. 755-766).
[25] ENISA: Exploring the opportunities and limitations of current Threat Intelligence Platforms (2018)
[26] Kozuch I.: Cyber Threat Intelligence:How To Turn Quantity Into Quality (Apr 2018), https://www.peerlyst.com/posts/cyber-threat-intelligence how-to-turn-quantity-into-quality-itay-kozuch.
[27] Azevedo, R., Medeiros, I., & Bessani, A. (2019, August). PURE: Generating quality threat intelligence by clustering and correlating OSINT. In *2019 18th IEEE International Conference On Trust, Security And Privacy In Computing And Communications/13th IEEE International Conference On Big Data Science And Engineering (TrustCom/BigDataSE)* (pp. 483-490). IEEE.

[28] HAIL A TAXIL ,2018.[Online].Available:http://hailataxi.com/,2018.[Accessed 25-January-2018].
[29] Lu, X., Zhou, X., Wang, W., Lio, P., & Hui, P. (2020). Domain-oriented topic discovery based on features extraction and topic clustering. *IEEE Access*, *8*, 93648-93662.
[30] Rodríguez, M., Betarte, G., & Calegari, D. (2021, November). A Process Mining-based approach for Attacker Profiling. In *2021 IEEE URUCON* (pp. 425-429). IEEE.
[31] Venkatesan, S., Sugrim, S., Izmailov, R., Chiang, C. Y. J., Chadha, R., Doshi, B., ... & Buchler, N. (2018, October). On detecting manifestation of adversary characteristics. In *MILCOM 2018-2018 IEEE military communications conference (MILCOM)* (pp. 431-437). IEEE.
[32] Brynielsson, J., Franke, U., Tariq, M. A., & Varga, S. (2016, September). Using cyber defense exercises to obtain additional data for attacker profiling. In *2016 IEEE Conference on Intelligence and Security Informatics (ISI)* (pp. 37-42). IEEE.
[33] Mavroeidis, V., Hohimer, R., Casey, T., & Jesang, A. (2021, May). Threat actor type inference and characterization within cyber threat intelligence. In *2021 13th International Conference on Cyber Conflict (CyCon)* (pp. 327-352). IEEE.
[34]"ProgrammableSearchEngine."https.google.com/cse?cx=003248445720253387346:turlh5vi4xc (accessed Jul. 28, 2022).
[35] K. Bandla, "APT Notes." Jul. 26, 2022. Accessed: Jul. 27, 2022. [Online]. Available: https://github.com/kbandla/APTnotes
[36] "About FireEye | FireEye Developer Hub." Accessed: Nov. 23, 2023. [Online]. Available: https://fireeye.dev/docs/about/fireeye/
[37] "Rapid7 - Practitioner-First Cybersecurity Solutions." Accessed: Nov. 16, 2023. [Online]. Available: https://www.rapid7.com/
[38] "ClearSky Cyber Security – ClearSky Cyber Security." Accessed: Nov. 23, 2023. [Online]. Available: https://www.clearskysec.com/
[39] "AlienVault - Open Threat Exchange." Accessed: Nov. 23, 2023. [Online]. Available: https://otx.alienvault.com/
[40] "#1 in Cloud Security & Endpoint Cybersecurity | Trend Micro." Accessed: Nov. 16, 2023. [Online]. Available: https://www.trendmicro.com/en_us/business.html
[41] "Symantec Enterprise Cloud." Accessed: Nov. 23, 2023. [Online]. Available: https://www.broadcom.com/products/cybersecurity
[42] "Leader in Cybersecurity Protection & Software for the Modern Enterprises - Palo Alto Networks." Accessed: Nov. 16, 2023. [Online]. Available: https://www.paloaltonetworks.com/
[43] "DDoS Mitigation Software & Tools - Arbor DDoS Platform." Accessed: Nov. 23, 2023. [Online]. Available: https://www.netscout.com/arbor-ddos
[44] "Global Cybersecurity & Privacy." Accessed: Nov. 23, 2023. [Online]. Available: https://www.pwc.com/gx/en/issues/cybersecurity.html
[45] "Cybersecurity and Zero Trust Leader | Zscaler." Accessed: Nov. 16, 2023. [Online]. Available: https://www.zscaler.com/
[46] "CrowdStrike: Stop breaches. Drive business." Accessed: Nov. 16, 2023. [Online]. Available: https://www.crowdstrike.com/
[47] "Enterprise Cybersecurity Solutions, Services & Training | Proofpoint US." Accessed: Nov. 23, 2023. [Online]. Available: https://www.proofpoint.com/us
[48] "Kaspersky Cyber Security Solutions for Home and Business | Kaspersky." Accessed: Nov. 23, 2023. [Online]. Available: https://me-en.kaspersky.com/
[49] "Digital Shadows | Cyberseer." Accessed: Nov. 23, 2023. [Online]. Available: https://www.cyberseer.net/technologies/digital-shadows/
[50] "Novetta – Info Security Index." Accessed: Nov. 23, 2023. [Online]. Available: https://infosecindex.com/companies/novetta/

[51] “Who we are - Fox IT.” Accessed: Nov. 23, 2023. [Online]. Available: https://www.fox-it.com/nl-en/who-we-are/
[52] “Cyber Security Software and Anti-Malware | Malwarebytes.” Accessed: Nov. 23, 2023. [Online]. Available: https://www.malwarebytes.com/
[53] “Threat Intelligence Platform | Group-IB Cybersecurity Products.” Accessed: Nov. 23, 2023. [Online]. Available: https://www.group-ib.com/products/threat-intelligence/
[54] “Sites-Cisco-Site.” Accessed: Nov. 16, 2023. [Online]. Available: https://learningnetworkstore.cisco.com/specials/?src=wl_cer_sam&gclid=Cj0KCQjwkOqZBhDNARIsAACsbfJbZnFNr42-%209HesIRqHTbE1tFL0LGE2UpZLNXXVbHe_rl49ygL3zBcaAnv5EALw_wcB
[55] “LookingGlass Cyber Solutions - Crunchbase Company Profile & Funding.” Accessed: Nov. 23, 2023. [Online]. Available: https://www.crunchbase.com/organization/lookingglass-cyber-solutions
[56] “Leader in Cyber Security Solutions | Check Point Software.” Accessed: Nov. 16, 2023. [Online]. Available: https://www.checkpoint.com/
[57] “Cybersecurity Consulting, Cybersecurity Services-InterSec, Inc.” Accessed: Nov. 23, 2023. [Online]. Available: https://www.intersecinc.com/
[58] ] “RiskIQ - Crunchbase Company Profile & Funding.” Accessed: Nov. 23, 2023. [Online]. Available: https://www.crunchbase.com/organization/riskiq
[59] “Information Security Blogs - Infosecurity Magazine.” Accessed: Nov. 16, 2023. [Online]. Available: https://www.infosecurity-magazine.com/blogs/
[60] “Security Affairs - Read, think, share … Security is everyone’s responsibility.” Accessed: Nov. 23, 2023. [Online]. Available: https://securityaffairs.com/
[61] “Krebs on Security – In-depth security news and investigation.” Accessed: Nov. 23, 2023. [Online]. Available: https://krebsonsecurity.com/
[62] “WeLiveSecurity - CypherHunter.” Accessed: Nov. 23, 2023. [Online]. Available: https://www.cypherhunter.com/en/p/welivesecurity/
[63] “Securelist | Kaspersky’s threat research and reports.” Accessed: Nov. 23, 2023. [Online]. Available: https://securelist.com/
[64] “Microsoft Security Blog | Digital Security Tips and Solutions.” Accessed: Nov. 23, 2023. [Online]. Available: https://www.microsoft.com/en-us/security/blog/
[65] “Cyber Threat Intelligence and Cyber Risk Quantification Company | ThreatConnect.” Accessed: Nov. 23, 2023. [Online]. Available: https://threatconnect.com/
[66] “Threatpost | The first stop for security news.” Accessed: Nov. 16, 2023. [Online]. Available: https://threatpost.com/
[67] “CISA Certification | Certified Information Systems Auditor | ISACA.” Accessed: Nov. 16, 2023. [Online]. Available: https://www.isaca.org/credentialing/cisa?utm_source=google&utm_medium=cpc&utm_campaign=%20CertBAU&utm_content=sem_CertBAU_certification-cisa-asia-%20productgoogle&cid=sem_2006816&Appeal=sem&gclid=Cj0KCQjwkOqZBhDNARIsAACsbfIkFPnHw4Vho3_vPIzC3z-NVL0X1gfeApQe8p-%20pNVS1rNRnk9i4gkwaAn9PEALw_wcB